\documentclass[12pt,a4paper,twoside,titlepage,openright,openbib]{report}          

\usepackage{StylDokumentu}
\usepackage{Skroty}

\usepackage[toc,page]{appendix}
\usepackage{threeparttable}
\usepackage{tocbibind} 

\let\chaptertitle=\chapter
\let\introformatting=\relax
\let\outroformatting=\relax

 \makeatletter

\def\clap#1{\hbox to 0pt{\hss #1\hss}}%

\def\ligne#1{%
  \hbox to \hsize{%
    \vbox{\centering #1}}}%

\def\haut#1#2#3{%
  \hbox to \hsize{%
    \rlap{\vtop{\raggedright #1}}%
    \hss
    \clap{\vtop{\centering #2}}%
    \hss
    \llap{\vtop{\raggedleft #3}}}}%

\def\bas#1#2#3{%
  \hbox to \hsize{%
    \rlap{\vbox{\raggedright #1}}%
    \hss
    \clap{\vbox{\centering #2}}%
    \hss
    \llap{\vbox{\raggedleft #3}}}}%

\def\maketitle{%
  \thispagestyle{empty}\vbox to \vsize{%
    \haut{}{\@blurb}{}
    \vfill
    \ligne{\Large \@title}
    \vspace{5mm}
    \ligne{\Large \@author}
    \vspace{1cm}
    \vfill
    \vfill
    \bas{}{\@location, \@date}{}
    }%
  \cleardoublepage
  }

\def\date#1{\def\@date{#1}}
\def\author#1{\def\@author{#1}}
\def\title#1{\def\@title{#1}}
\def\location#1{\def\@location{#1}}
\def\blurb#1{\def\@blurb{#1}}

\makeatother
  \title{{\bf STUDY OF THE \eeeg{} DECAY USING WASA-at-COSY DETECTOR SYSTEM}}
  \author{{\small MAŁGORZATA \textsc{HODANA}\\[10em]
	THESIS ADVISOR:\\
    PROF. DR. HAB. PAWEL MOSKAL}
  }
  \date{2012}
  \location{Cracow}
  \blurb{%
    A DOCTORAL DISSERTATION\\
    PERFORMED IN THE RESEARCH CENTRE J\"{U}LICH, GERMANY\\
    AND IN THE INSTITUTE OF PHYSICS\\
    OF THE JAGIELLONIAN UNIVERSITY\\ 
    SUBMITTED TO THE FACULTY OF PHYSICS, ASTRONOMY\\
    AND APPLIED COMPUTER SCIENCE\\
    OF THE JAGIELLONIAN UNIVERSITY\\[5em]
    }%

\begin{document}
\maketitle

\cleardoublepage
\thispagestyle{empty}
\vspace*{\fill}
\begingroup
\begin{center}Dla Moich Rodziców \end{center}
\endgroup
\vspace*{\fill}

\tableofcontents	
\clearpage		

\unitlength = 1mm

\pagestyle{fancy}                         
\renewcommand{\chaptermark}[1]{\markboth{\thechapter.\ #1}{}}
\renewcommand{\sectionmark}[1]{\markright{\thesection\ \boldmath{#1}\unboldmath}}
\fancyhf{}                                
\fancyhead[RE,LO]{\nouppercase{\leftmark}}                                          

\fancyfoot[FRE,FLO]{\rightmark}
\fancyfoot[FLE,FRO]{\textbf{\thepage}}
\renewcommand{\headrulewidth}{0.5pt}      
\renewcommand{\footrulewidth}{0.5pt}      

\chaptertitle*{Introduction}
\addcontentsline{toc}{chapter}{\protect\numberline{}Introduction}
\introformatting
  
The term {\it meson} stems from the Greek word {\it mesos} which means {\it middle}. It was used by Hideki Yukawa to name a particle with a mass between the mass of an electron and a proton\footnote{Today called the $\pi$ meson}. At present, under this term, strongly interacting hadrons with a baryon number equal to zero are included. 

The elementary components of mesons are quarks and antiquarks: fermions with 
baryon number $\pm \nicefrac{1}{3}$. In the lowest energy state of quark-antiquark pairs, where the spins are anti-parallel, they create a pseudoscalar meson of negative parity and zero orbital angular momentum. Nine of these mesons, having spin equal to zero, form the pseudoscalar nonet, a member of which is the \e{} meson.

The \e{} meson was discovered in the 1960s at the Berkeley Bevatron \cite{Pevsner:1961pa} in the ${\pi^{+}d \rightarrow pp \pi^{+}\pi^{-}\pi^{0}}$ reaction. Since then, many efforts have been made to investigate and understand its inner properties. Experimentally, the mass of this meson was found to be ${547.853 \pm 0.024 \, MeV}$ \cite{pdg}. In terms of the SU(3)-flavour group, the \e{} meson can be represented as the superposition of the singlet $\eta_{1}$ and the octet $\eta_{8}$ state characterized by the mixing angle\footnote{The value of the mixing angle was determined to ${-15.5^{\circ} \pm 1.3^{\circ}}$  \cite{Gilman:1987ax,Bramon:1997va}.} $\theta$
\begin{equation}
|\eta\rangle = cos \theta |\eta_{8}\rangle - sin \theta |\eta_{1}\rangle,
\end{equation}
where
\begin{equation}
|\eta_{1}\rangle = 1/\sqrt{3}(u\bar{u} +d\bar{d} + s\bar{s}) \quad \mbox{and} \quad |\eta_{8}\rangle = 1/\sqrt{6}(u\bar{u} +d\bar{d} - 2s\bar{s}). 
\end{equation}
Unlike the octet state, the singlet state can be either a quark-antiquark combination or a pure gluon configuration. In order to describe the \e{} mass, both, the mixing and the dynamics of gluons have to be taken into account \cite{Bass:1999is,Bass:2008fr,Moskal:2002jm}.

For the \e{} meson is a short-lived, neutral particle, it is not possible to investigate its structure via the classical method of particle scattering. To learn about its quark wave function, one studies those decay processes of this meson, in which a pair of photons is produced, at least one of them being virtual. The virtual photons have a non-zero mass and convert into lepton-antilepton pairs. The squared four-momentum transferred by the virtual photon corresponds to the squared invariant mass of the created lepton-antilepton pair. Therefore, information about the quarks' spatial distribution inside the meson can be achieved from the lepton-antilepton invariant mass distributions by comparison of empirical results with predictions, based on the assumption that the meson is a point-like particle. The last can be obtained from the theory of Quantum Electrodynamics. The deviation from the expected behavior in the leptonic mass spectrum expose the inner structure of the meson. This deviation is characterized 
by a form factor. 
It is currently not possible to precisely predict the dependence of the form factor on the four-momentum transferred by the virtual photon in the framework of Quantum Chromodynamics theory. Therefore, to conduct calculations, assumptions about the dynamics of the investigated decay are needed.

The knowledge of the form factors is also important in studies of the muon anomalous magnetic moment, ${a_{\mu} = (g_{\mu}-2)/2}$, which is the most precise test of the Standard Model and, as well, may be an excellent probe of new physics. The theoretical error of calculation of $a_{\mu}$ is dominated by hadronic corrections and therefore limited by the accuracy of their determination. 
Especially, the hadronic light-by-light scattering contribution to $a_{\mu}$ includes two meson-photon-photon vertices and therefore also depends on the form factors \cite{Bijnens:1999jp}.
At present, the discrepancy between the  $a_{\mu}$ prediction based on the Standard Model and its experimental value \cite{Bennett:2006fi} is equal to ${(28.7 \pm 8.0) \cdot 10^{-10}}$  ($3.6\sigma$) \cite{Davier:2010nc}. 

The goal of this work is to extract the electromagnetic transition form factor for the \e{} meson through the study of its decay to the \eeg{} final state. 
For this aim the \e{} mesons were produced in proton-deuteron collisions in the \pdhe{} reaction. The measurement was performed using the WASA\cite{Adam:2004ch} detector system and the proton beam of the Cooler Synchrotron COSY \cite{Maier:1997ax}. The $4\pi$ geometry of the WASA-at-COSY detector and its availability to work with high luminosities makes it a suitable tool for such studies \cite{Kupsc:2009zz}. Already at the previous location of the WASA detector (see Sec.\ref{sec:wasa}) the studies on leptonic decays of the \e{} meson were performed \cite{Berlowski:2007ys} leading to the branching ratio estimate for the \eeeg{} decay\footnote{The branching ratio for the decay \eeeg{} was determined to $({7.8 \pm 0.5_{\mbox{stat}}\pm0.8_{\mbox{syst}}})\times10^3$} \cite{Berlowski:2008zz}.

In the following chapter the theoretical aspects and the results of the previous measurements are presented. 
Next, in the second chapter, the reader will find a description of the experimental setup relevant for this work. 
The third chapter contains information about analysis tools.
In the fourth chapter, the experimental conditions and methods of the track reconstruction are described.  
In the fifth chapter the analysis chain is presented and the selection criteria leading to the extraction of the \eeeg{} signal are described. 
Finally, results, a summary of this work and an outlook is given.

\outroformatting

\chaptertitle{Towards the Form Factor}
\introformatting

The \e{} meson lifetime ($5\times10^{-19}~s$) is relatively long since all its strong, electromagnetic and weak decays are forbidden in the first-order. 
The permitted weak \e{} decays in the Standard Model are expected to occur at the level of $10^{-13}$ and below \cite{Nefkens:2002sa}.
The strong decays $\eta \to 2\pi$ and $\eta \to 4\pi$ are forbidden due to P and CP invariance. The later one
also due to the small available phase space \cite{Moskal:2011ve}. 
Most of the \e{} decays detected so far, involve photon(s) and thus proceed
through electromagnetic interactions.
The first order electromagnetic decays as $\eta \to \pi^0\gamma$ or $\eta \to 2\pi^0\gamma$ break
charge conjugation invariance. The decay $\eta \to \pi^+\pi^-\gamma$ is also suppressed, because charge
conjugation conservation requires odd (and hence nonzero) angular momentum in
the $\pi^+\pi^-$ system.  
The remaining, purely hadronic \e{} decays (\eppp{} and \etp{}), violate G–parity, and have therefore small branching ratios, comparable with the one of the \gaga{} decay \cite{Ametller:2001nq}. Thus it is reasonable to assume that they are also electromagnetic. The measured life time of the \e{} meson confirms this assumption.
The most common decay modes of this meson with corresponding branching ratios are presented in Tab.\ref{t:etaBR}. With the bold font the decay being the subject of this work is marked.

\begin{figure}[!h]
  \subfigure[ In the theory of QED ]{
	\label{fig:fd_qed}
	\includegraphics[width =0.49\textwidth]{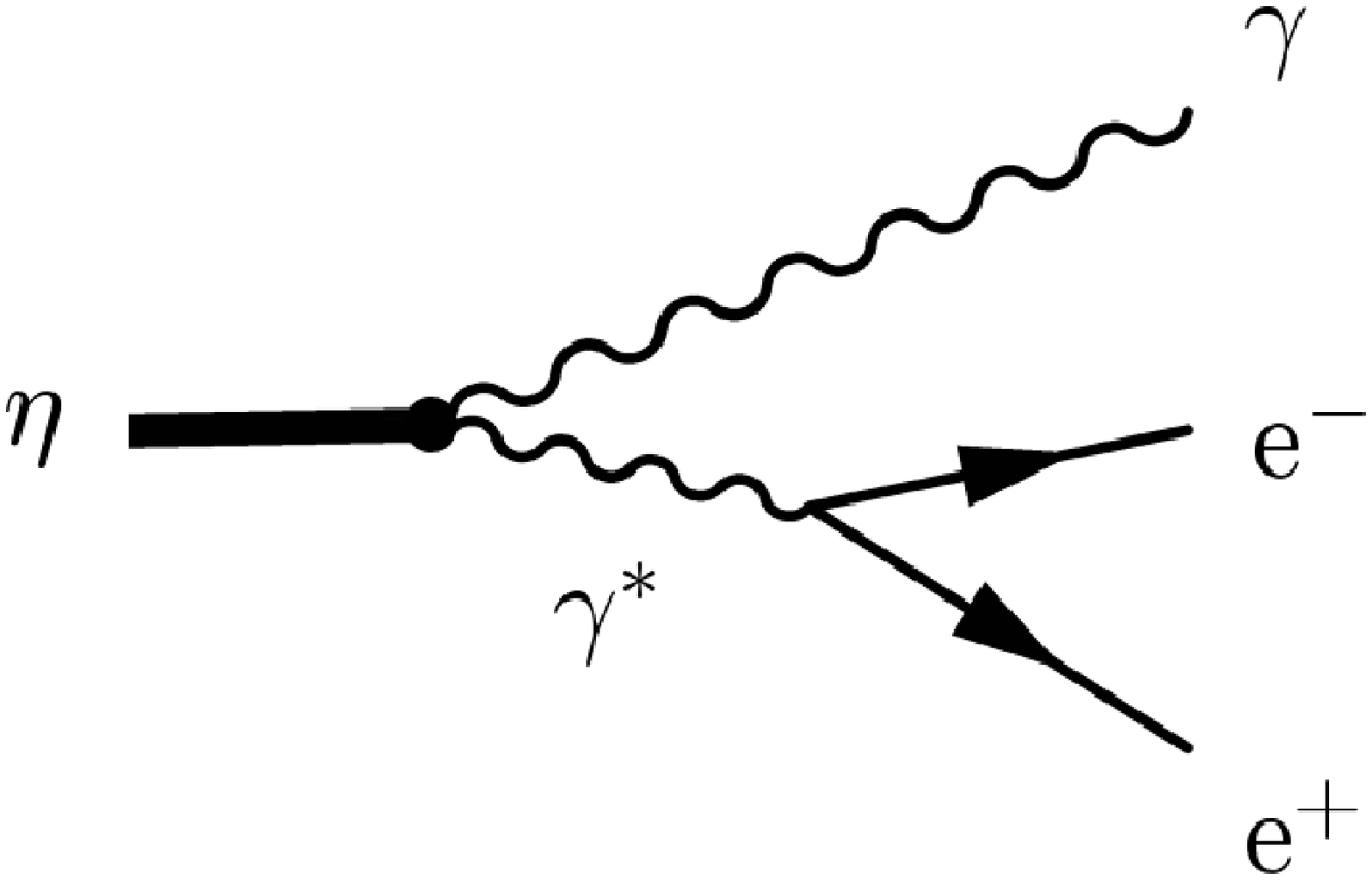}
  }
  \subfigure[In the VMD model ] {
	\label{fig:fd_vmd}
	\includegraphics[width =0.49\textwidth]{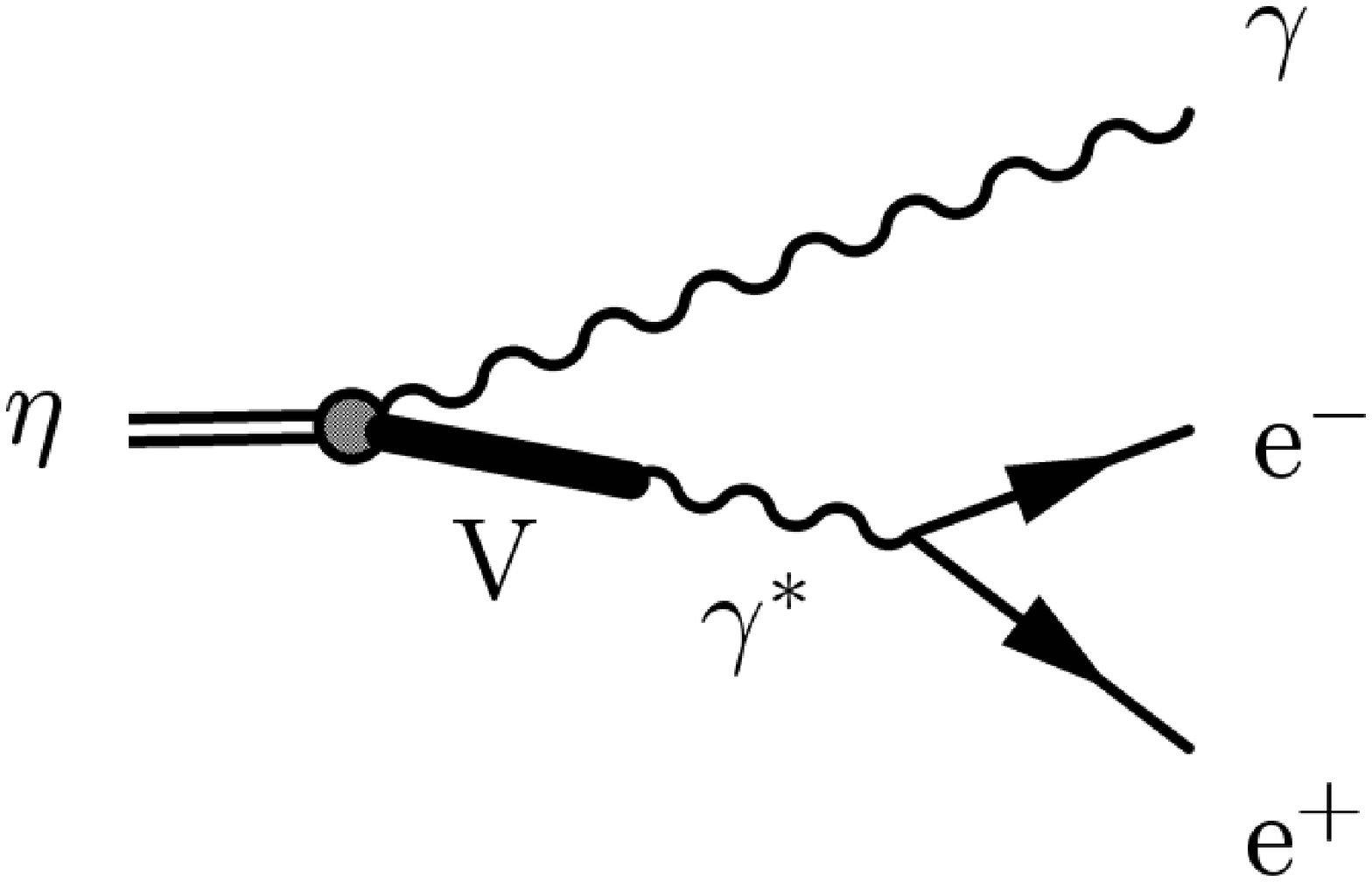}
  }
  \caption[ Feynman diagrams for \eeeg decay ]{Feynman diagrams showing the \eeeg{} conversion decay in two theoretical approaches. The letter V on the Feynman diagram for the VMD model denotes the neutral vector mesons (that is $\rho,\,\omega,\,\phi$). }
  \label{fig:fey}
  \end{figure}

The \eeeg{} decay is a single Dalitz decay, in which the meson decays in a virtual photon and a real photon as shown in the left panel of \fig{}\ref{fig:fey}.
The virtual photon converts into an \ee{} pair and therefore, this decay is also referred to as a conversion decay. The squared four-momentum of this virtual photon, \qsq{}, is equal to the mass squared of the \ee{} pair:
 \begin{equation}
  q^{2} = M_{e^{+}e^{-}}^{2} 
		= ( E_{e^{+}} + E_{e^{-}} )^{2} 
		  - ( {\bf p}_{e^{+}} + {\bf p}_{e^{-}} )^{2} > 0.
\label{eq:q2}
\end{equation}
The simple mechanism of the exchange of the virtual photon, causes the \eeeg{} decay to be very special. It makes possible, with relatively high statistics, to study the electromagnetic structure of this neutral meson.
\begin{table}[!h]
  \centering
  \begin{tabular}{ l | r@{.} c l r@{.} l }
\hline	\hline  \\ [-1.5ex]
  Decay Mode 	& \multicolumn{5}{c}{Branching Ratio [$\%$]} \\[0.3em]	\hline	\hline \\ [-1.5ex]
  \gaga{}   	&	39&30 &$\pm$ 	&0&20  	\\[0.3em]
  \etp{}   		&	32&56 &$\pm$ 	&0&23  	\\[0.3em]
  \eppp{}	 	&	22&73 &$\pm$ 	&0&28  	\\[0.3em]
  \eppg{} 		&	 4&60 &$\pm$ 	&0&16  	\\[0.3em]
  \eeeg{} 		&{\bf0}&{\bf70} 	&{\bf$\pm$} &{\bf0}&{\bf07}  	\\
  \multicolumn{1}{c|}{\vdots} & \multicolumn{5}{c}{\vdots} \\
\hline	\hline
  \end{tabular}
  \caption[Branching ratios]{Branching ratios for the most frequent decay modes of the \e{} meson \cite{pdg}. The branching ratio of the decay studied in this work is marked with bold font. }
\vspace*{0.4cm}
  \label{t:etaBR}
\end{table}

  \section{The Form Factor}\label{sec:ff}
The study of the electromagnetic structure of particles started at the beginning of the 20th century with Rutherford's scattering experiment and a discovery of the nucleon. Since then, the concept of the form factor plays an important role in scattering theory
and appears when the scatterer is not a structure-less particle.

In the case of the transition between the \e{} meson and two photons, the QED calculations give a differential cross section, in the limit where the meson is a point-like particle. The dilepton mass spectrum in the framework of the QED was firstly derived by N.M. Kroll and W. Wada in the 1950s for the  ${\pi^{0}\rightarrow e^{+}e^{-}\gamma}$ decay \cite{Kroll:1955zu}.
It can be expressed in the following form\cite{Landsberg:1986fd}:
\begin{equation}
  \frac{ d\Gamma_{l^{+}l^{-}\gamma} }{ dq^{2} \cdot \Gamma_{\gamma\gamma} }=
	\frac{ 2\alpha }{ 3\pi q^{2} }
	\sqrt{ 1- \frac{ 4M_{l}^{2} }{ q^{2} } }
	\left( 1 + \frac{ 2M_{l}^{2} }{ q^{2} } \right)
	\left( 1 - \frac{ q^{2} }{ M_{P}^{2} } \right)^{3},  
\label{eq:FFland}
\end{equation}
where ${\it l}$ stands for the lepton (mion or electron), ${\it M_{l}}$ is the lepton mass, ${\it M_{P}}$ is the mass of the pseudoscalar meson and ${\it q}^{2}{\it = M }^{2}_{\it l^{+}l^{-}} $ is the effective mass squared of the leptonic pair.

That is, however, only a rough approximation of reality, for all mesons are made up from quarks and gluons. To resolve this problem, all effects caused by their inner structure are introduced as an additional term in the decay amplitude, the transition form factor, ${\it F(q}^{2})$:
	\begin{equation}
  \frac{ d\Gamma_{l^{+}l^{-}\gamma} }{ dq^{2} }=
  \left[ \frac{d\Gamma}{dq^{2}} \right]_{pointlike} \cdot
  | F(q^{2}) |^{2}.
	\label{eq:QEDxFF}
	\end{equation}
It is a function of the square of the transferred four-momentum, \qsq{}, or equivalently, the square of the dilepton mass.
The term {\it transition} is general for processes where a neutral meson A decays into a neutral meson B and a photon. In this case namely, the form factor reflects the effects of the electromagnetic structure arising at the $A \rightarrow B$ {\bf transition} vertex.
The \eeeg{} is special at this point since the vertex incorporates only one meson, the one which is decaying. Therefore, the corresponding transition form factor defines the electromagnetic properties of this particular meson.

The kinematic limits for the transition form factor are determined by the masses of the particles participating in the 
${A \rightarrow B + \gamma^{*} \rightarrow B + {\it l^{+} + l^{-}} }$ process. We have:
	\begin{equation}
	(2M_{l})^{2} \leq q^{2} \leq (M_{A} - M_{B})^{2}.
	\label{eq:q2limit}
	\end{equation}
 
 Experimentally, the form factor can be determined by comparison of the measured ${\it l^{+}l^{-} }$ differential cross section with the QED calculation for a point-like particle. 

  \section{Vector Meson Dominance Model}\label{ssec:vmd}
In the Vector Meson Dominance Model, VMD, a photon is represented by a superposition of neutral vector meson states. It means that it fluctuates between an electromagnetic and a hadronic state. This approach is based on the equivalence of spin, parity and charge conjugation quantum numbers of neutral vector mesons and the photon. Hence, the coupling of photons to hadrons is determined by the intermediate neutral vector mesons as shown in the right panel of \fig{}\ref{fig:fey}. 

\begin{figure}[!h]
  \begin{center}
	\includegraphics[width =0.55655\textwidth]{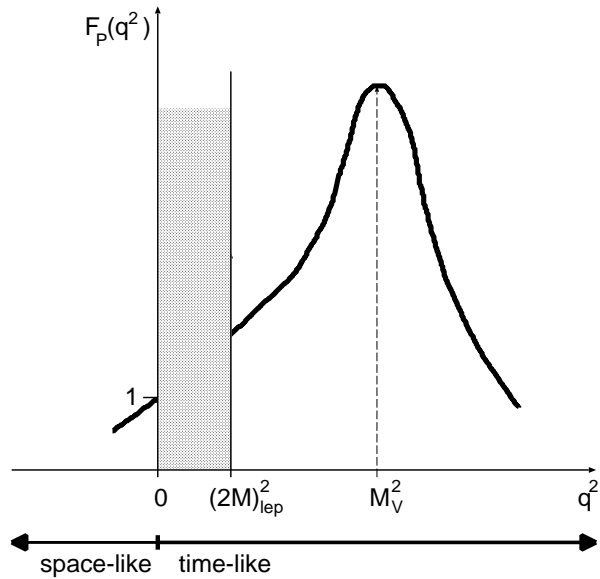}
  \caption[ ]{ The qualitative behavior of the electromagnetic transition form factor as a function of $q^{2}$. The shaded area is the region kinematically prohibited (see Eq.~\ref{eq:q2limit}). Picture is adapted from \cite{Landsberg:1986fd}. }
  \label{fig:Fq_q}
  \end{center}
  \end{figure}

According to the isobar model which describes resonances by the Breit-Wigner formula \cite{McGowan:1995eb}, the form factor in the VMD model takes the following form 

	\begin{equation}
		F(q^{2}) = \frac{1}{\sum_{V}\frac{g_{V\eta\gamma}}{g_{V\gamma}}}\sum_{V}\frac{g_{V\eta\gamma}}{g_{V\gamma}}\frac{ M_{V}^{2} }{ M_{V}^{2} - q^{2} - iM_{V}\Gamma_{V}(q^{2}) }
				\simeq \frac{1}{1- \frac{q^{2}}{ M_{V}^{2}} },
	\label{eq:vdm_ff2}
	\end{equation}
where the summation index $V$ runs over the $\rho$, $\omega$ and $\phi$ vector mesons with couplings to the photon ($g_{V\gamma}$) and to the $\eta\gamma$ ($g_{V\eta\gamma}$), and the $\Gamma_{V}({ \it q}^{2})$ corresponds to the total vector meson width \cite{Budnev:1979fz, Landsberg:1986fd, Soyeur:1996cp, Bramon:1981sw, Kaptari:2008nb}. 
The charge distribution inside of the meson is given by the Fourier transform of Eq.~\ref{eq:vdm_ff2}.
The qualitative behavior of the electromagnetic transition form factor in the range of \qsq{} is depicted in \fig{}\ref{fig:Fq_q}. 
It should be noted here that study of the electromagnetic transition form factor in ${A \rightarrow B  {\it l^{+}  l^{-}} }$ decays is limited to the time-like region, where the squared four-momentum of the virtual photon, \qsq{} is greater than $(2M_l)^2$. In this case the mechanism of photon-hadron interaction is especially well pronounced since the squared four-momentum, \qsq{}, approaches the squared mass of the vector meson (${\it q}^{2} \approx {\it M_{V}^{\text 2}}$). The virtual meson reaches its mass shell, i.e. becomes real and then decays to a lepton pair. It results in a strong resonance enhancement of the form factor of a meson. Then, at ${\it q}^{2} > {\it M_{V}^{\text 2}}$, the form factor begins to diminish (see \fig{}\ref{fig:Fq_q}).

The theoretical uncertainty of the VMD form factor in the \eeeg{} decay was estimated to be on the level of $5 - 10$ percent \cite{Landsberg:1986fd}.

\section{Previous Experiments}
The observed \qsq{} distribution is fitted using a single-pole formula with 
parameter $\Lambda_{P}$ related to the mass of the vector meson \cite{Ametller:1991jv}:
	\begin{equation}
		F_{P}(q^{2}) = \left( 1 - \frac{q^{2}}{\Lambda_{P}^{2}} \right)^{-1} \equiv \left( 1 - b_{P}^{2}q^{2} \right)^{-1}.
	\label{eq:vdm_ff}
	\end{equation}
In the limit of small \qsq{}, the form factor slope parameter, ${\it b_{P} } \equiv 1/\Lambda_{P}^{2}$ is associated with the size of the pseudoscalar meson,
${\it b_{P} }= \langle r_{P}^{2} \rangle/6 $ \cite{Landsberg:1986fd}.

The currently available data on the form factor measurements, are gathered in Tab.~\ref{t:FF}. 
\begin{table}[!h]
\begin{center}
\begin{threeparttable}
{\footnotesize
  \begin{tabular}{l|r|c|c}
  \hline \hline	
           &              & Slope parameter            & Characteristic mass\\
Experiment & $N_{events}$ & $\Lambda^{-2}$ [$GeV^{-2}$]&  $\Lambda$ [$GeV$]\\

  \hline\hline &&&\\[-0.8em]
 Rutherford& \multirow{3}{*}{50\tnote{a}}& \multirow{3}{*}{$-0.7 \pm 1.5$}& \multirow{3}{*}{$-$}\\
  Laboratory\cite{Jane:1975hn} &		&					& \\[0.2em]
  \hline &&&\\[-0.8em]
  SND\cite{Achasov:2000ne}    & 109\tnote{a}		   & $1.6 \pm 2.0$	&	$-$   \\[0.2em]
  \hline  &&&\\[-0.8em]
  CB/TAPS\cite{Berghauser:2011zz}    & 1345\tnote{a}	& $1.92 \pm 0.35 \pm 0.13$	&	$0.720 \pm 0.06 \pm 0.05$   \\[0.2em]
  \hline  &&&\\[-0.8em]
  Lepton-G\cite{Dzhelyadin:1980kh}& 600\tnote{b}	& $1.9 \pm 0.4$	& $0.724 \pm 0.076$  \\[0.2em]
  \hline  &&&\\[-0.8em]
  NA-60\cite{Arnaldi:2009wb}& 9000\tnote{b}	& $1.95 \pm 0.17 \pm 0.05$	& $0.716 \pm 0.031 \pm 0.009$  \\[0.2em]
  \hline &&&\\[-0.8em]
  VMD	\cite{Ametller:1991jv}&	\multicolumn{1}{c|}{$-$}& $ 1.78$			& 		$0.75$	\\ 
  \hline  \hline

  \end{tabular}
}
  \begin{tablenotes}
  \begin{footnotesize}
  \item [a] \ee{} measured
  \item [b] $\mu^{+}\mu^{-}$ measured
  \end{footnotesize}
  \end{tablenotes}
  \end{threeparttable}
  \caption[Published results on form factor slope]{ Published results of measurements of the $\eta$ transition form factor with statistics of at least 50 events. In the last row, the VMD model prediction is shown. }
  \label{t:FF}
\end{center}
\end{table}
  \begin{figure}[!h]
\centering
  \subfigure[ Result of the CB/TAPS measurement of the \eeeg{} decay \cite{Berghauser:2011zz} ]{
	\label{fig:eeg}
	\includegraphics[width =0.4\textwidth, height=0.47\textwidth]{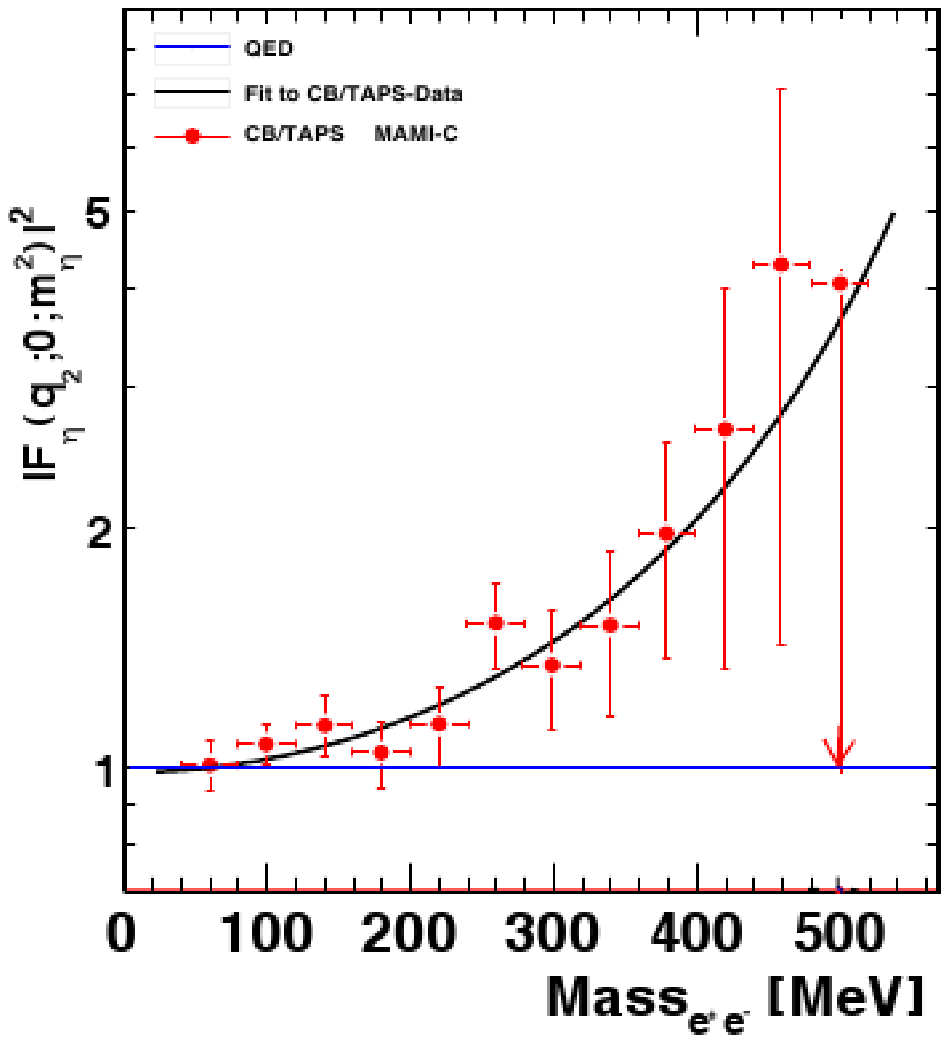}
  }
  \subfigure[ Results of the Lepton-G (open circles) and the NA60 (triangles) measure-\newline ments of the $\eta \rightarrow \mu^{+}\mu^{-}\gamma$ decay. The solid and dashed-dotted lines are fits to the NA60 data while the dotted line is the VMD model prediction. Picture is taken from \cite{Arnaldi:2009wb}] {
	\label{fig:mmg}
	\includegraphics[width =0.45\textwidth, height=0.49\textwidth]{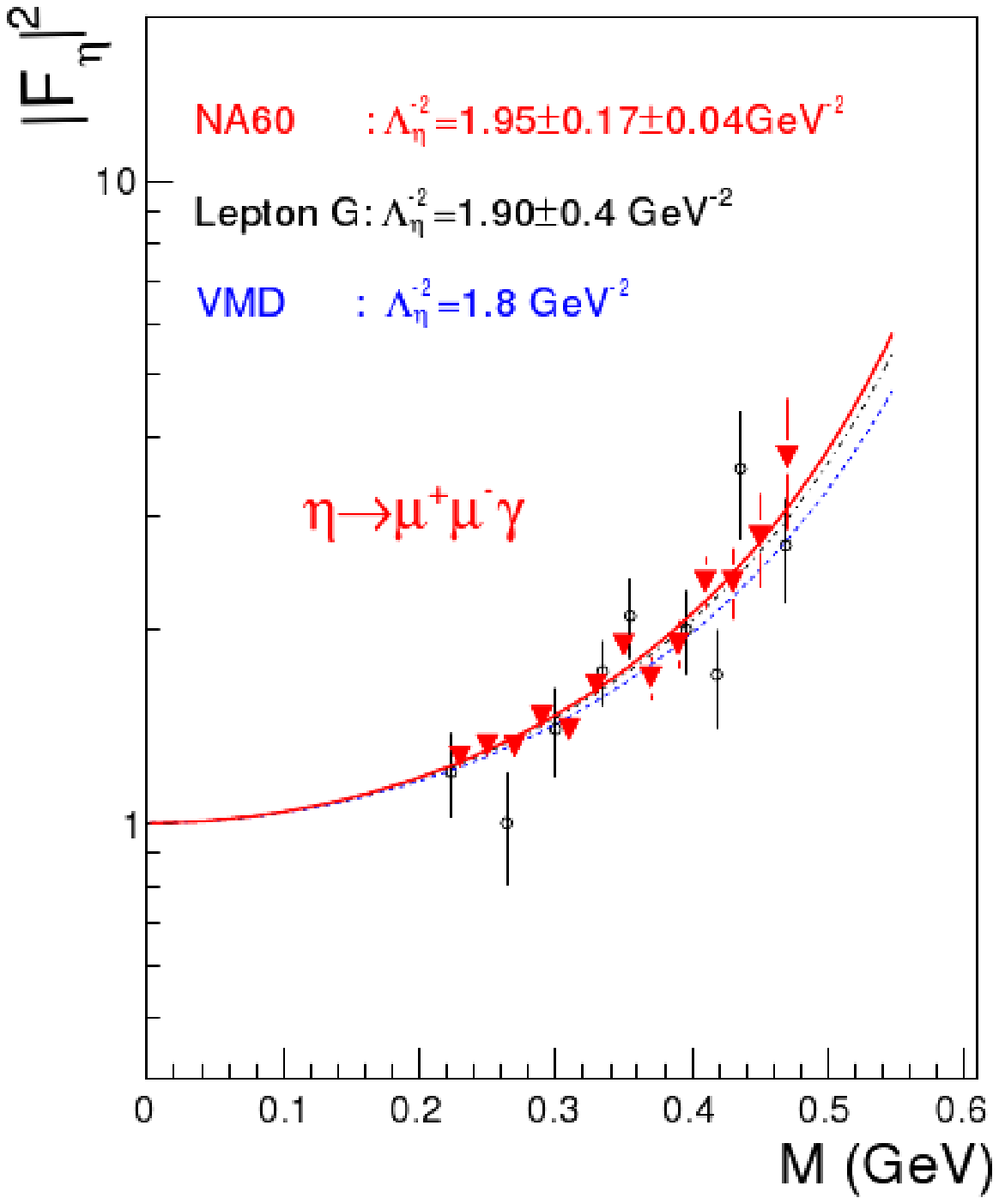}
  }
  \caption[ Recent measurements of the $F_{P}(q^{2})$]{ The squared transition form factor of the \e{} meson as a function of the lepton pair mass, obtained in the CB/TAPS, Lepton-G and NA60 experiments for \eeeg{} and $\eta \rightarrow \mu^{+}\mu^{-}\gamma$ decays. }
\vspace*{9cm}
  \label{fig:dotExp}
  \end{figure}
Nearly 9000 of $\eta \rightarrow \mu^{+}\mu^{-}(\gamma)$ events were reconstructed in the NA60 experiment using data taken in 2003 for In-In collisions \cite{Arnaldi:2009wb}. It has to be noted here that in this experiment, the photon was not measured and therefore the meson was not fully reconstructed but obtained from unfolding the $\mu^{+}\mu^{-}$ mass spectrum.
The same decay channel was studied with the Lepton-G experiment \cite{Dzhelyadin:1980kh}. However, since those measurements concern muons, the lower kinematic limit of the squared four-momentum transfer is much higher than in case of electrons and there is no information below ${ {\it q}^{2}{\it \equiv M = ( }2 { \it M_{\mu}) }^{2} = (0.23~GeV/c^{2})^{2}}$ (see Eq.\ref{eq:q2limit}). This can be seen in \fig{}\ref{fig:mmg} where both results are shown.

As regards the \eeeg{} decay, in two experiments the number of reconstructed \eeeg{} mesons exceeded the level of hundred events. One measurement was carried out with the SND detector on the VEPP-2M collider in Novosibirsk, in the years 1996 and 1998 \cite{Achasov:2000ne}. Only 109 of \eeeg{} events were reconstructed, resulting in the determination of the form factor slope ${\it b_{P} }= (1.6 \pm 2.0)~GeV^{-2}$. The most recent result, obtained from the analysis carried out in parallel to this work, concerns the measurement performed at the MAMI-C accelerator using the combined Crystal Ball (CB) and TAPS detectors. Although there is no magnetic field available for particle tracking and hence, the sign of the charged particle cannot be determined, the $1345$ of \eeeg{} mesons produced in the $\gamma p \rightarrow p \eta \rightarrow p e^+ e^- \gamma $ reaction were reconstructed and a value of ${\it b_{P} }= (1.92 \pm 0.35 \pm 0.13)~GeV^{-2}$ has been determined \cite{Berghauser:2011zz}. The form factor, extracted from CB/TAPS data, together with the single-pole approximation, is shown in \fig{}\ref{fig:eeg}.

\outroformatting

\chaptertitle{Experimental Setup}
\introformatting

  \section{The COSY Storage Ring}\label{sec:cosy}
The data used for the analysis in this work, were taken at the Research Center J\"{u}lich, in Germany. The measurement was carried out using the WASA\footnote{Wide Angle Shower Apparatus}\cite{Adam:2004ch} detector installed at the COoler SYnchrotron COSY \cite{Maier:1997zj, Maier:1997ax}. The COSY ring is a 184 m circumference accelerator which provides beams of protons and deuterons (also polarized), in the momentum range from $0.3$~GeV/c to $3.7$~GeV/c \cite{Lorentz:2011zz}. 

\begin{figure}[]
  \begin{center}
	\includegraphics[width =\textwidth]{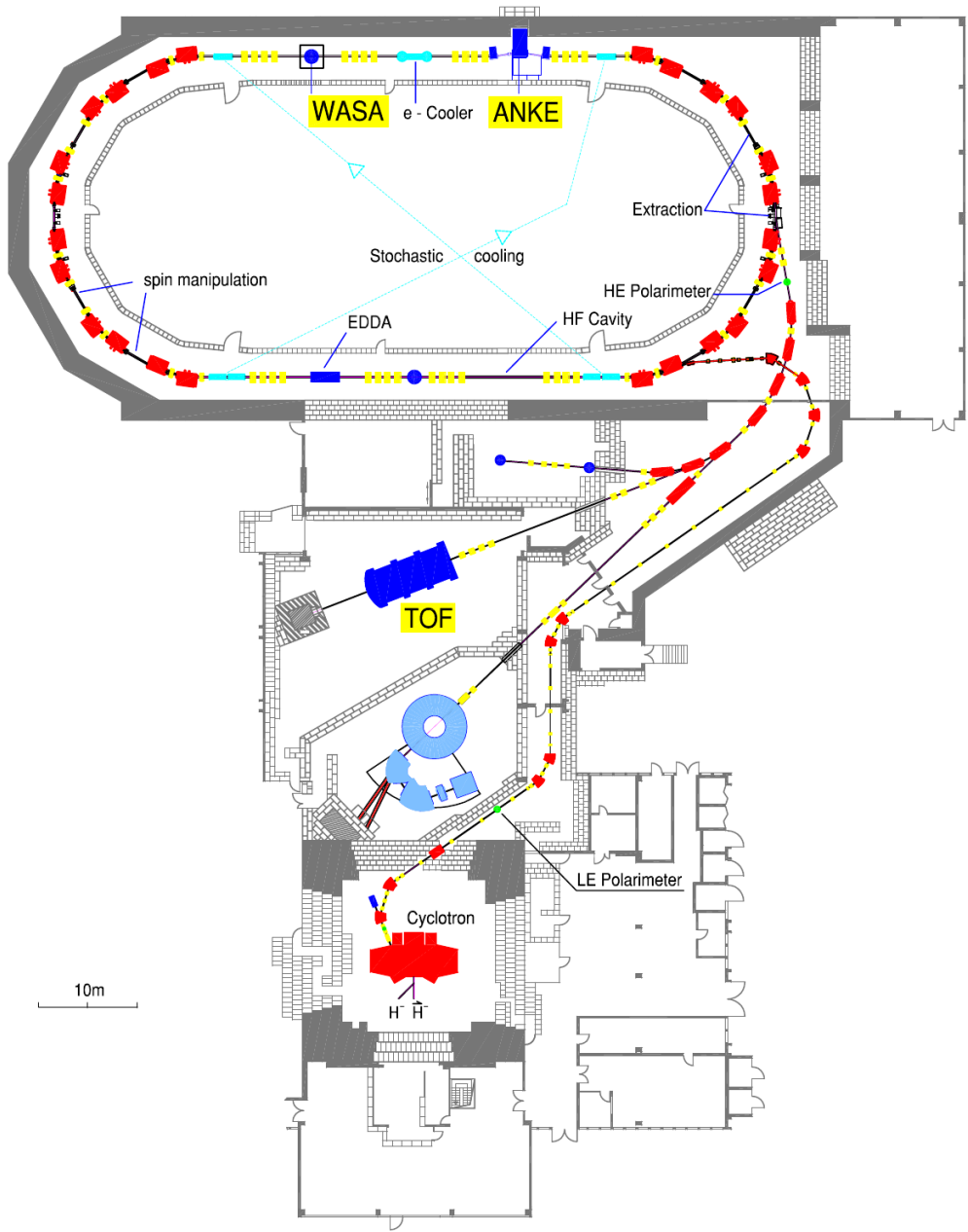}
  \caption[ The COSY accelerator ]{ Schematic view of the COSY facility. The WASA detector \cite{Adam:2004ch} is situated in the COSY ring, next to the ANKE experiment \cite{Barsov:2001xj} and opposite to the EDDA experiment \cite{Schwarz:1998xf}. Further down in the picture the location of the TOF detector \cite{tof} is marked. Below it, there is the JULIC Cyclotron.}
  \label{fig:cosy}
  \end{center}
  \end{figure}

A floor plan of the COSY facility is shown in \fig{}\ref{fig:cosy}. The JULIC cyclotron delivers either $H^{-}$ or $D^{-}$ ions pre-accelerated up to the momentum of $0.3$~GeV/c. 
Up to $10^{11}$ particles injected by the cyclotron can be then stored in the COSY ring which, in the case of internal pellet targets, allows for luminosities of $10^{32}~cm^{-2}~s^{-1}$.

For reducing the beam momentum spread and to compensate the mean energy loss, three methods of beam cooling are available. Electron cooling is applied in case of protons with momenta up to $0.6~GeV$. To obtain well focused beams of particles with momenta above $1.5$~GeV/c, stochastic cooling is used \cite{Prasuhn:2000eu}. Stochastic cooling allows to achieve a beam momentum resolution, $\Delta p/p$, below $10^{-4}$. For experiments with thick targets of more than $10^{15}~atoms/cm^{2}$, also the barrier bucket cavity method is applied \cite{grs08}.

  \section{The WASA detector}\label{sec:wasa}

The WASA detector is one of the internal detectors of COSY. Up to 2005, it was operating at the CELSIUS storage ring in Sweden \cite{Bargholtz:2008ze}. After the shutdown of the accelerator, the detector was moved to J\"{u}lich and mounted in COSY where it has been taking data since 2007 \cite{psbw06, Adolph:2008vn, Adlarson:2011bh, Adlarson:2011xb, Adlarson:2011tt}. 

The physics program aims at pursuing the knowledge of hadron structure and symmetry breaking in QCD, in the sector of the up, down and strange quarks \cite{Adam:2004ch}.
The main interest is to study rare \e{}, $\eta'$ and $\omega$ decays, which provides an understanding of structure of matter and hadron dynamics.
  \begin{figure}[!h]
  \begin{center}
	\includegraphics[width =0.48\textwidth, angle=-90]{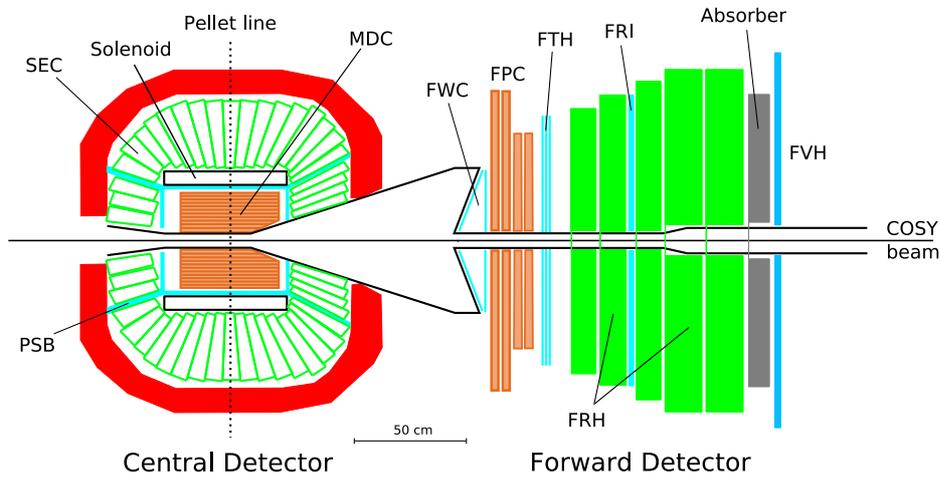}
  \caption[ Overview of the WASA detector ]{  Overview of the WASA detector. The names abbreviations of the detector elements are explained in the text. See also the List of Acronyms.}
  \label{fig:wasa}
  \end{center}
  \end{figure}
For that purpose, WASA is equipped with a set of plastic scintillators, straw tubes and an electromagnetic calorimeter which covers almost full $4\pi$ solid angle in the laboratory frame. The arrangement of the detector components is optimized to tag a reaction via the recoil particle going to the forward part of the detector, and to register meson decay products in the central part of the detector. Therefore, the two parts of WASA  are called forward (FD) and central (CD) detector, respectively.

The overview of the detector's layout is shown in \fig{}\ref{fig:wasa}.  

	\subsection{Central Detector}\label{sec:cd}
The central detector surrounds the interaction point. It is designed for the detection of particles being the products of the decays of produced mesons. The straw tube detector (Mini Drift Chamber - MDC) together with the solenoid, serves as a source of information about charged particles' momenta. The plastic scintillators (Plastic Scintillator Barrel - PSB) and a calorimeter (SEC) deliver information about energy deposited by particles. The calorimeter is used for identification of photons as well.
	  \subsubsection{Mini Drift Chamber}\label{ssec:mdc}
\begin{figure}[]
{\centering
 \subfigure[The MDC inside the Al-Be cylinder]{
	\label{fig:comp_mdc}
	\includegraphics[width =0.45\textwidth]{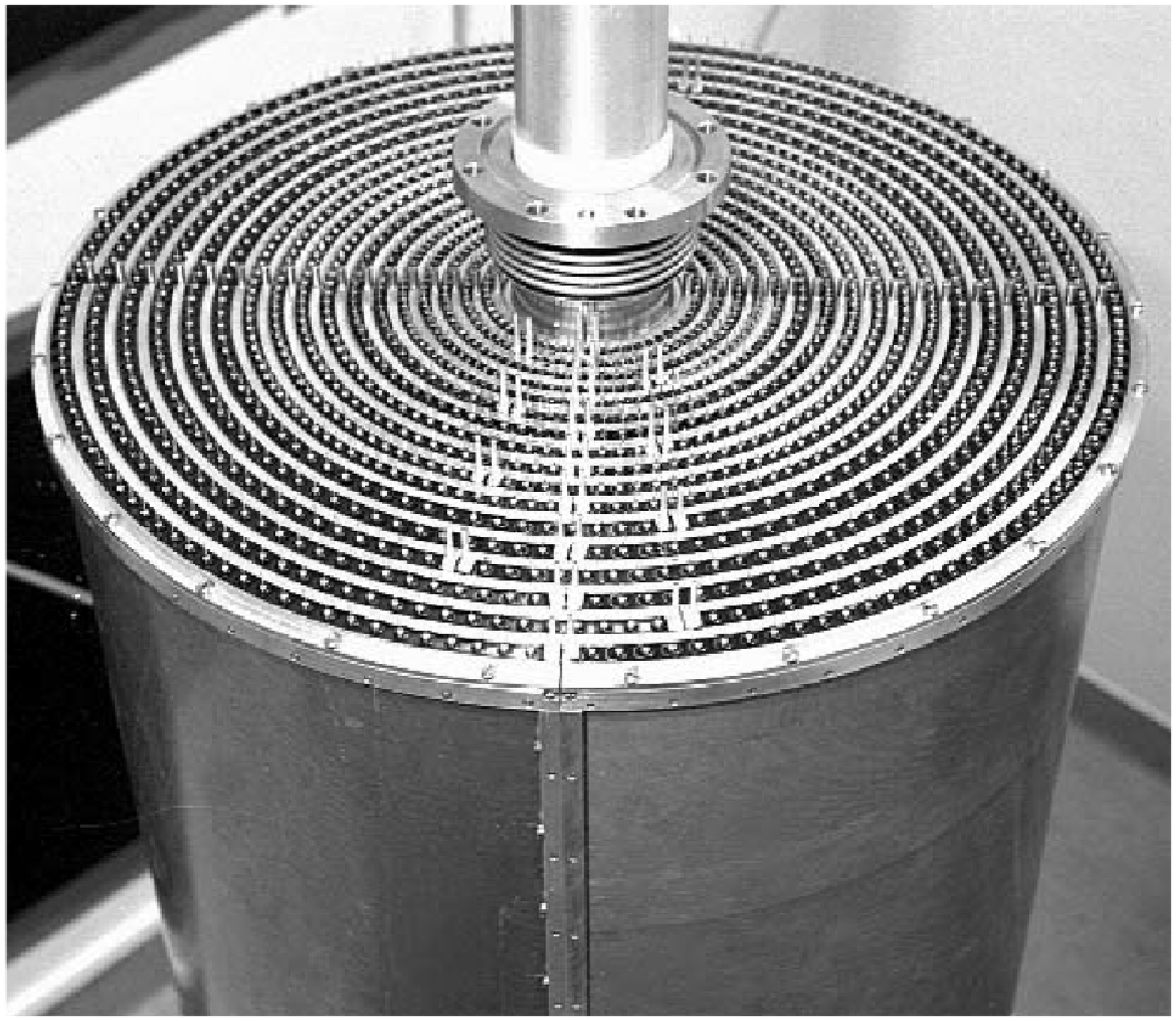}
  }
\\
 \subfigure[Schematics of the forward (\underline {left}), central (\underline {middle}) and backward (\underline {right}) parts of the PSB]{
	\label{fig:comp_psb}
	\includegraphics[width =\textwidth]{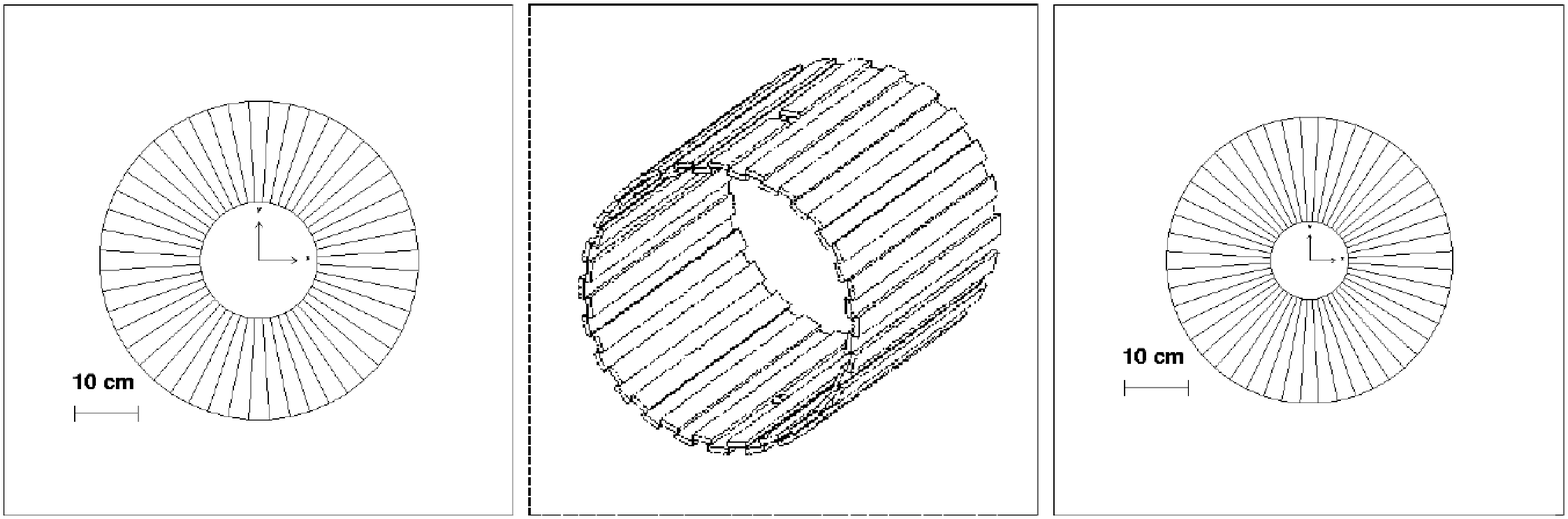}
  }
}
\\
 \subfigure[Schematic view of the SEC]{
	\label{fig:comp_sec}
	\includegraphics[width =0.5\textwidth]{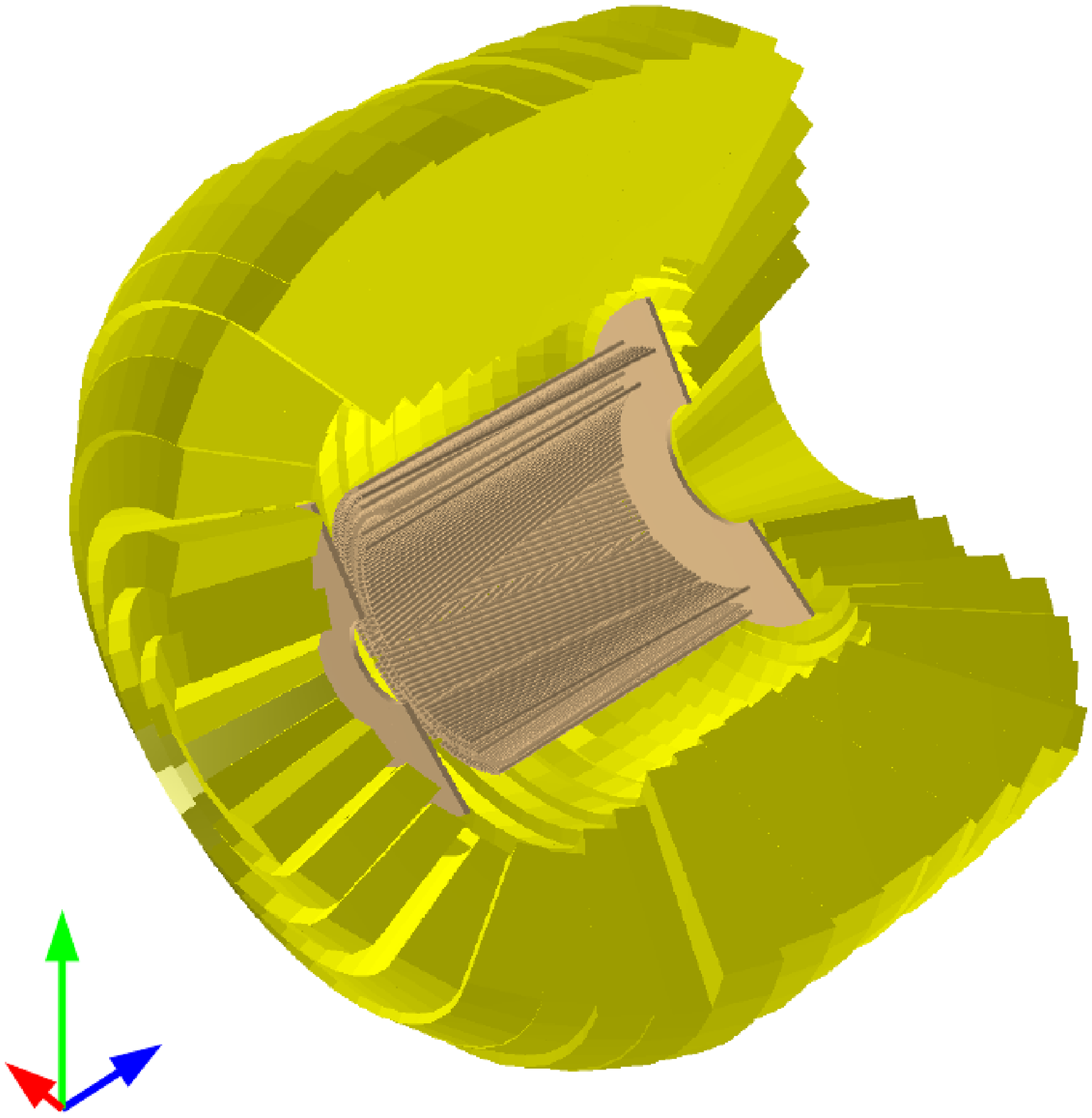}
  }
 \subfigure[Angular coverage of the SEC. The number of crystals in each ring is indicated by numbers above the picture ]{
	\label{fig:comp_sec_py}
	\includegraphics[width =0.5\textwidth]{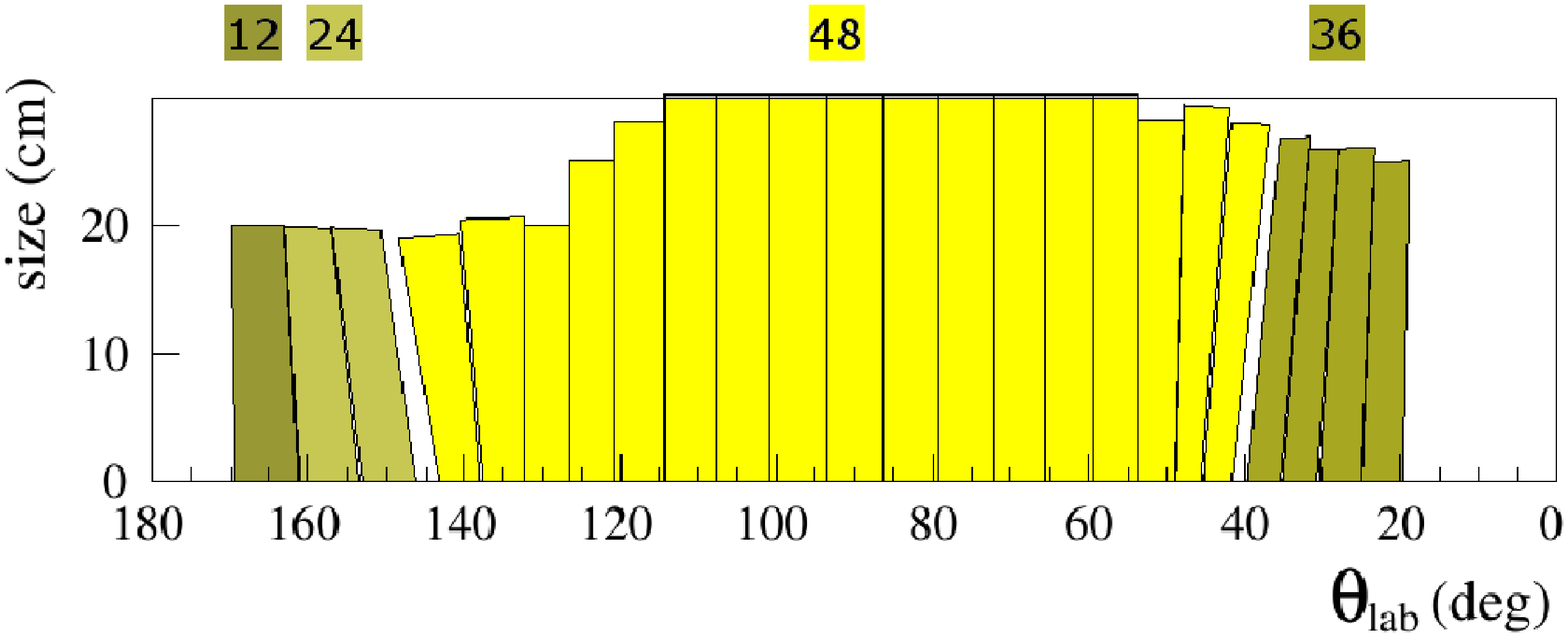}
  }
  \caption[Components of the WASA \cd{}]{Components of the \cd{} of the WASA setup.}
  \label{fig:cd_comp}
 \end{figure}

The MDC is placed around the beam pipe, inside the solenoid, and covers scattering angles from $24$ to $259$ degrees. It consists of 1738 straws arranged in 17 layers (see \fig{}\ref{fig:comp_mdc}). In nine of the layers the straws are aligned parallel to the beam axis while in the remaining layers they are placed with a small skew angle of $2$-$3$ degrees (see also \cite{jacphd}). 

Straws are made of $25~\mu m$ Mylar foil aluminized on the inner side. Sense wires are made of gold plated tungsten and have a diameter of $20~\mu m$. Thanks to the potential difference between the wire (kept at high voltage) and the grounded walls, electrons start to move towards the wire while ions in the direction of the wall. The resulting electrical current indicates that a particle was passing the tube.   

The gas mixture used to fill the tubes consist of $80\%$ argon and $20\%$ ethane. It was chosen so, that any particle passing through it causes its ionization. It provides linear correlation of the drift time to the drift distance with a relatively low operating voltage. 

The MDC operates in a magnetic field of the superconducting solenoid (SEC), under the action of which, the particles trajectories undergo bending. The MDC allows to extract particle trajectory parameters, the angles, momenta and the vertex position.

	  \subsubsection{Plastic Scintillator Barrel}\label{ssec:psb}
The three parts of the PSB surround the drift chamber (see \fig{}\ref{fig:comp_psb}). Each of the two end-cups is composed of 48 trapezoidal elements with a hole in the center designated for the beam pipe. The central part is comprised of 52 bars forming two layers with a small overlap between elements.  

Together with the MDC it is used for the identification of charged particles by the $\Delta E -p$ (energy-momentum) method. 

	  \subsubsection{Superconducting Solenoid}\label{ssec:sol}
The Superconducting Solenoid encloses the volume of the MDC and the PSB. The magnetic field produced by it has a maximum of $1.3$~T. It is cooled with liquid helium. It serves for the calculation of momenta of charged particles registered in the MDC.  
A detailed description of the solenoid can be found in \cite{rub99}.
	  \subsubsection{Scintillating Electromagnetic Calorimeter}\label{ssec:sec}
The \se{} (SEC) is the outermost component of the central detector. It consists of 1012 sodium-doped CsI scintillating crystals covering a polar angle from $20$ to $169$ degrees (see \fig{}\ref{fig:comp_sec_py}). The crystals are arranged in 24 layers perpendicular to the beam pipe in the center part and with the inclination growing with the distance from the center. The length of the crystals varies based on their location. The shortest ones, of $20$~cm length, are placed in the backward part, the ones of $25$~cm length in the forward part and the longest ones ($30$~cm) compose the central part (see \fig{}\ref{fig:comp_sec}). 

The calorimeter energy resolution for photons is given by $\frac{\sigma_{E}}{E} = \frac{5\%}{\sqrt{E/GeV}}$. The angular resolution is limited by the crystal size to $5$ degrees in the scattering angle \cite{Bargholtz:2008ze}.

A detailed description of the calorimeter can be found e.g. in Ref.~\cite{koc04}.

	\subsection{Forward Detector}\label{sec:fd}
The forward detector is used for the detection of charged recoil particles like protons, deuterons and helium ions. It consists of a set of plastic scintillators and a straw tube tracker, so that it allows for particle identification and four-momentum reconstruction of recoils, scattered in the range of the polar angle from $3$ to $18$ degrees. 

\begin{figure}[]
\begin{center}
 \subfigure[The Forward Window Counter, exploded view]{
	\label{fig:comp_fwc}
	\includegraphics[width =0.7\textwidth, height=6cm]{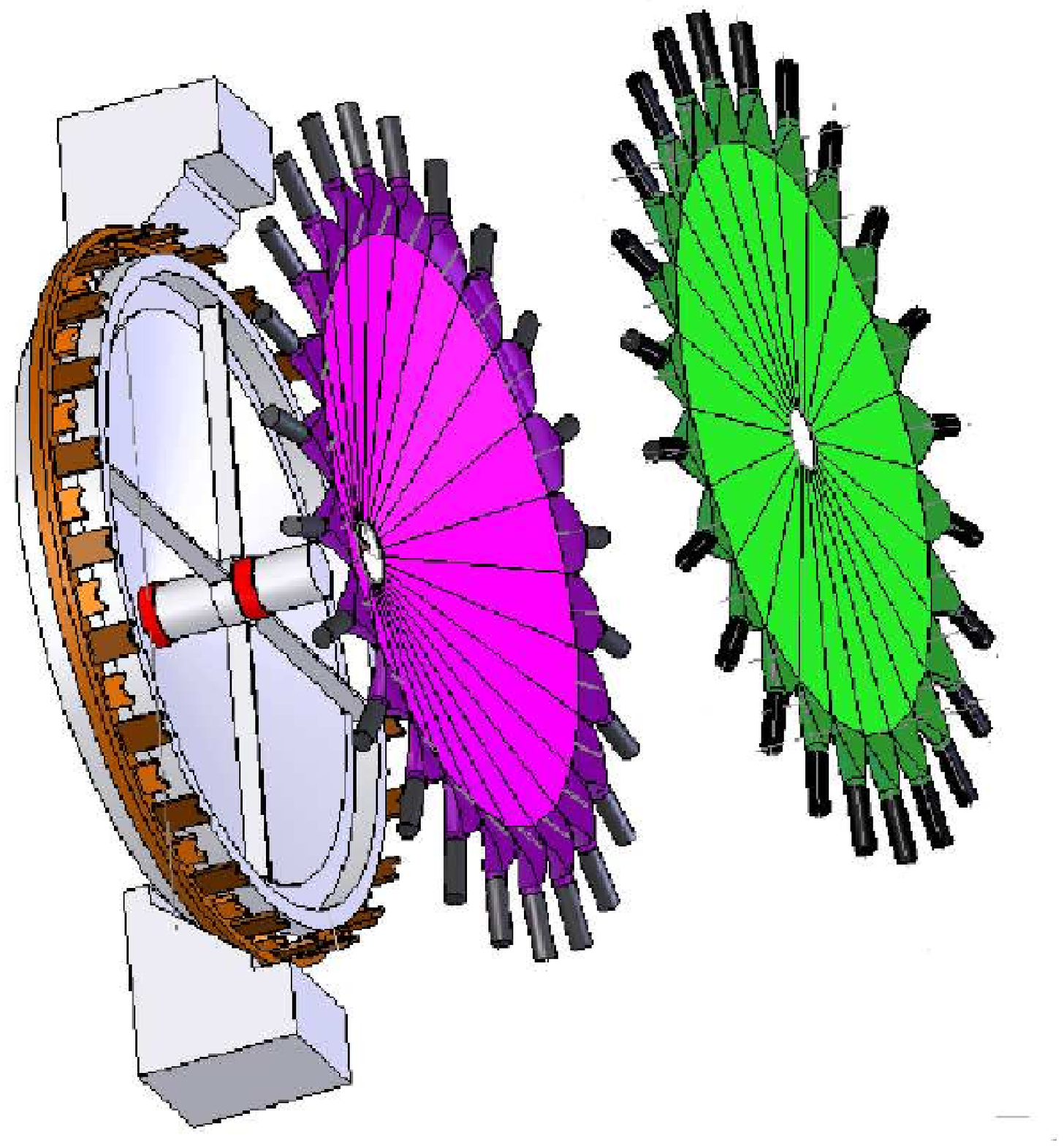}
  }
\\
 \subfigure[Schematic view on the arrangement of the \fpc{} modules ]{
	\label{fig:comp_fpc}
	\includegraphics[width =0.7\textwidth]{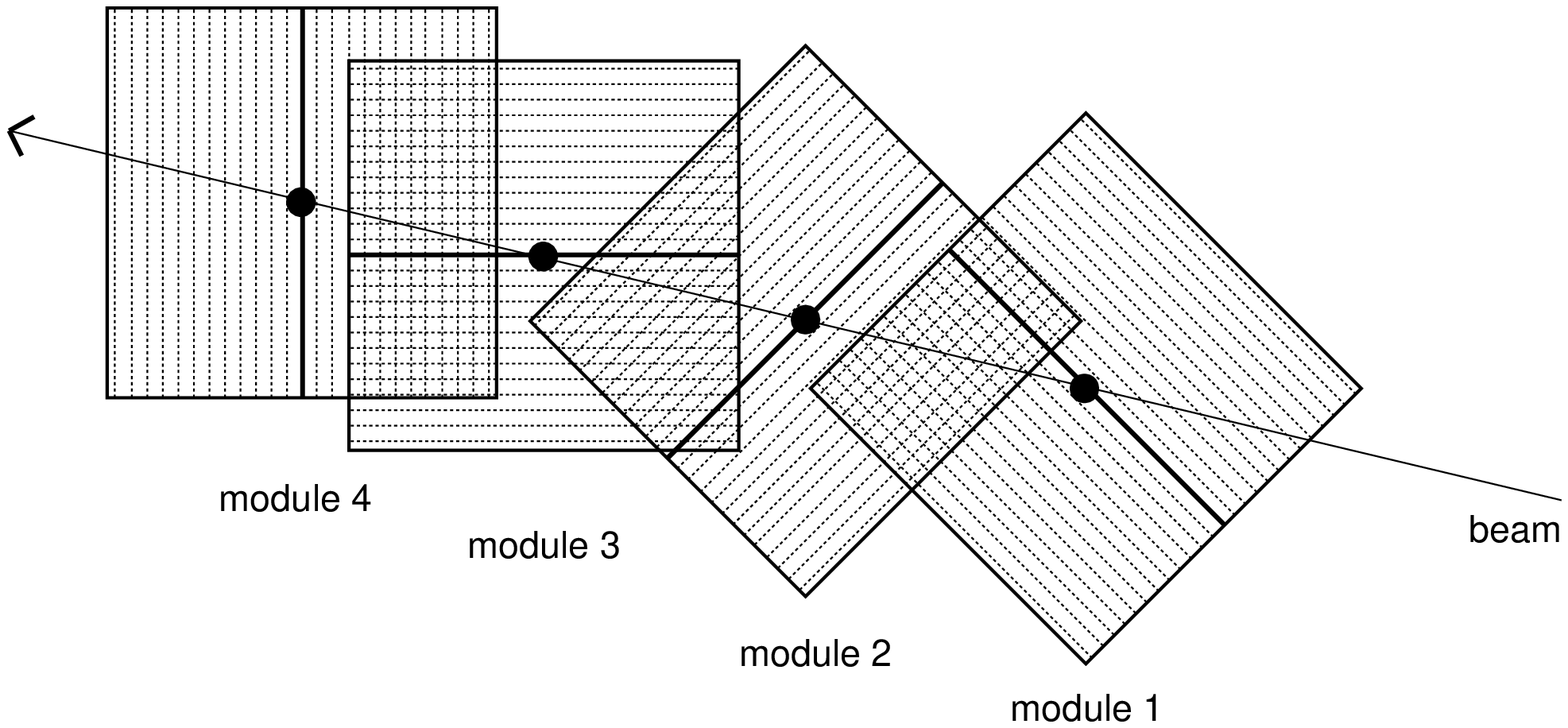}
  }
\\
 \subfigure[ \underline {Left}: three layers of the Forward Trigger Hodoscope hit by two particles. \underline {Right}: projection on the xy-plane, intersections of struck elements define pixels ]{
	\label{fig:comp_fth}
	\includegraphics[width =0.7\textwidth]{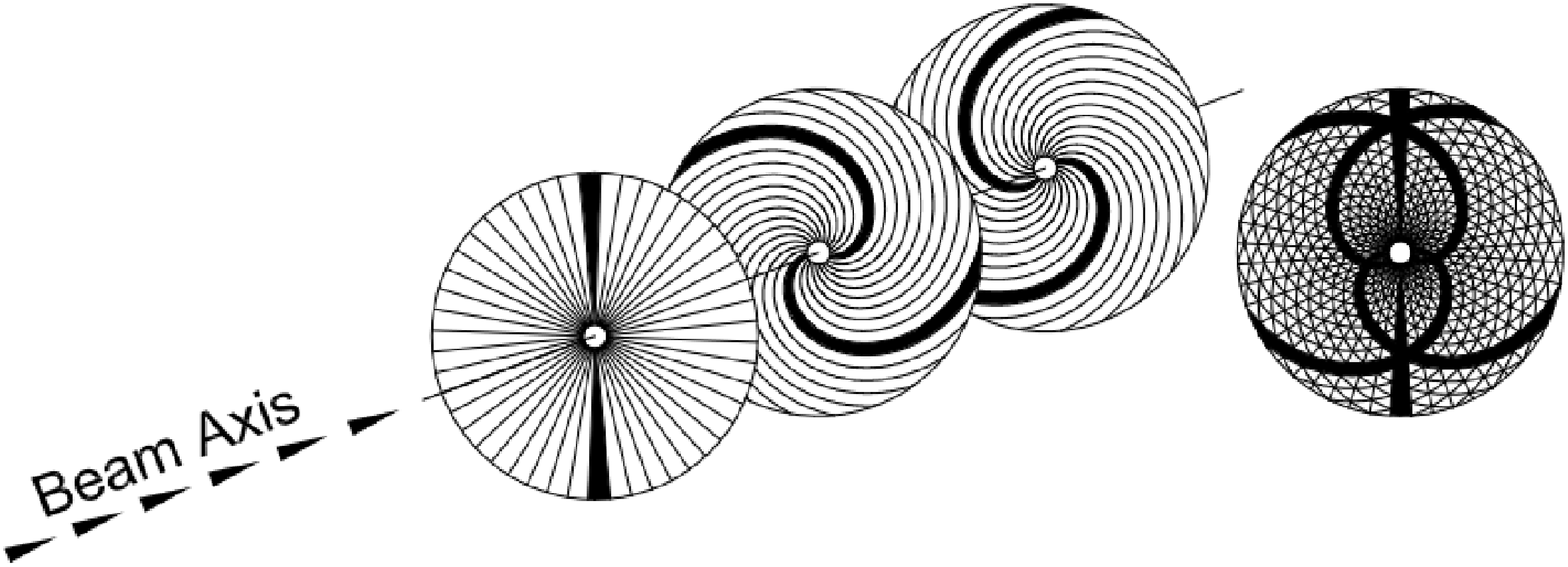}
  }
\end{center}
  \caption[Components of the WASA \fd{}]{Components of the \fd{} of WASA.}
  \label{fig:fd_comp}
 \end{figure}

	  \subsubsection{Forward Window Counter}\label{ssec:fwc}
The Forward Window Counter, FWC, is part of the forward detector placed closest to the interaction region. These are $48$ plastic scintillators arranged in two layers shifted by half an element with respect to each other.
In addition, the first layer is placed with a small inclination in order to mimic the shape of the exit window of the central detector (see \fig{}\ref{fig:comp_fwc} and \cite{pri11}). 
 
The FWC is one of the detectors involved in the identification of helium ions via the $\Delta E -E$ method. It is also a very important component of the trigger in experiments with \he{} as a recoil particle, in which case, the selection of events is based on the fact that the \he{} is characterized by a bigger energy loss in the FWC in comparison with other, lighter particles.
	  \subsubsection{Forward Proportional Chamber}\label{ssec:fpc}
Directly after the FWC, the Forward Proportional Chamber is located. It is made up of $1952$ straws grouped into sixteen layers, every of $92 \times 92~cm^{2}$ sensitive area. 
Each subsequent four layers correspond to one module. Straws in layers of each module are shifted by $\pm$ tube radius in respect to the straws of neighboring layers.
The modules are rotated with respect to the x-axis by $315$, $45$, $0$ and $90$ degrees, respectively, as shown in \fig{}\ref{fig:comp_fpc}.

The last two modules alone give already the information about the x and y coordinates of passing particle. Adding the first two modules allow in addition, to improve the spatial resolution, estimate the track coordinates and to reduce noise. 
A detailed description of the FPC can be found in Ref.~\cite{dyr97}.
	  \subsubsection{Forward Trigger Hodoscope}\label{ssec:fth}
The Forward Trigger Hodoscope consists of three layers of plastic scintillators. The first layer, closest to the interaction point, is built up from $48$ elements. Two subsequent ones consist of elements curved into archimedian spirals oriented clockwise and counter-clockwise (see \fig{}\ref{fig:comp_fth}).

A particle passing through the FTH, leaves a trace in the form of a hit in elements of consecutive layers. Three layers projected onto the xy-plane give an intersection point of struck elements, called pixel.

The first layer of the FTH is used to activate the trigger signal and further to determine the azimuthal angle of the particle trajectory. The FTH is also used in identification of the recoil particle(s) in the forward detector via the $\Delta E -E$ method. 
More detailed information about the design and performance of this detector can be found in Ref.~\cite{pau07}.
	  \subsubsection{Forward Range Hodoscope}\label{ssec:frh}
\begin{figure}[!b]
  \begin{center}
	\includegraphics[width =0.5\textwidth]{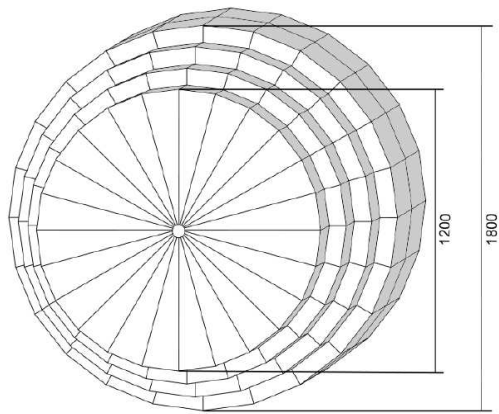}
  \caption[]{The Forward Range Hodoscope.}
  \label{fig:frh}
  \end{center}
  \end{figure}

The Forward Range Hodoscope consists of five layers made of $24$ plastic scintillators each (see \fig{}\ref{fig:frh}). The first three layers are of $110$~mm thickness while the last two are $40$~mm thicker. This detector is used for the identification of a recoil particle(s) with the $\Delta E -E$ method and together with the FWC and the FTH, it is used in the trigger to check the track alignment in the azimuthal angle.
 	  \subsubsection{Forward Veto Hodoscope}\label{ssec:fvh}
The Forward Veto Hodoscope is made of twelve, horizontally aligned bars of plastic scintillators each of which being $2$~cm thick and $13.7$~cm wide.
It is used to reject particles, punching through the Range Hodoscope and therefore, it increases the trigger selectivity.
	\subsection{The Pellet Target}\label{sec:pt}
\begin{figure}[!h] 
  \begin{center}
	\includegraphics[width =0.5\textwidth]{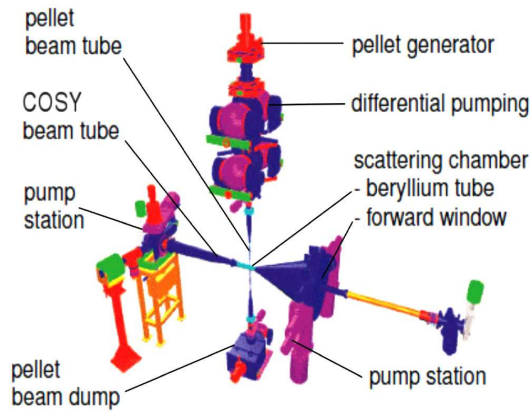}
  \caption[The WASA pellet target]{ Schematic view of the WASA Pellet Target system.}
  \label{fig:target}
  \end{center}
  \end{figure}

The special - Pellet Target - system was developed for the WASA experiment to satisfy the conditions required by a $4\pi$ detector \cite{Trostell:1995in}. It was designed to keep as less of the material inside of the detector as possible and therefore most of it, is located outside the detector. Only the $2$ meters long, $7$~mm narrow pellet beam tube, crosses the scattering chamber, delivering pellets to the impact point (see \fig{}\ref{fig:target}).  
A target in form of pellets was chosen to achieve luminosities of the order of $10^{32}~cm^{-2}~s^{-1}$, needed to study rare decays of light mesons, the beam target in the form of a gas or a cluster jet is not enough good collimated.

In order to obtain a stream of pellets of the same size and with the same distance separating them, the high-purity liquid jet is broken up by vibrations of a thin glass nozzle located in the pellet generator. 

The achieved divergence of the target beam is on the order of $0.04^{\circ}$, the single pellet size amounts to about $35~\mu$m. Together with the effective areal target thickness greater than $10^{15}~atoms \cdot cm^{-2}$, the luminosities of $10^{32}~cm^{-2}~s^{-1}$ are becoming feasible.

	\subsection{Data Acquisition System}\label{sec:daq}
\begin{figure}[!h] 
  \begin{center}
	\includegraphics[width =\textwidth]{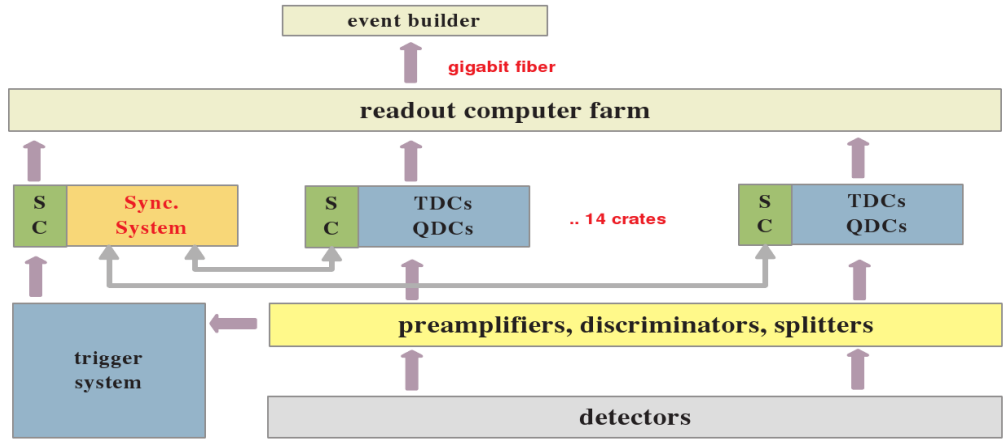}
  \caption[The WASA DAQ system]{ Structure of the Data Acquisition system.}
  \label{fig:daq}
  \end{center}
  \end{figure}

The Data Acquisition system (DAQ) collects and process signals from the detector elements, in order to make them available for further analysis.
The structure of the Data Acquisition system is shown in \fig{}\ref{fig:daq}.
\interfootnotelinepenalty=10000
Signals from 3800 straws and 1570 photomultiplier tubes connected with detector elements, are distributed and adapted by electronic modules of the lowest layer of the Data Acquisition system ( pre-amplifiers, splitters, discriminators).
Next, the conversion of analogue signals is done in fourteen crates of the digitization layer. There, the digitized signal is marked with a timestamp as well, and queued in FIFOs\,\footnote{That is a type of data structure, to which subsequent data is added to the end of the queue and data for the processing is taken from the beginning of the queue ('First In First Out')}. 

The timestamp is broadcasted by a special module (master module) of the main synchronization system. To this module also, the first level trigger is connected. 
Invoked by the trigger, the master module computes the event number and sends it, together with the timestamp, to all fourteen crates, where digitized information, also marked with the timestamp, wait in FIFOs. While the crates are processing received information, the master module blocks the trigger. 

Data with a matching timestamp, marked with an event number, are passed to the readout computer farm and to the event builder. More detailed information on the performance and the operation of the Data Acquisition system can be found in \cite{Hejny:2007sv,Kleines:2006cy}.

\outroformatting

\chaptertitle{Analysis Tools}
\introformatting
\section{Root}
The software package Root \cite{Brun:1997pa} is the most used tool in this work. It is a successor of the, FORTRAN implemented, Physics Analysis Workstation (PAW)\cite{Brun:1989vg}, developed in the European Organization for Nuclear Research (CERN) \cite{cern}. All the histogramming, fitting, calculations were done using its object oriented, C++ operated structure.
\section{Event Generator}
For the purpose of studies made in this work, the Pluto++ \cite{Frohlich:2007bi} event generator was used. It is a collection of C++ classes based on the ROOT~\cite{Antcheva:2009zz} environment.

\begin{figure}[!h]
  \begin{center}
	\includegraphics[width =0.6\textwidth]{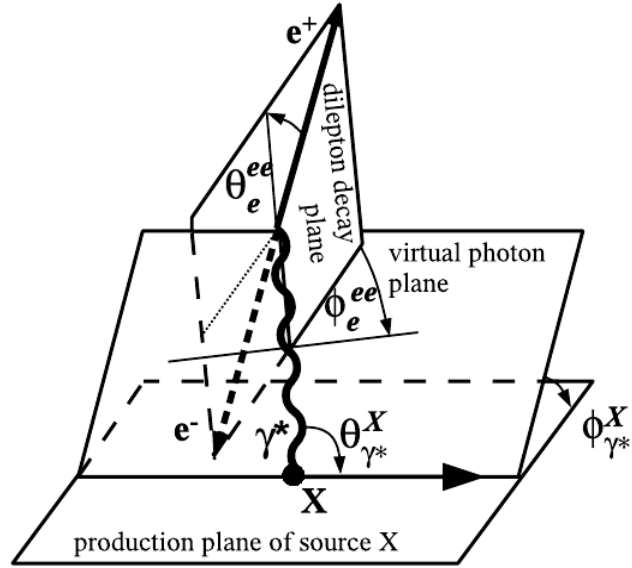}
  \caption[  ]{The picture shows four angles which need to be taken into account while considering $\gamma^{*} \rightarrow e^{+}e^{-}$ conversion. More details are given in the text. Picture is taken from \cite{Frohlich:2006cz}.}
  \label{fig:p_angles}
  \end{center}
  \end{figure}

The event generator generates the pseudoscalar meson Dalitz decays with a virtual photon (described by the $\phi_{\gamma}^{*}$ and  $\theta_{\gamma}^{*}$ angles) being isotropically distributed in space, carrying a momentum determined by its invariant mass and the mass of the meson (see \fig{}\ref{fig:p_angles}). 

The azimuthal angle, $\phi_{e}^{ee}$, of the dilepton decay plane around the photon direction is isotropic, while the helicity angle - $\theta^{ee}_{e}$ is distributed according to $1+cos^{2}\theta^{ee}_{e}$\cite{Bratkovskaya:1995kh}.

The invariant mass spectrum of lepton-antilepton pairs is given according to \cite{Landsberg:1986fd} by equation~\ref{eq:FFland}, see Sec.~\ref{sec:ff}.
 The Pluto++ event generator enables also modifications in the spectrum due to the presence of a given form factor.

\section{Detector Simulation}
For the reproduction of the detector response, the WASA Monte Carlo software is used. 
It is based on the GEANT package~\cite{:1994zzo}. 

The description of the detector components (theirs dimensions, positions, type of the material, magnetic field) has been implemented in the GEANT framework while the reaction kinematics (the initial four-momenta of particles) is delivered by the event generator. 

The WASA Monte Carlo takes into account such effects like energy loss, multiple scattering and conversion in the detector material. In addition, it is possible to improve the matching between data and simulations via smearing of the simulated observables according to the known resolution of detectors.    

\begin{figure}[!h]
  \begin{center}
	\includegraphics[width =0.45\textwidth]{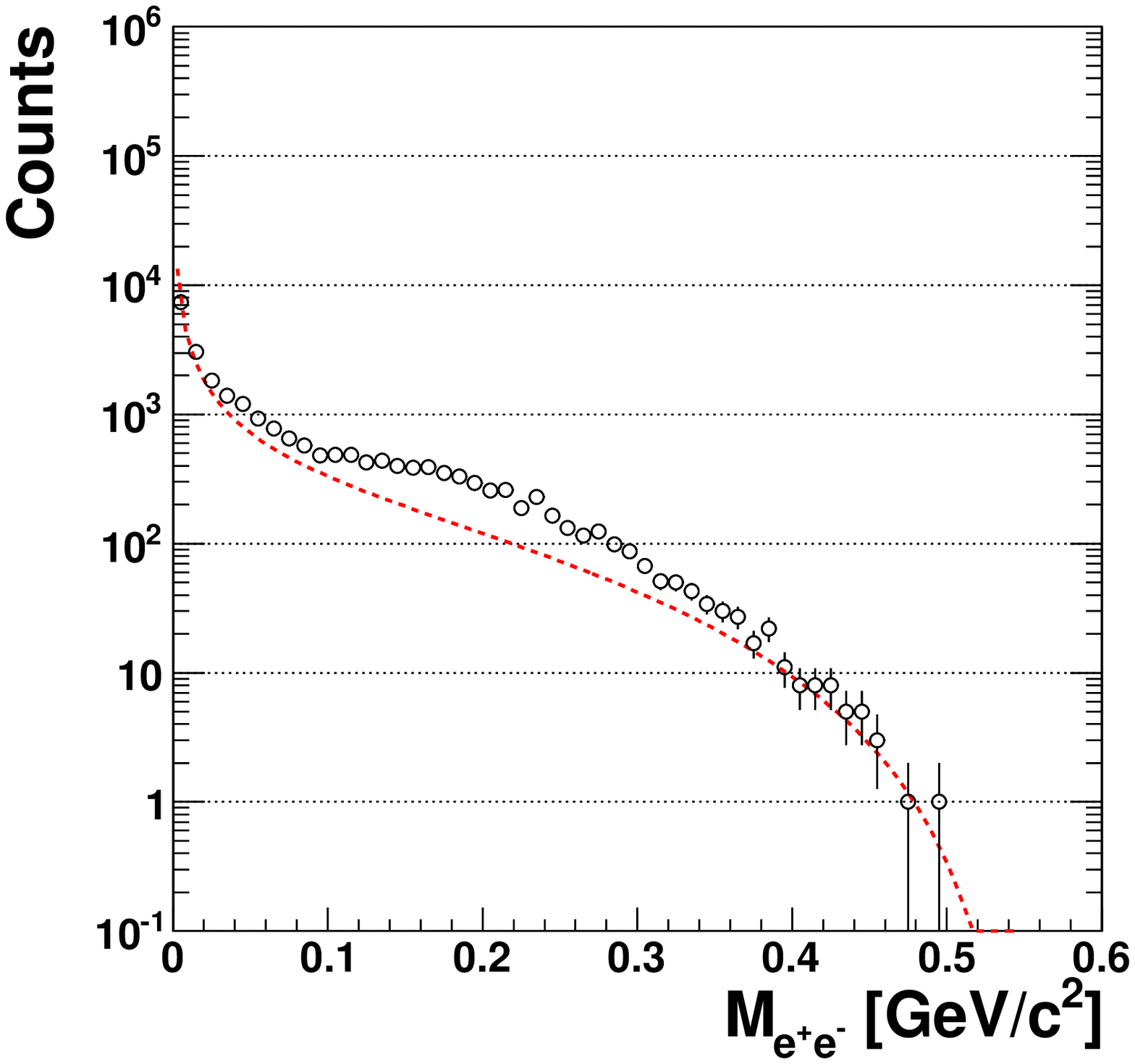}
	\includegraphics[width =0.45\textwidth]{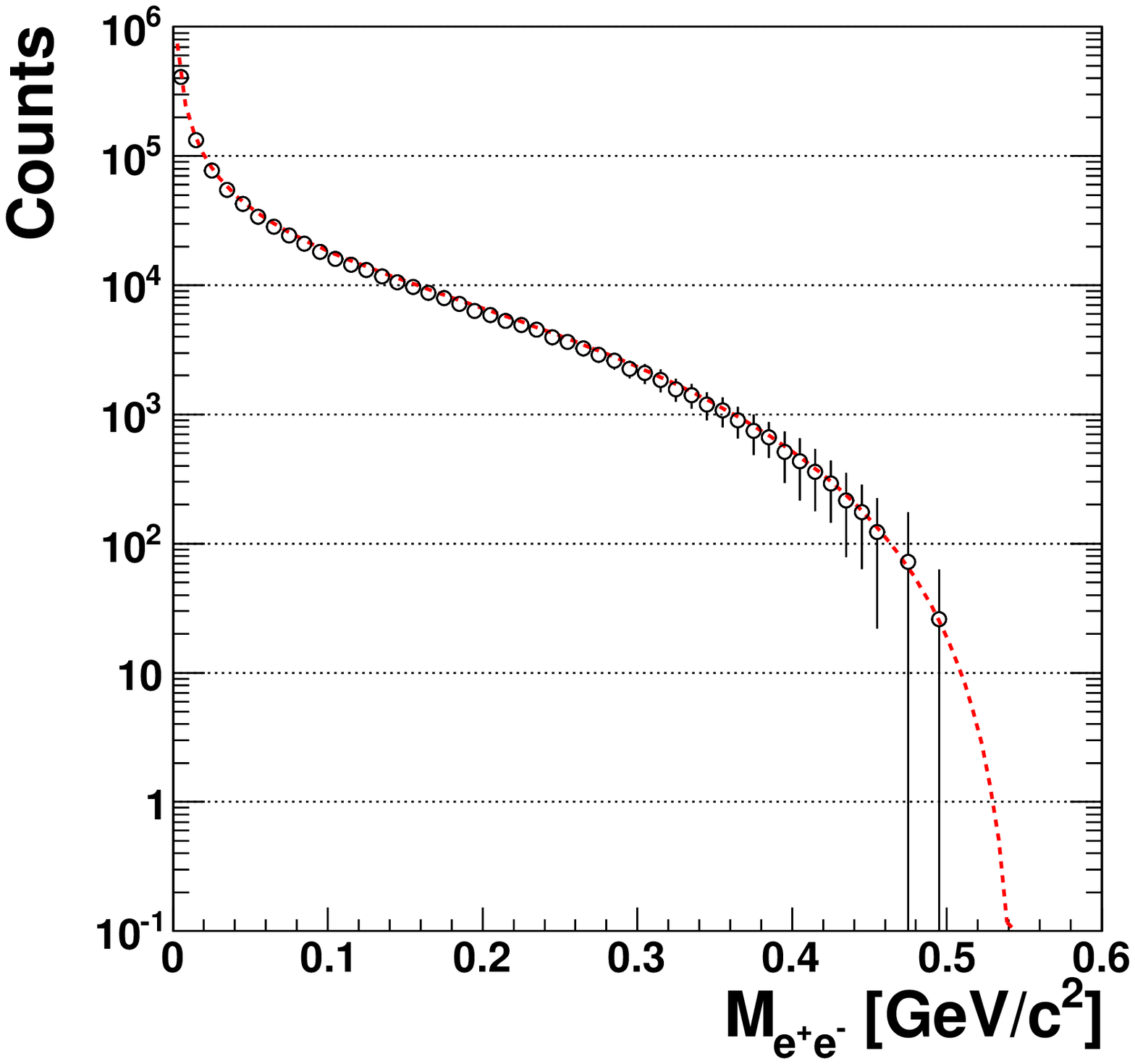}
  \caption[ Leptonic mass spectrum in simulations ]{ Simulated and then reconstructed lepton-antilepton mass spectrum (\underline {left}), acceptance corrected (\underline {right}). $10^{6}$ events (circles) were generated using the Pluto++ event generator with a form factor equal to one and then processed by the WASA Monte Carlo software which reproduces the experimental environment. The dotted, red line shows the calculation according to the QED.}
  \label{fig:p_ff}
  \end{center}
  \end{figure}

Within preparations to work with experimental data, the algorithms created for the purpose of the form factor extraction were tested on the Monte Carlo sample of $10^{6}$ events of the \reac{} reaction. \fig{}\ref{fig:p_ff} shows the distribution of the invariant mass of the \ee{} pairs generated with Pluto++ and then reconstructed by the WASA analysis software before and after acceptance corrections. The correctness of developed procedure manifests itself on the right panel of this figure, where all points are situated on the QED line, as expected. Having constructed a properly working method of the determination of the variable of interest, one can start a thorough analysis of experimental data.  

\chaptertitle{First Stage Event Reconstruction}

\section{Trigger Conditions}
\label{sec:RunInfo}
\introformatting

The \e{} mesons were produced in the \pdhe{} reaction with a beam momentum of $1.7$~GeV/c which corresponds to an excess energy of $60$~MeV. The cross section for the \e{} meson production at this energy amounts to $0.412 \pm 0.016\, \mu b$ \cite{Bilger:2002aw}. That leads to an event rate which makes it possible to use an unbiased trigger. It means, that the trigger logic was fully based on signals coming from the components of the forward part of the detector and therefore, no requirements on the meson decay products were used.

The requirement was at least one charged particle with a minimum energy loss of $5$~MeV in the
Forward Window Counter. This particle must have had signature also in the first layers of the Forward Trigger Hodoscope and the Forward Range Hodoscope. Hits in all those three detectors had to be aligned with the same azimuthal angle.

\outroformatting

\section{Track Reconstruction}
  Signals invoked in the detectors by particles passing through them are merged into tracks using reconstruction algorithms. Merging is done within given boundaries regarding time, energy deposits and angular information. Their default values have been chosen to provide the best quality of reconstruction \cite{Adam:2004ch}. Nevertheless, it must be checked and, if necessary, also tuned for each experimental run, especially when coming to study of some specific reaction channel characterized by small branching ratio.
  \section{Track of the Recoil Particle}
\label{sec:RecFD}
\introformatting
In the case of the \pdhe{} reaction, the recoil particle is the helium ion going in the forward direction. The reconstruction of tracks in the \fd{} 
begins with combining time coincident hits in adjacent detector elements into larger groups called clusters.
This is necessary since particles passing a detector close to the border of a given element, can cause a signal in the adjacent one.  
Having all hits assigned to the cluster, the geometrical overlap between clusters formed in different layers of the detector can be checked. 
The procedure starts from taking the \fth{} pixel (see Fig.\ref{fig:comp_fth}) and searches the detector in order to find overlapping clusters. 
The \fth{} also serves as a source of the initial information about the angular coordinates. The reconstruction procedure assumes a vertex located 
in the center of a nominal beam and target interaction region. Next, the angular information of the track can be improved based on signals from the \fpc{}. The time assigned to the track is 
taken as the average time of hits contributing to the \fth{} pixel.

\outroformatting

  \section{Tracks of Meson Decay Products}
\label{sec:RecCD}
\introformatting
Particles coming from the decay of the \e{} meson are registered in the \cd{}.
The track reconstruction starts in the calorimeter. All hits are combined into clusters according to their position, energy and relative time. Typically, the allowed limits for the hits to be enclosed in a cluster are: i) a time difference with the central module of $50~ns$ and, ii) a minimum energy of $2~MeV$.

The starting crystal, the center of the cluster, is the one with the highest energy deposit (minimum of $5~MeV$) and its time defines the time of the cluster. Hits in crystals adjoining it, are added to it, if they were not included already in another cluster.

The formed cluster's energy, is the sum of energies from the contributing crystals. It has to be more than $5~MeV$.
The cluster position is given by the energy weighted mean of the positions of crystals constituting it. If there are no matching clusters found in the \mdc{} and \psb{}, the cluster is assigned as coming from a neutral particle. More detailed description of the cluster reconstruction in the \se{} can be found in \cite{Bednarski:2011vm}. 

A cluster in the \psb{} constitutes a single hit or, in case of overlapping elements, two hits if the deposited energy  amounts to at least $1~MeV$ and their time difference is less than $10~ns$.

Particles in the magnetic field follow a helical trajectory. Therefore, hits registered in the \mdc{} are described with helices. Hits belonging to one helix create a cluster. The procedure of extracting all the helix parameters, consists of two main steps. First, hits are projected onto the {\it xy} plane (perpendicular to the beam axis) and fitted with circles.
Next, a straight line is fitted to the hits in the {\it Rz} plane\footnote{R denotes the helix radius}. Description of the algorithm can be found in Ref.~\cite{kmpt}.

Having grouped the information in all three subdetectors of the \cd{}, the track assignment can be done. That is, clusters formatted in the \mdc{}, the \psb{} and the \se{} are checked with regard to their belonging to one track. The geometric overlap of clusters is used as a criterion. 

\begin{figure}[!t]
 \subfigure[The difference in the azimuthal angle between the \mdc{} and the \psb{}. The cut was chosen to $|\Delta\phi|<20^{\circ}$ ]{
	\label{fig:d_MDC-PSB}
	\includegraphics[width =0.5\textwidth]{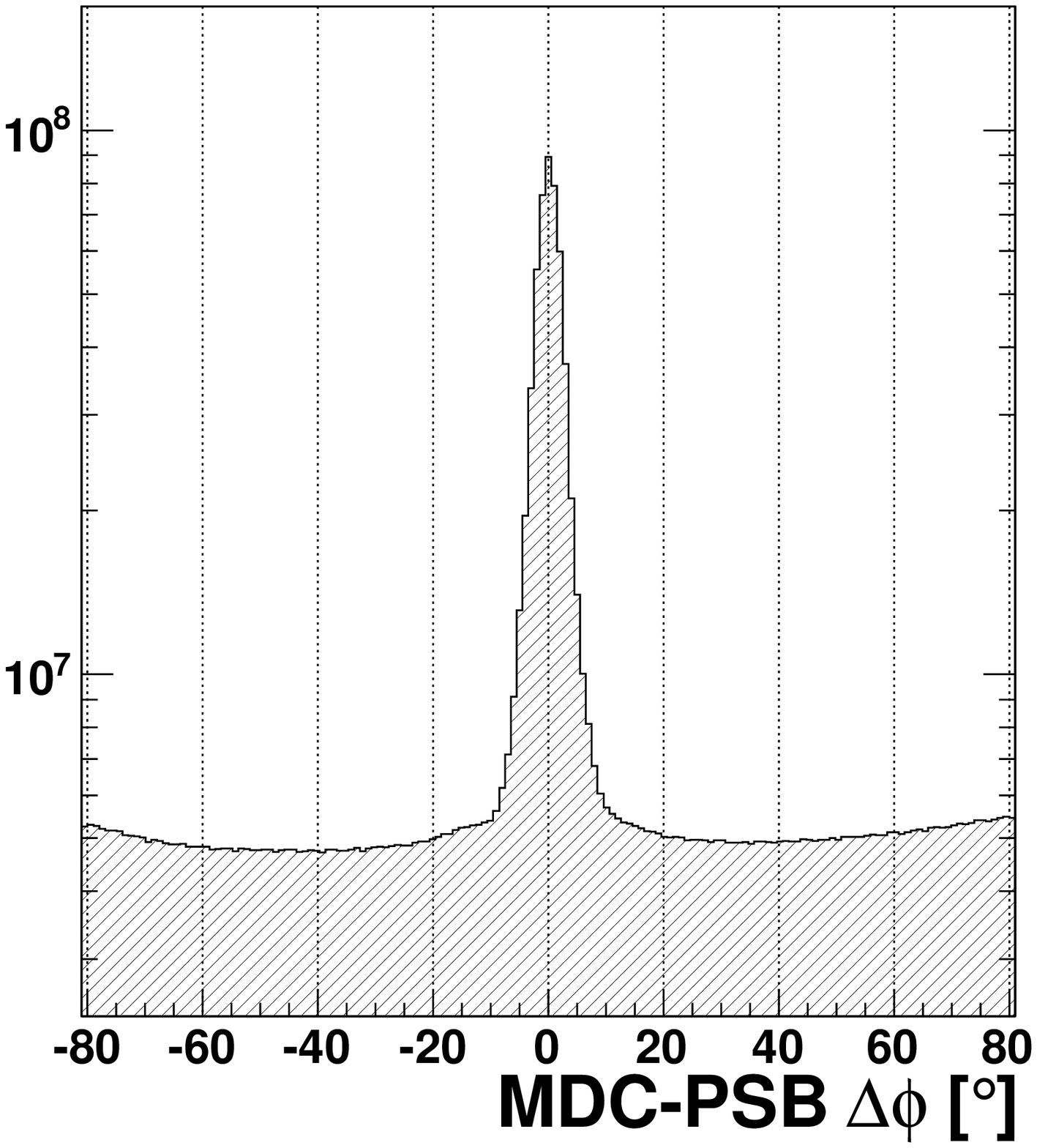}
  }
 \subfigure[The opening angle between the \mdc{} and the \se{}{} was chosen to be less than $25^{\circ}$  ]{
	\label{fig:d_MDC-SEC}
	\includegraphics[width =0.5\textwidth]{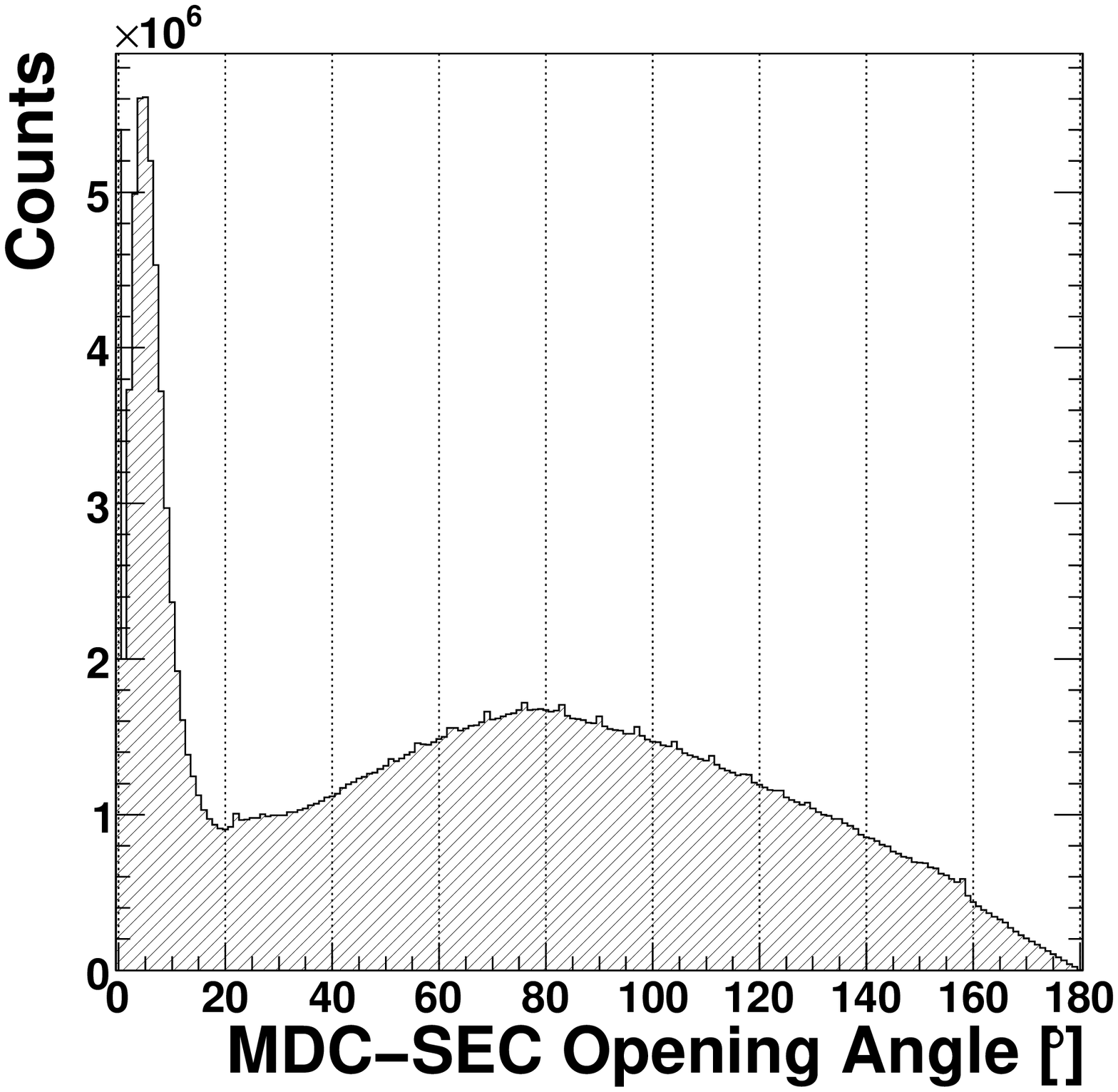}
  }
  \caption[Track assignment in the \cd{}]{ Experimental distributions used to choose the matching conditions in the \cd{}. The track assignment in the \cd{} consist in checking for geometric overlaps of chamber's cluster with clusters in the \psb{} and the \se{}. }
\vspace*{10cm}
  \label{fig:TrackAss}
\end{figure}

In case of the \mdc{} and the \psb{} the criterion is difference of the azimuthal angles $\Delta \phi$ between the exit coordinate of the helix and position of the cluster in the Plastic  Scintillator  Barrel. The experimental distribution of $\Delta \phi$ is shown in \fig{}\ref{fig:d_MDC-PSB}. A window of $|\Delta \phi| < 20^{\circ}$ was conservatively applied.

Matching between the \mdc{} and the \se{} is done by building a straight line, tangential to the helix in its exit point, and calculating its position at the calorimeter surface. The angular difference between this position and the position given by the calorimeter cluster is the matching criterion. The relevant, experimental distribution is shown in \fig{}\ref{fig:d_MDC-SEC}. Small peaks seen over the whole range of the x-axis are caused by the detector granularity. A maximum opening angle of $25^{\circ}$ is selected. 

\outroformatting

\outroformatting

\chaptertitle{Extraction of the Signal Channel}
\introformatting
\graphicspath{{Fifth/Rysunki/}}

In case of the \reaction{} reaction, particle identification consists in recognizing three tracks and an additional cluster in the SEC. The \he{} ion is identified using the forward part of the WASA detector while $e^{+}$, $e^{-}$ and $\gamma$ particles coming from the \e{} decay, are detected and reconstructed in the central part of the detector.  

 \section{Kinematics - Phase Space }
\label{sec:kin}
\introformatting

\begin{figure}[!h]
  \centering
  \subfigure[ Generated spectrum showing the cosine of the \e{} scattering angle in the center of mass system (CMS)]{
	\label{fig:cosEtaTh}
	\includegraphics[width =0.45\textwidth]{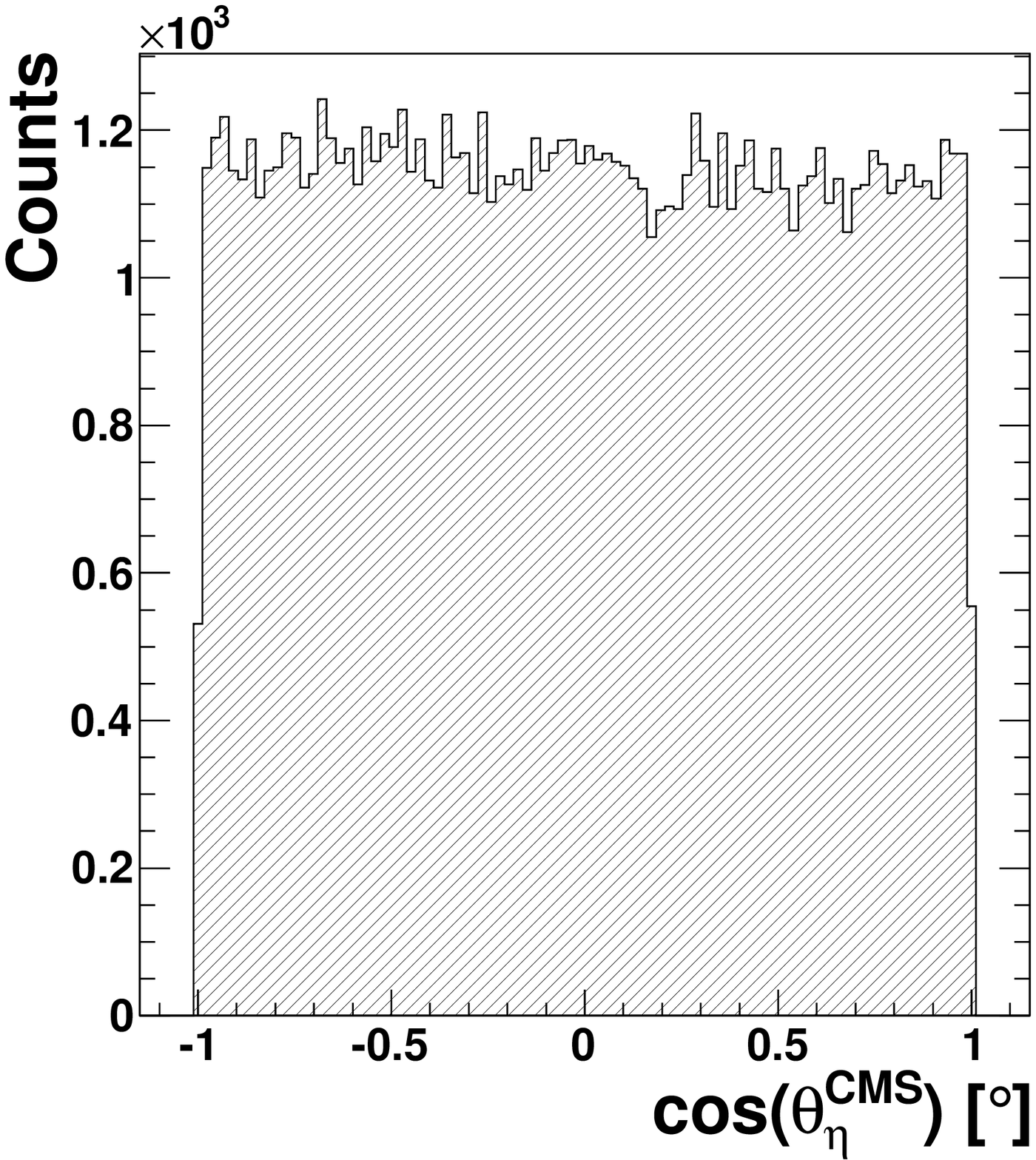}
  }
  \subfigure[Generated spectrum showing the \he{} scattering angle, in the center of mass system (CMS) as a function of the \he{} scattering angle in the laboratory system (LAB)]{
	\label{fig:HeTh}
	\includegraphics[width =0.45\textwidth]{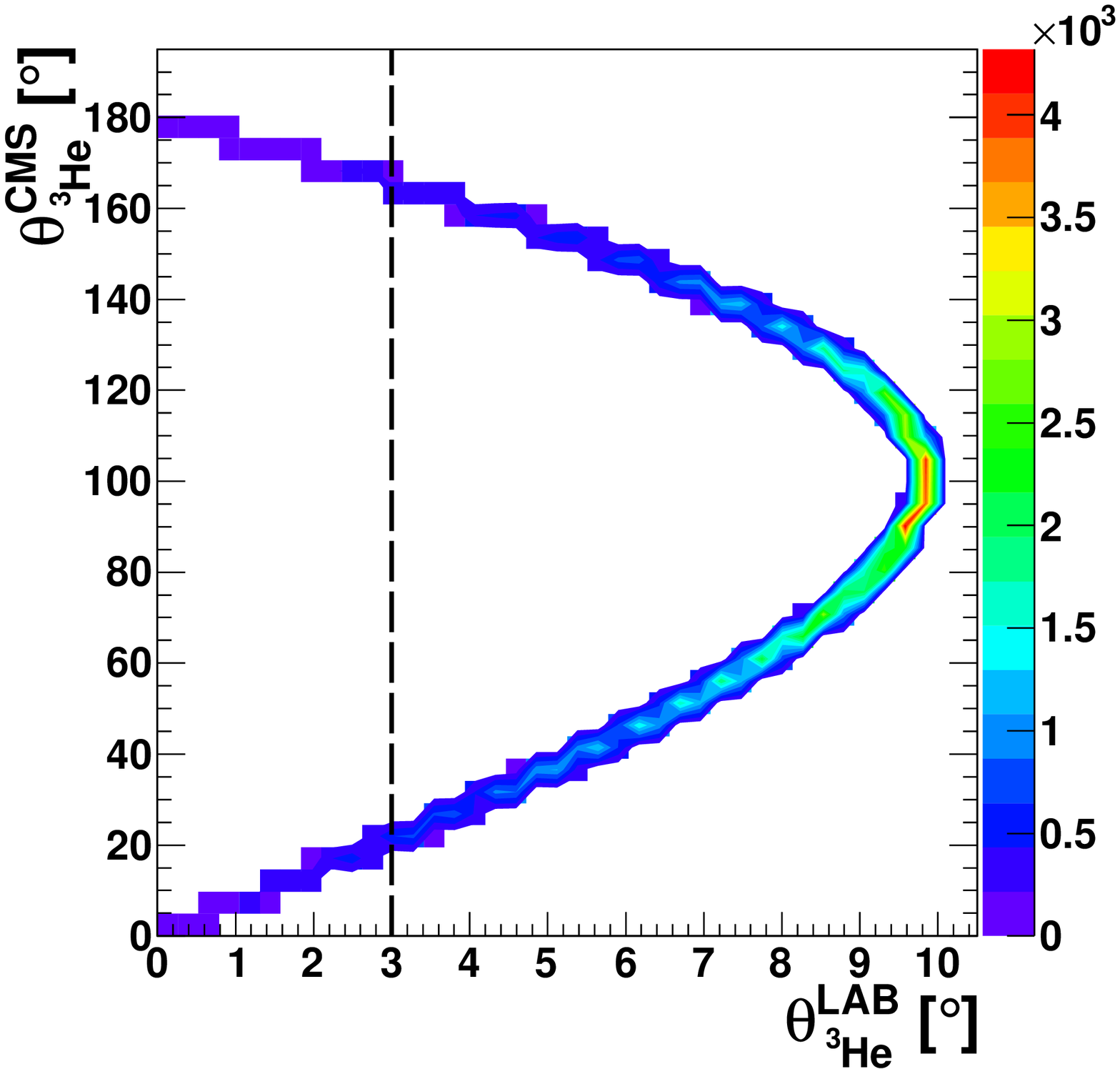}
  }
	\caption[]{ The output of the Pluto event generator for the \pdhe{} reaction. About 5\% of generated events is outside of the \fd{} geometrical acceptance shown in (\subref{fig:HeTh}) as a dashed line. }
	\label{fig:ThCMS}
  \end{figure}

In the used Monte Carlo events generator, Pluto++, it is assumed that the phase space in \pdhe{} production is homogeneously and isotropically populated. In \fig{}\ref{fig:cosEtaTh} one can see the flat distribution of the 
cosine of the \e{} scattering angle in the center of mass system as it comes from the generator whereas, in \fig{}~\ref{fig:HeTh} the helium scattering angle in the center of mass system as a function of the \he{} scattering angle in the laboratory system is shown.
The dashed line corresponds to the geometrical acceptance of the \fd{}. Helium particles which were scattered in the laboratory system under a~$3^{\circ}$ angle, are not seen in the detector. One may notice also, that \he{} particles produced in the \pdhe{} reaction at a beam momentum of $1.7$~GeV/c are emitted up to the $\sim10^{\circ}$ of scattering angle, so well below $18^{\circ}$ which is the upper geometrical limit of the \fd{}. 

The overall geometrical acceptance of the WASA detector for registering all four particles in a given event, coming from the \reac{} reaction, amounts to almost $80\%$.

\outroformatting

 \section{Identification of Helium }
\label{sec:id_hel}
\introformatting

\he{} is detected and reconstructed in the \fd{}. Already on the preselection level, helium ions were initially identified. 
However, this identification was done using a rough calibration \cite{zlo09}. Therefore, additional checks are made. 

\begin{figure}[!h]
\centering
 \subfigure[Energy deposited in FWC1]{
	\label{fig:EdepWC1}
	\includegraphics[width =0.45\textwidth]{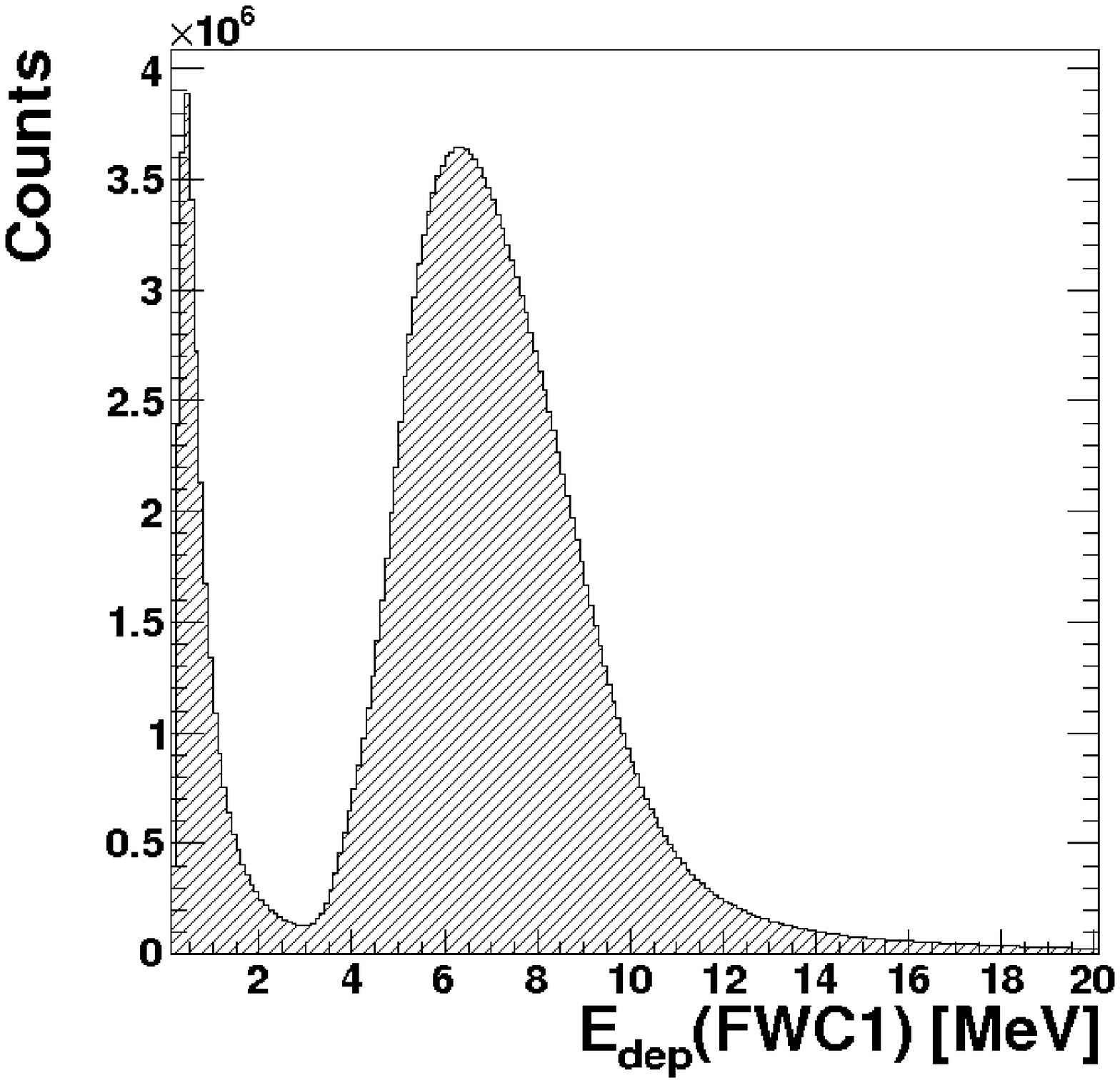}
  }
 \subfigure[Energy deposited in FTH2]{
	\label{fig:EdepTH2}
	\includegraphics[width =0.45\textwidth]{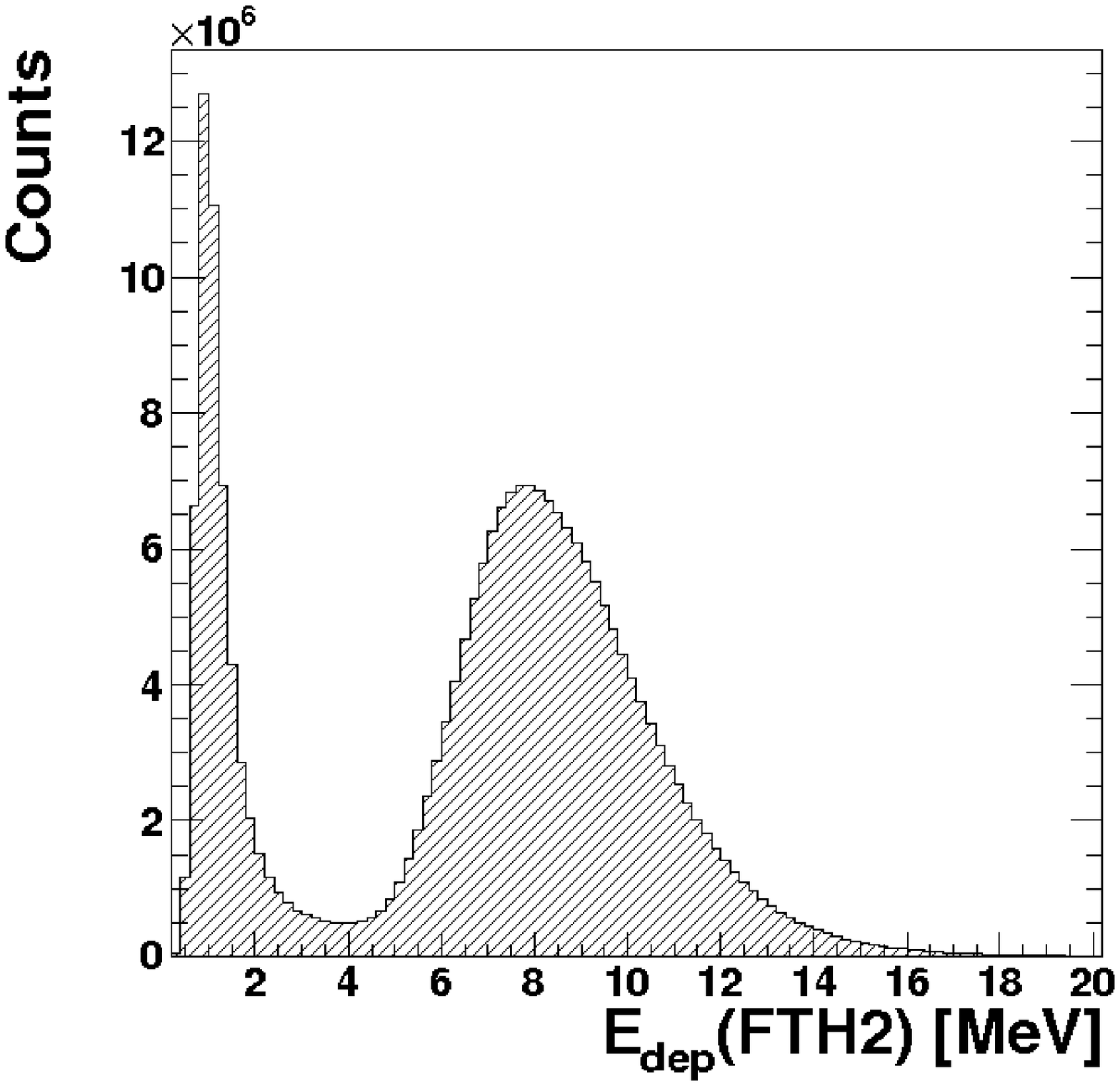}
  }
  \caption{  Energy depositions in experiment for the first layer of the \fwc{} (\subref{fig:EdepWC1}) and the second layer 
of the \fth{} (\subref{fig:EdepTH2}). Cut on minimal deposited energy was chosen to $3~ MeV$ in the FWC1 and $4~MeV$ in the FTH2. }
  \label{fig:Edep}
\end{figure}

In the first step, appropriate minimal energy deposits in each of the detectors are set. This is done based on experimental 
spectra of energies deposited in each layer of the detector as shown in \fig{}\ref{fig:Edep}.

\begin{figure}[!h]
\centering
	  \includegraphics[width =0.6\textwidth]{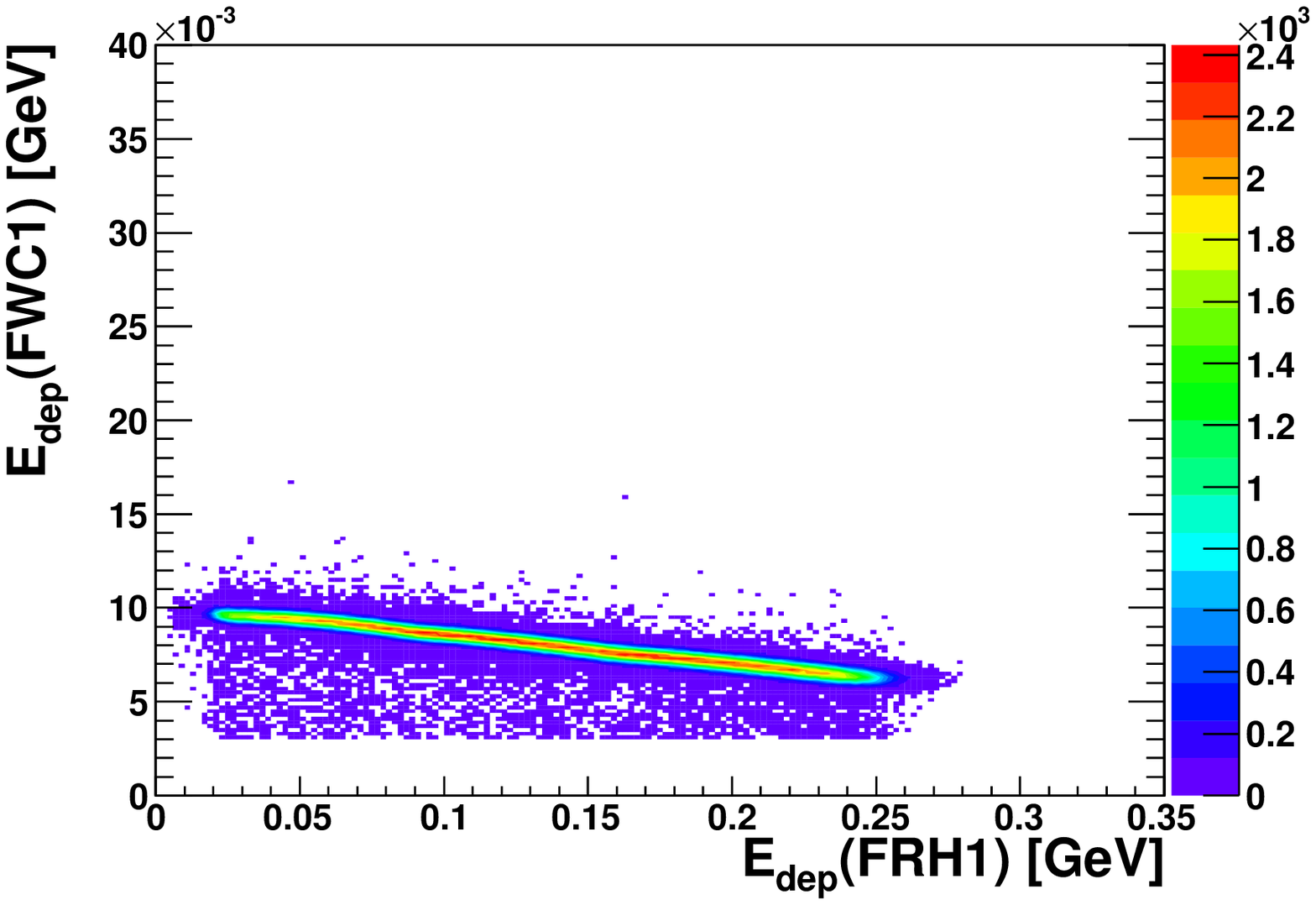}\\
	  \includegraphics[width =0.6\textwidth]{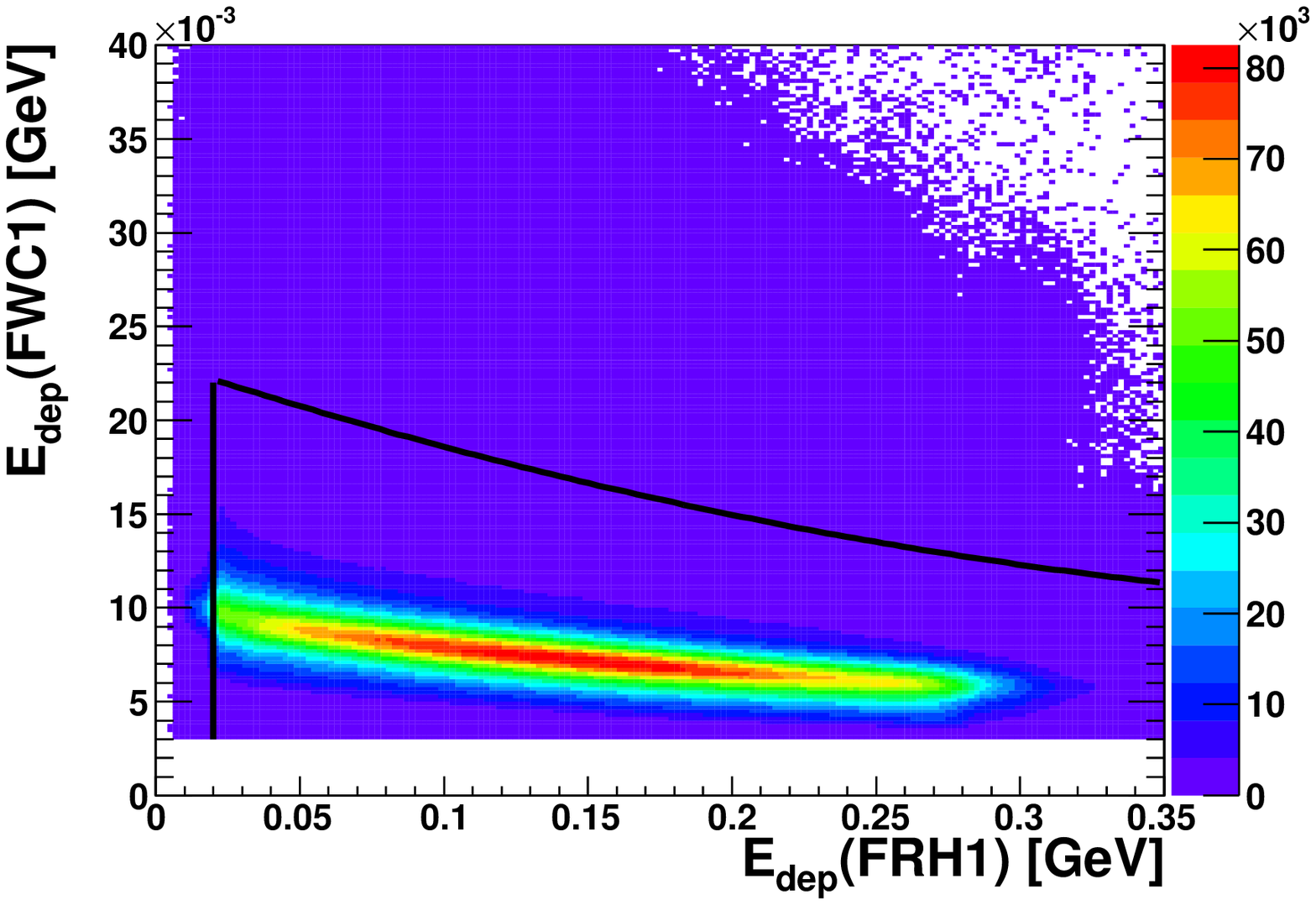}
	\caption{ $\Delta{}E-E$ plots of energy deposited in the first layer of the \fwc{} and the first layer of the \frh{} used 
for the final helium selection. In the upper panel simulated data are presented while the bottom plot shows experimental data. Particles, 
having energy depositions within boundaries shown in the picture with black lines, are considered further as helium ions. }
	\label{fig:Edep_WC1_RH1}
\end{figure}
Candidate for a helium track must have also the scattering angle within geometrical boundaries of the \fd{}.

Finally, to select tracks of \he{} particles, the $\Delta{}E-E$ method is used. Multilayered architecture of the \fd{} allows 
to perform identification, based on energy losses in different layers of the detector. Correlation of energy deposited in the first 
layer of the \fwc{} and the first layer of the the \frh{} is shown in \fig{}\ref{fig:Edep_WC1_RH1}.

In order to convert deposited energy into kinetic energy a set of parameters were derived from the Monte Carlo simulations. 
This correction parameters are needed since due to the occurrence of additional energy losses in the detector material, the sum 
of deposited energies can be smaller than the true kinetic energy.

For this purpose, the Monte Carlo simulation of single particle tracks was used and the relative difference of the reconstructed deposited 
energy and the true kinetic energy was parametrized as a function of the deposited energy. 

In case of the \pdhe{} reaction with the beam momentum of $1.7~GeV/c$, the kinetic energy in units of GeV may be described by equation \ref{eq:HelEk}:
\hspace{0.5cm}
\begin{equation}
\label{eq:HelEk}
  E_{kin}(E_{dep},\theta)  = ( c_{0} + c_{1} E_{dep} + c_{2} E_{dep}^{2} - c_{3} E_{dep}^{3} ) 							
						     ( a_{0} - a_{1}  cos\theta ),
\end{equation}
where $c_{0}=0.199876,\,c_{1}=0.640187,\, c_{2}=1.60489,\,  c_{3}=2.28946,\, a_{0}=1.42133$,  $a_{1}=0.364951$ and $E_{dep}$ is the energy 
deposited in the \frh{} expressed in GeV and $\theta$ is the scattering angle \cite{bas09}.

\outroformatting

 \section{Identification of Photon }
\label{sec:IDgam}
\introformatting

\begin{figure}[!h]
\centering
	  \includegraphics[width =0.5\textwidth,height=0.45\textwidth]{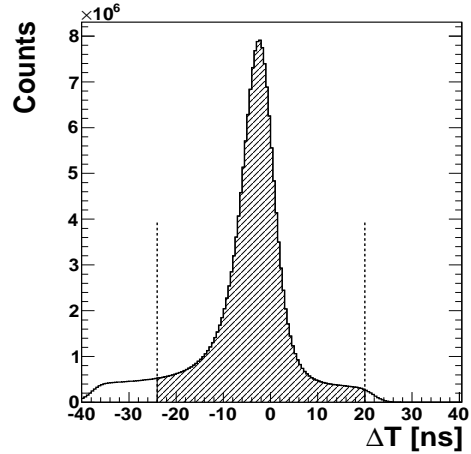}
	\caption{ Experimental distribution of the time difference, $\Delta T$, between \he{} in the \fd{} and the neutral particle in the \cd{}. 
                  Dashed line shows the time cut used in the analysis. }
	\label{fig:TimeG}
  \end{figure}

\begin{figure}[!h]
  \centering
  \subfigure[The experiment]{
	\label{fig:so_exp}
	\includegraphics[width =0.45\textwidth,height=0.4\textwidth]{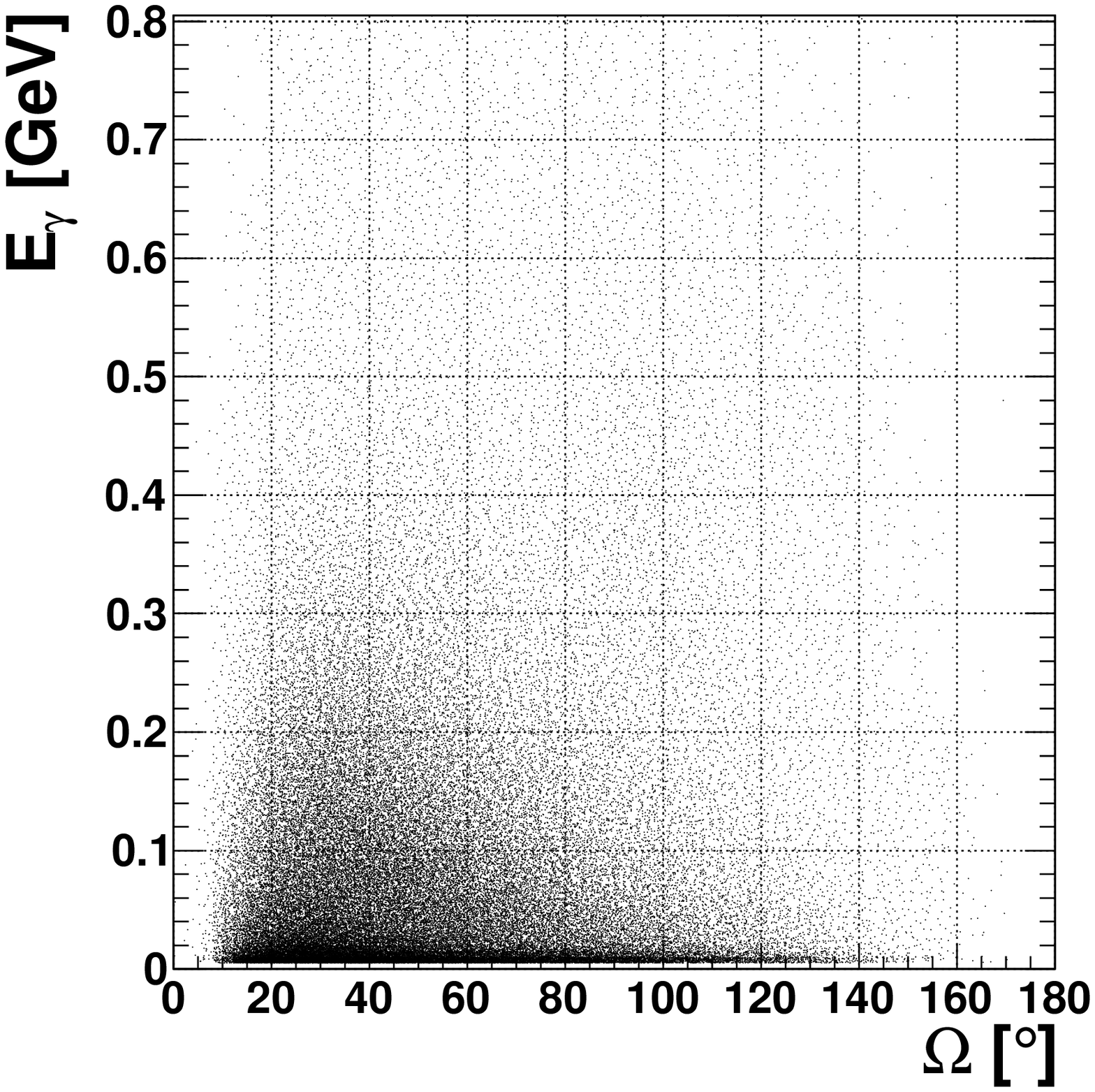}
	}
  \subfigure[Simulation of the \reac{} reaction]{
	\label{fig:so_eeg}
	\includegraphics[width =0.45\textwidth,height=0.4\textwidth]{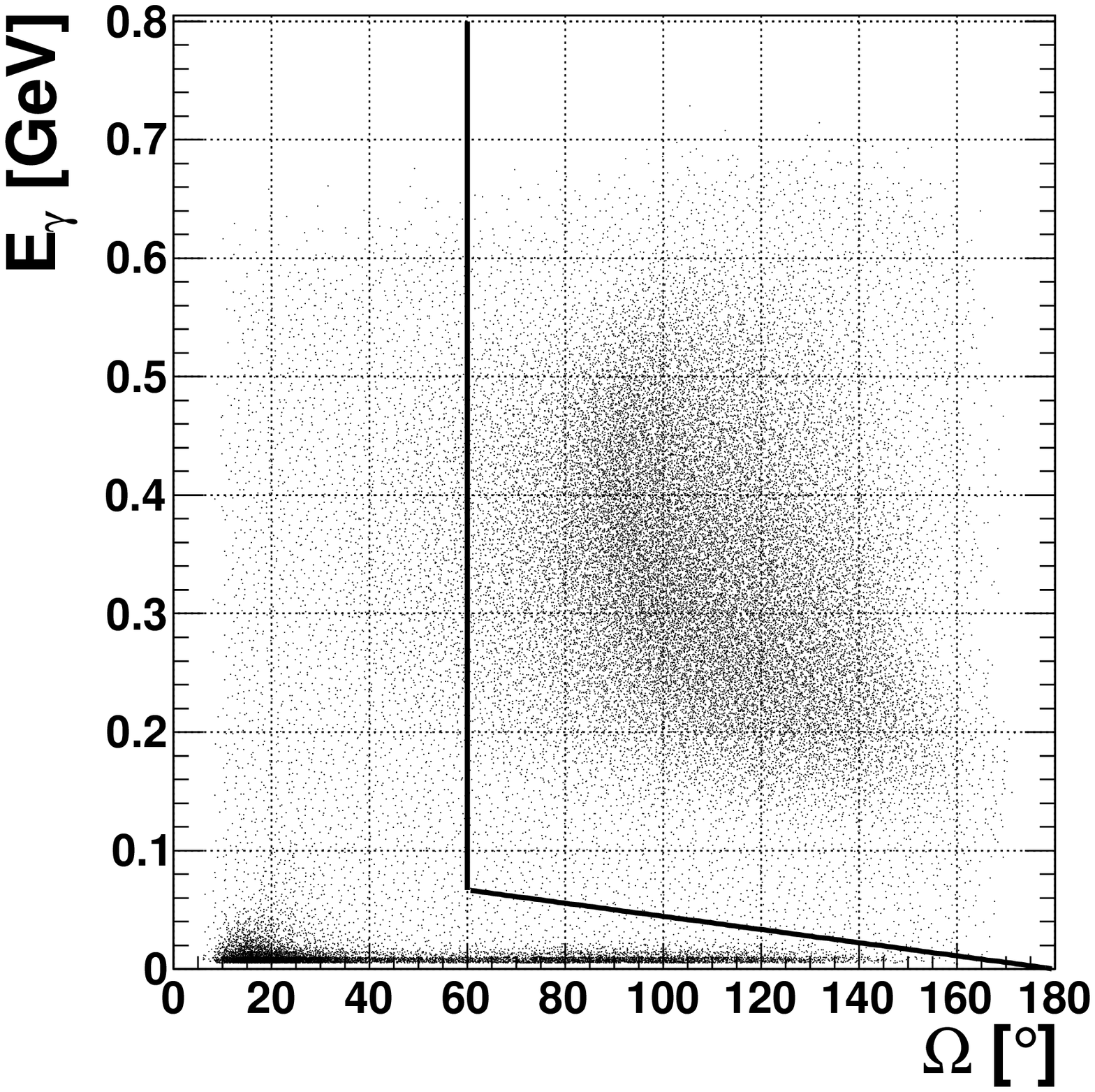}
	}
	\caption{ Spectra of cluster's energy vs. the angle it creates with the nearest track. 
	}
	\label{fig:SpOff}
  \end{figure}

Selection of a neutral particle in the \cd{}, starts with checking its time correlation with the \he{}, selected in the \fd{}. 
Time cut of \mbox{[-24,20] ns} was chosen (see \fig{}\ref{fig:TimeG}).
Cluster in the SEC without an associated track in the MDC can originate not only from the $\gamma$ coming from the impact point 
but also from the interaction of charged particle in the detector material. So-called split-off, is characterized by small energy 
and a small angle to the nearest charged particle. Energy of the photon candidate presented as a function of the angle it creates 
with the nearest charged particle, $\Omega$, is shown in \fig{}\ref{fig:SpOff}. The Monte Carlo distribution in the right panel consists 
of events from the simulated \reaction{} reaction. The signal is visible in the upper right part of the picture. 
At the $\Omega \sim 20^{\circ}$ 
an enhancement caused by split-offs appears. The same distribution from the experiment, shown in the left panel, is 
strongly contaminated. In order to cut out events with false photon candidates, the restriction on the ${\Omega > 60^{\circ}}$ 
and ${E_{\gamma} > 0.1 - 0.00055*\Omega}$ was applied as indicated by lines in \fig{}\ref{fig:so_eeg}.

The fact that products of the \gsg{} decay fly back-to-back in the meson rest frame, is applied in further selection on the 
remaining set of neutral particles reconstructed in \cd{}. That is, each of photon candidate is checked for the azimuthal angle 
it creates with the virtual photon in the \e{} rest frame. The four-momentum of the virtual photon is reconstructed based 
on four-momentum vectors of $e^{+}$ and $e^{-}$. In the \e{} meson rest frame, real and virtual photon create the opening 
angle $\Delta\phi_{\gamma\gamma*}$ of $180^{\circ}$. 

\begin{figure}[!hbt]
\centering
	  \includegraphics[width =0.6\textwidth,height=0.5\textwidth]{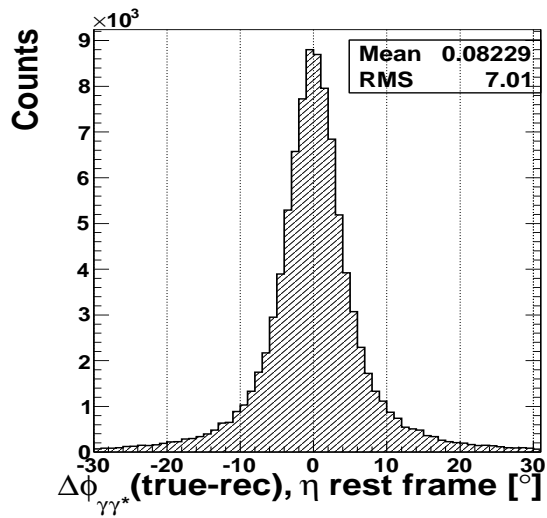}
	\caption{ Simulated distribution of the difference between true and reconstructed values of the azimuthal 
                  angle between real and virtual photon in the \e{} rest frame, 
                  $\Delta\phi_{\gamma\gamma*}=\Delta\phi_{\gamma\gamma*}^{true}-\Delta\phi_{\gamma\gamma*}^{reconstructed}$. }
	\label{fig:dPRes}
  \end{figure}

\begin{figure}[!hbt]
 \subfigure[ Without particle identification in SEC]{
	\label{fig:dPa}
	\includegraphics[width =0.5\textwidth,height=0.45\textwidth]{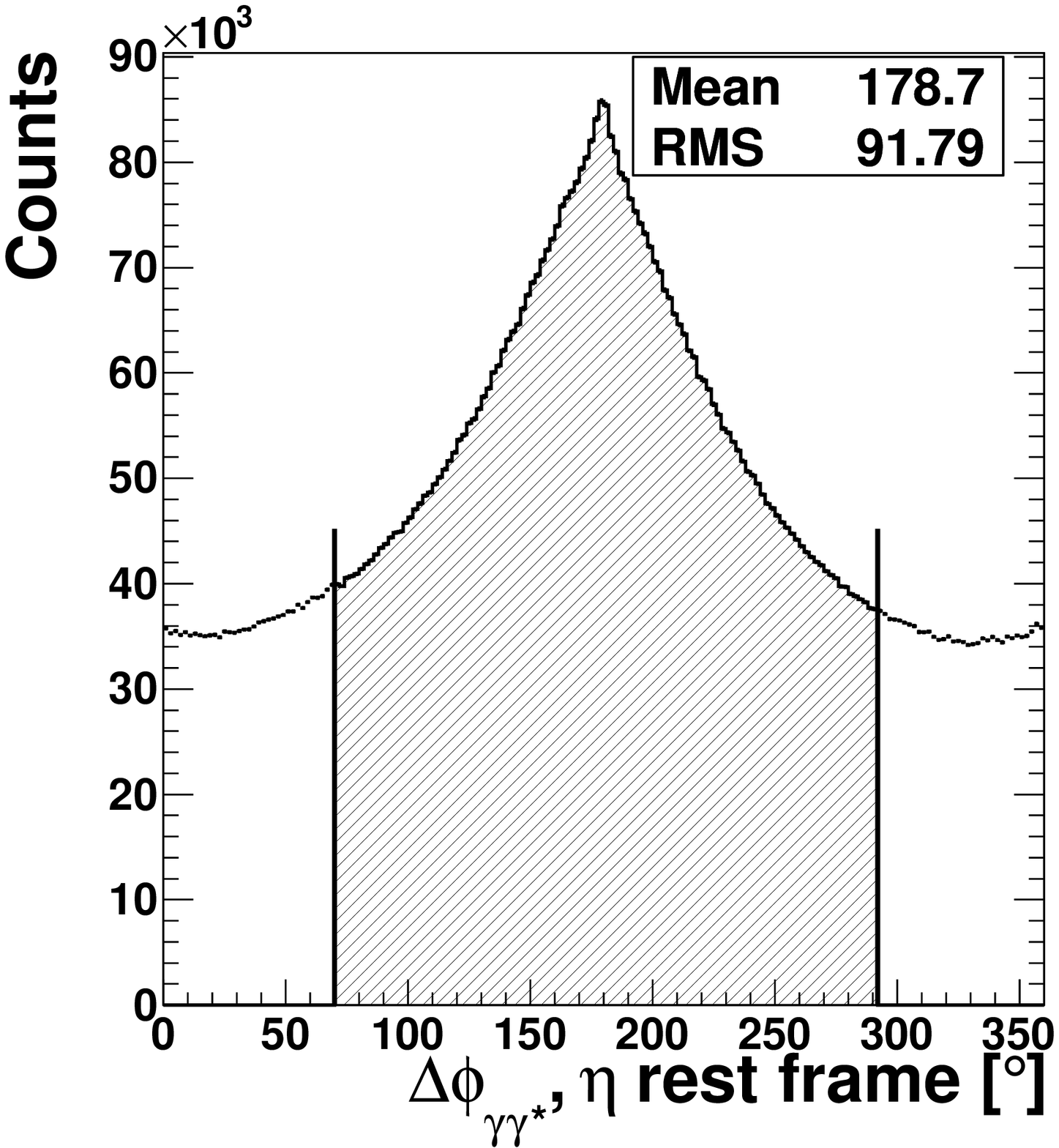}
  }
 \subfigure[ With particle identification in SEC]{
	\label{fig:dPsec}
	\includegraphics[width =0.5\textwidth,height=0.45\textwidth]{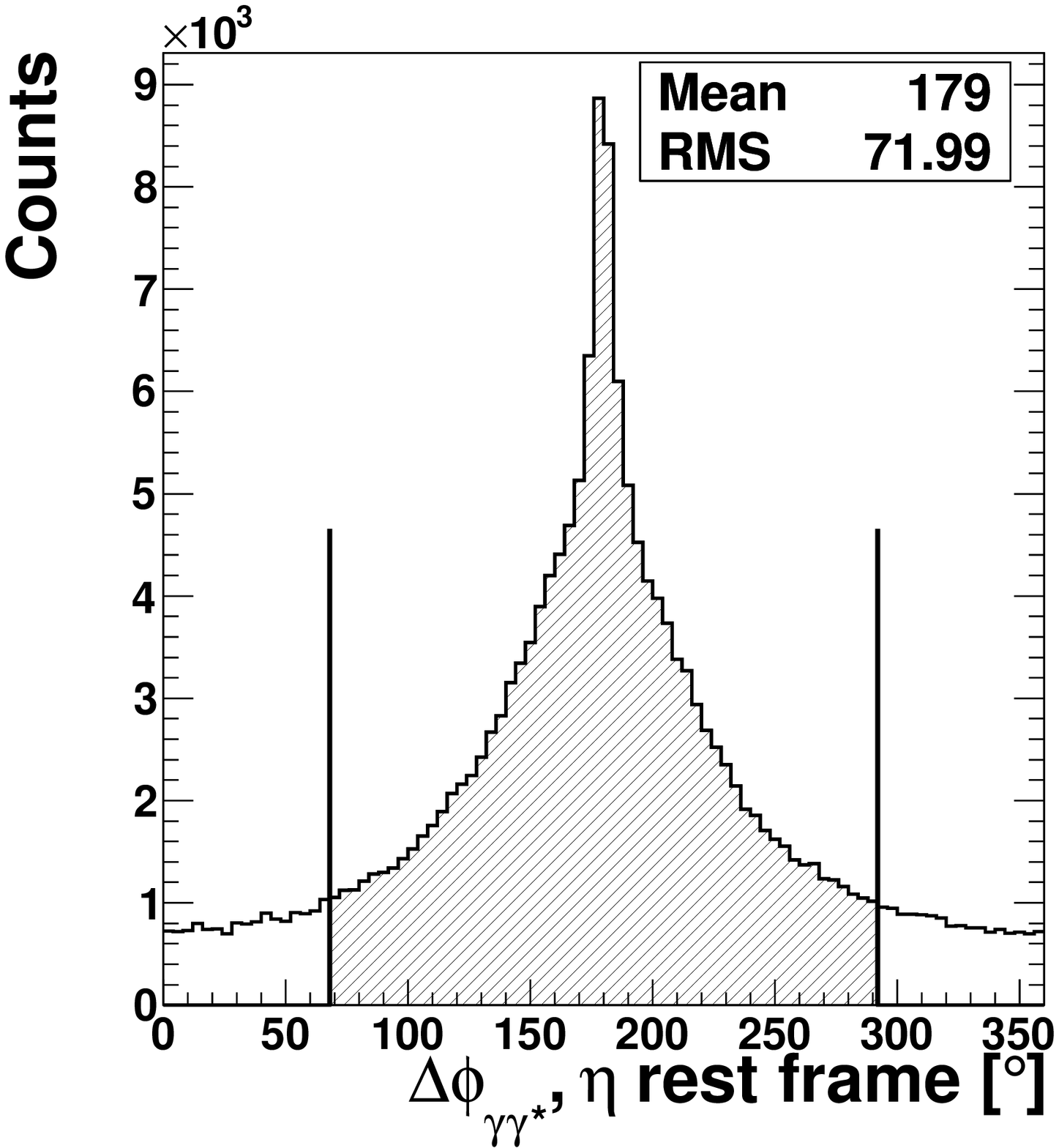}
  }
  \caption{ Experimental distributions of the difference in the azimuthal angle between real and virtual photon $\Delta\phi_{\gamma\gamma*}$, 
	    in the \e{} rest frame. Shaded area corresponds to the window accepted in the analysis.}
  \label{fig:dP}
\end{figure}

In order to check the $\Delta\phi_{\gamma\gamma*}$ reconstruction resolution, the histogram in \fig{}\ref{fig:dPRes} was filled with values 
calculated as $\Delta\phi_{\gamma\gamma*}^{true} - \Delta\phi_{\gamma\gamma*}^{reconstructed}$. 
The FWHM of the distribution is ${\sim9^{\circ}}$. 
\fig{}\ref{fig:dP} shows experimental distributions of $\Delta\phi_{\gamma\gamma*}$  with and without usage of the charged 
particle identification (~see Section~\ref{sec:IDlep}, \fig{}\ref{fig:dIdentSEC_OA}~). If in a given event more than one 
photon candidate was identified, then it was also included in the picture. The cut was chosen conservatively in the range 
of $[70,290]$ degrees of $\Delta\phi_{\gamma\gamma*}$ difference as shown by vertical lines in \fig{}\ref{fig:dP}.
Application of the \ee{} identification causes, that the signal channel is more pronounced and, therefore, the $\Delta\phi_{\gamma\gamma*}$ 
distribution becomes sharper around $180^{\circ}$.

\begin{figure}[!b]
 \subfigure[ All photon candidates]{
	\label{fig:Krot0}
	\includegraphics[width =0.5\textwidth,height=0.45\textwidth]{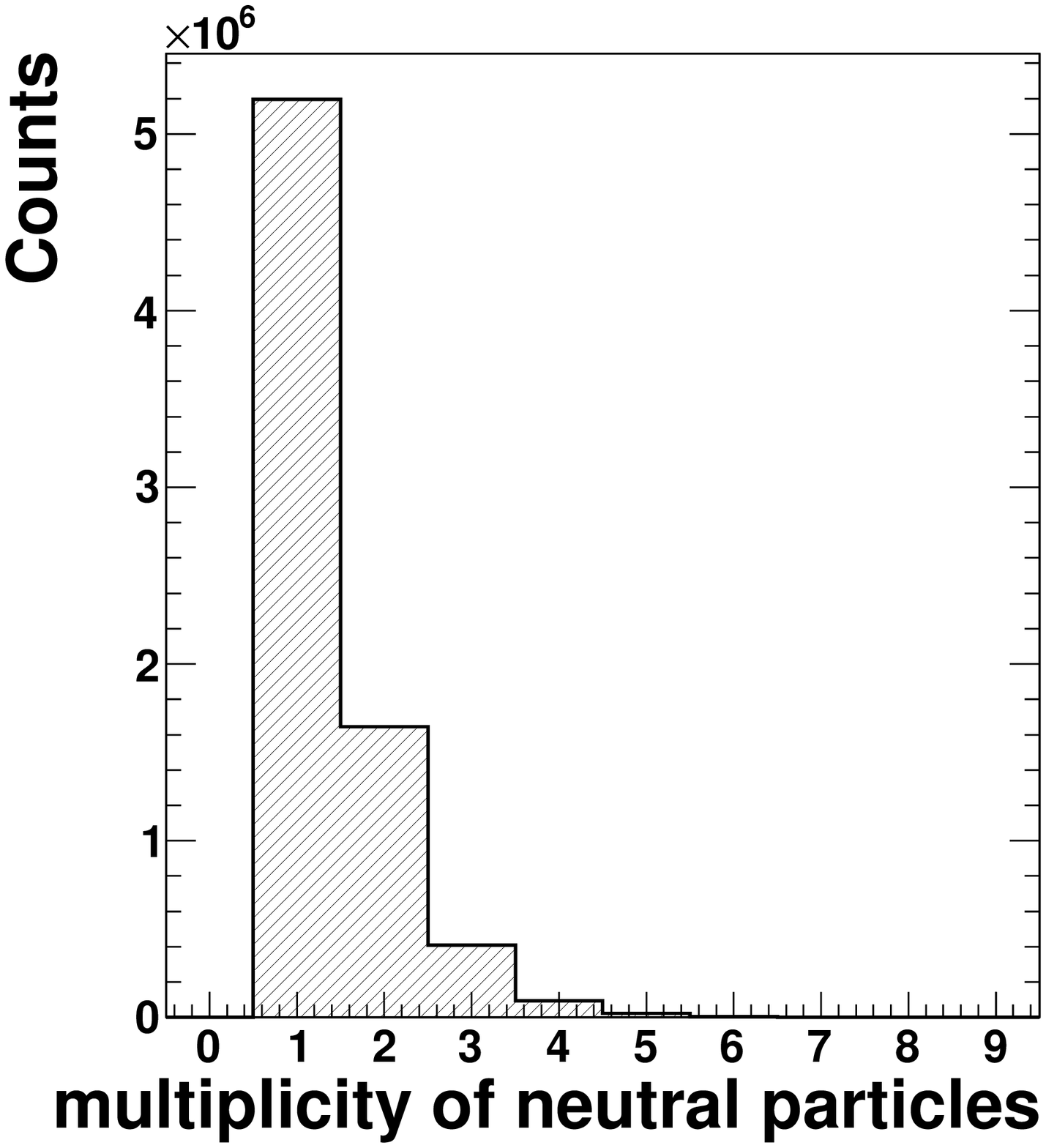}
  }
 \subfigure[ Photons candidates within $\Delta\phi_{\gamma\gamma*}$ cut ]{
	\label{fig:Krot1}
	\includegraphics[width =0.5\textwidth,height=0.45\textwidth]{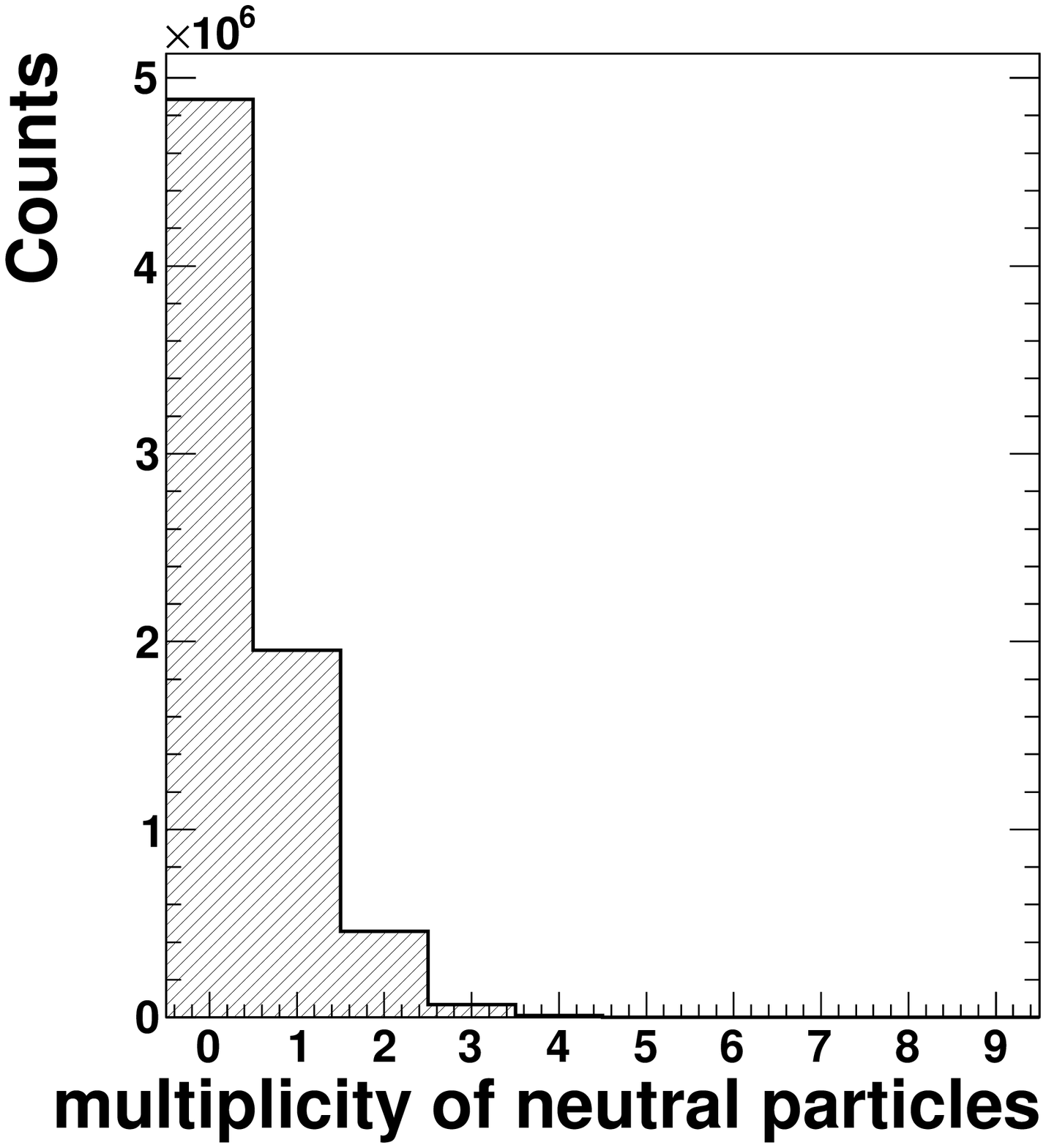}
  }
  \caption{ Multiplicity of candidates for neutral particles. }
  \label{fig:Krot}
\end{figure}

Multiplicity of neutral particles is shown in \fig{}\ref{fig:Krot}. 
The initial situation is presented in \fig{}\ref{fig:Krot0}. Here, the only demand was to have at least one neutral particle in each event. 
The situation after applying selection cuts described above is depicted in \fig{}\ref{fig:Krot1}. Most of photons candidates do not fulfill 
selection criteria. For further analysis, those events, where only one photon matches the $\Delta\phi_{\gamma\gamma*}$ criteria, are accepted.

\outroformatting

 \section{Identification of Leptons }
\label{sec:IDlep}
\introformatting
Selection of leptons aims at reducing the background from reactions with charged pions such as the \eppg{} and the \mbox{$\displaystyle \eta \rightarrow \pi^+ \pi^- \pi^0 \,$}.

\begin{figure}[!h]
 \subfigure[For positively charged particles registered\newline in the \cd{}]{
	\label{fig:LepTimeP}
	\includegraphics[width =0.5\textwidth,height=0.5\textwidth]{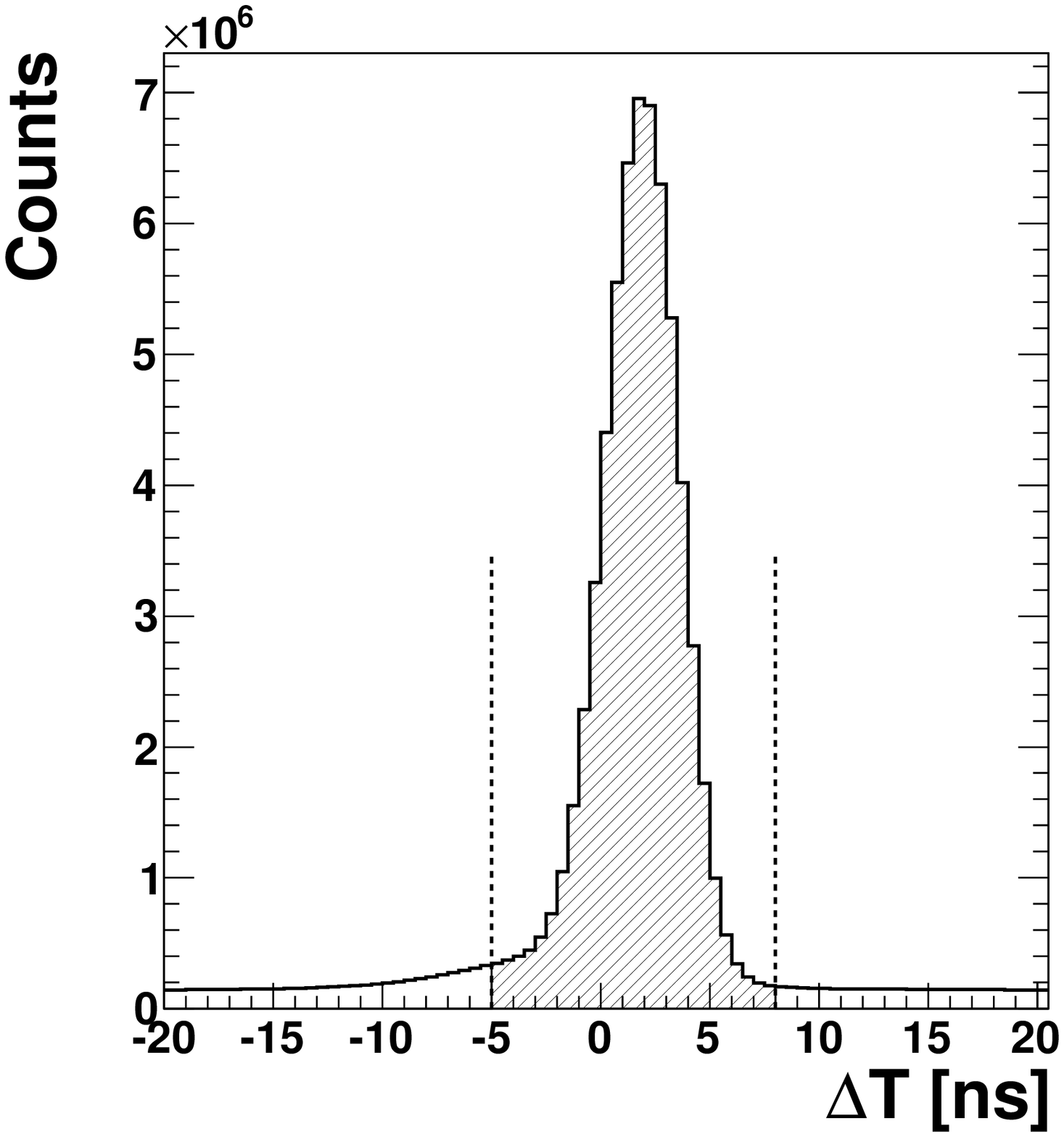}
  }
 \subfigure[For negatively charged particles registered in the \cd{}]{
	\label{fig:LepTimeE}
	\includegraphics[width =0.5\textwidth,height=0.5\textwidth]{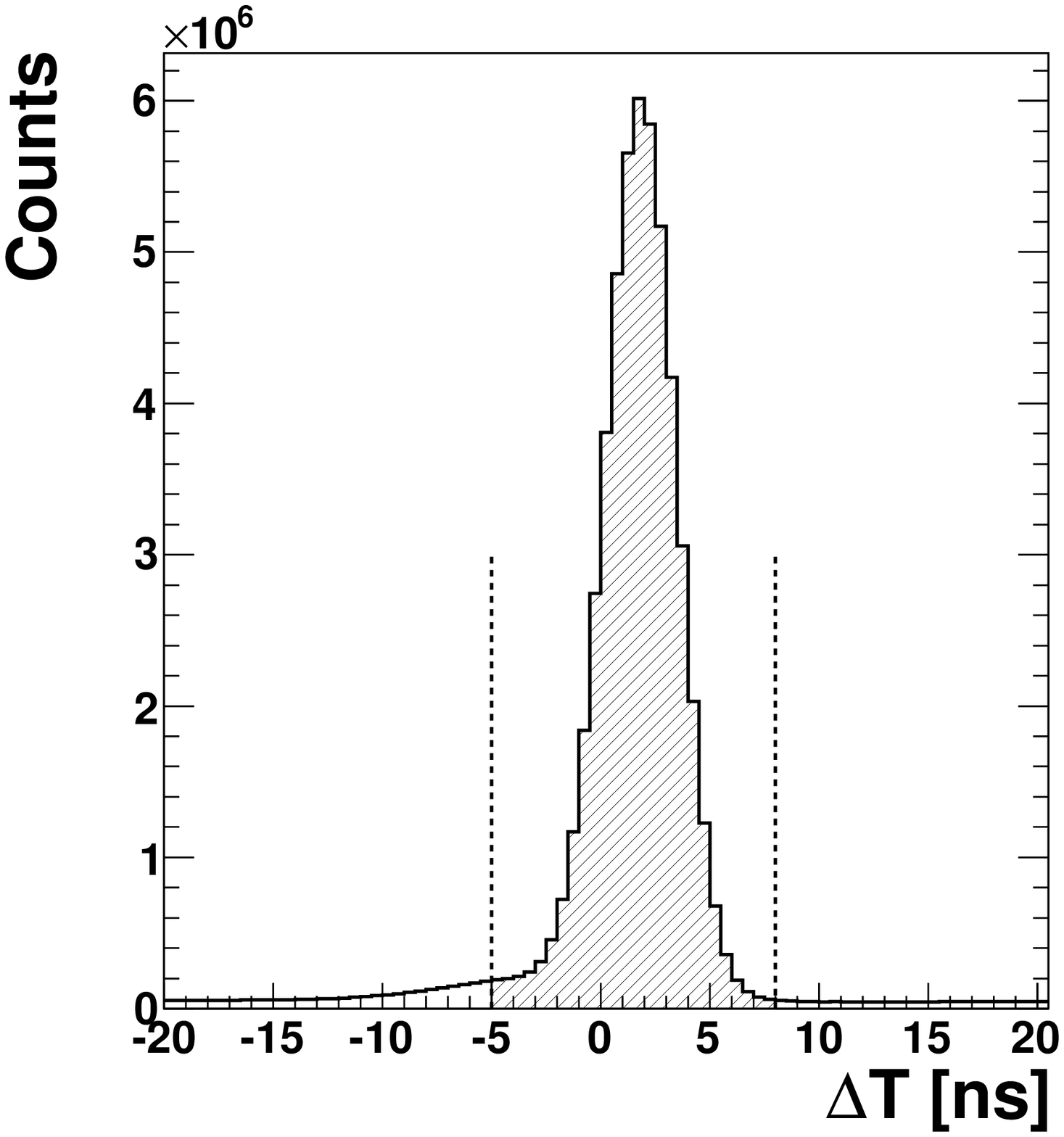}
  }
  \caption{ Experimental distribution of the time difference between charged particles in the \cd{} and \he{} ions identified in the \fd{}. Dashed lines show time cut used in analysis.}
  \label{fig:LepTime}
\end{figure}

In the first step, to make sure that tracks from oppositely charged particles, reconstructed in the \cd{}, come from the \mbox{$\displaystyle pd \rightarrow {}^{3}He\, \eta \to  {}^{3}He\,\gamma \gamma^*$}\\ \mbox{$\displaystyle\to {}^{3}He\,\gamma\,  e^- e^+ $} reaction, they are checked for time coincidences with the \he{} identified in the \fd{}. Corresponding spectrum is shown in \fig{}\ref{fig:LepTime}. A cut on time difference $\Delta T$ of [-5,8] ns was chosen.

\begin{figure}[!t]
\centering
	  \includegraphics[width =0.6\textwidth,height=0.55\textwidth]{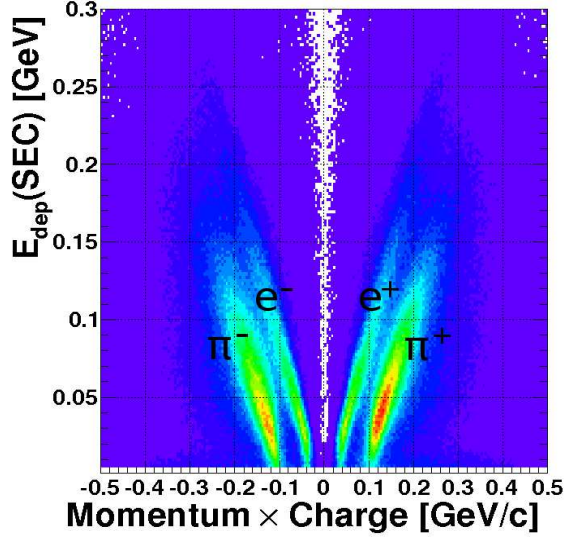}
	\caption{ Simulated spectrum of the energy deposited in the calorimeter as a function of the product of particles' momentum and charge. Lepton bands arrange themselves close to the momentum-$E_{dep}$ diagonal, while pions, carrying same momentum, have relatively smaller energy and they group below lepton bands. }
	\label{fig:sIdentSEC}
\end{figure}

The charged particle identification in the \cd{} aims at selecting tracks coming from $e^{+}$ and $e^{-}$ particles. For this purpose, correlation between energy deposited in the \se{} and particle's momentum is used. Four densely populated areas are visible in  \fig{}\ref{fig:sIdentSEC}. As marked in the picture, they correspond to leptons and pions with opposite charge.

\begin{figure}[!htb]
 \subfigure[Before restriction on $\Omega^{+,-}$]{
	\label{fig:dIdentSEC_b}
	\includegraphics[width =0.5\textwidth,height=0.48\textwidth]{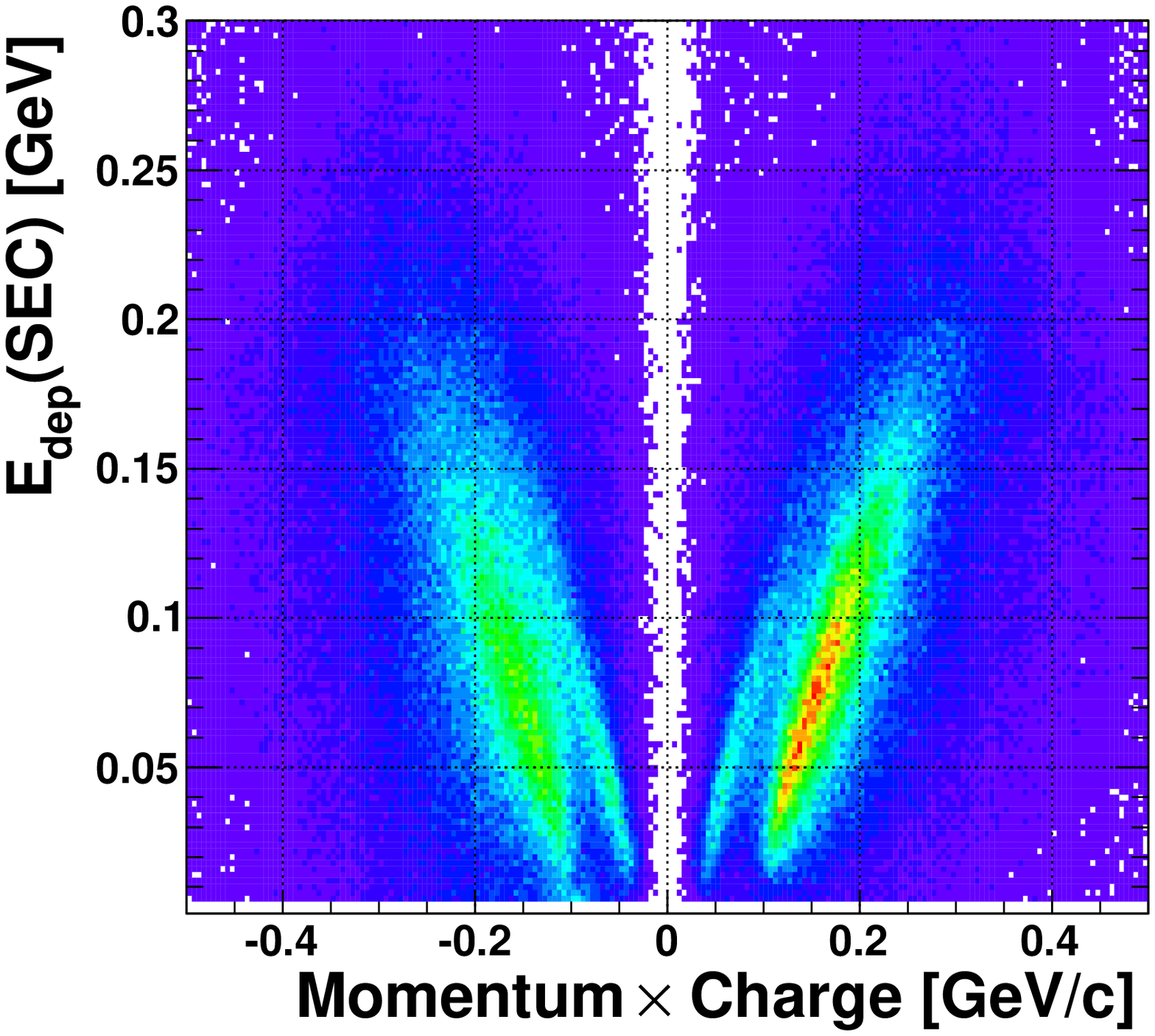}
  }
 \subfigure[$\Omega^{+,-} < 10^{\circ}$ ]{
	\label{fig:dIdentSEC_OA}
	\includegraphics[width =0.5\textwidth,height=0.48\textwidth]{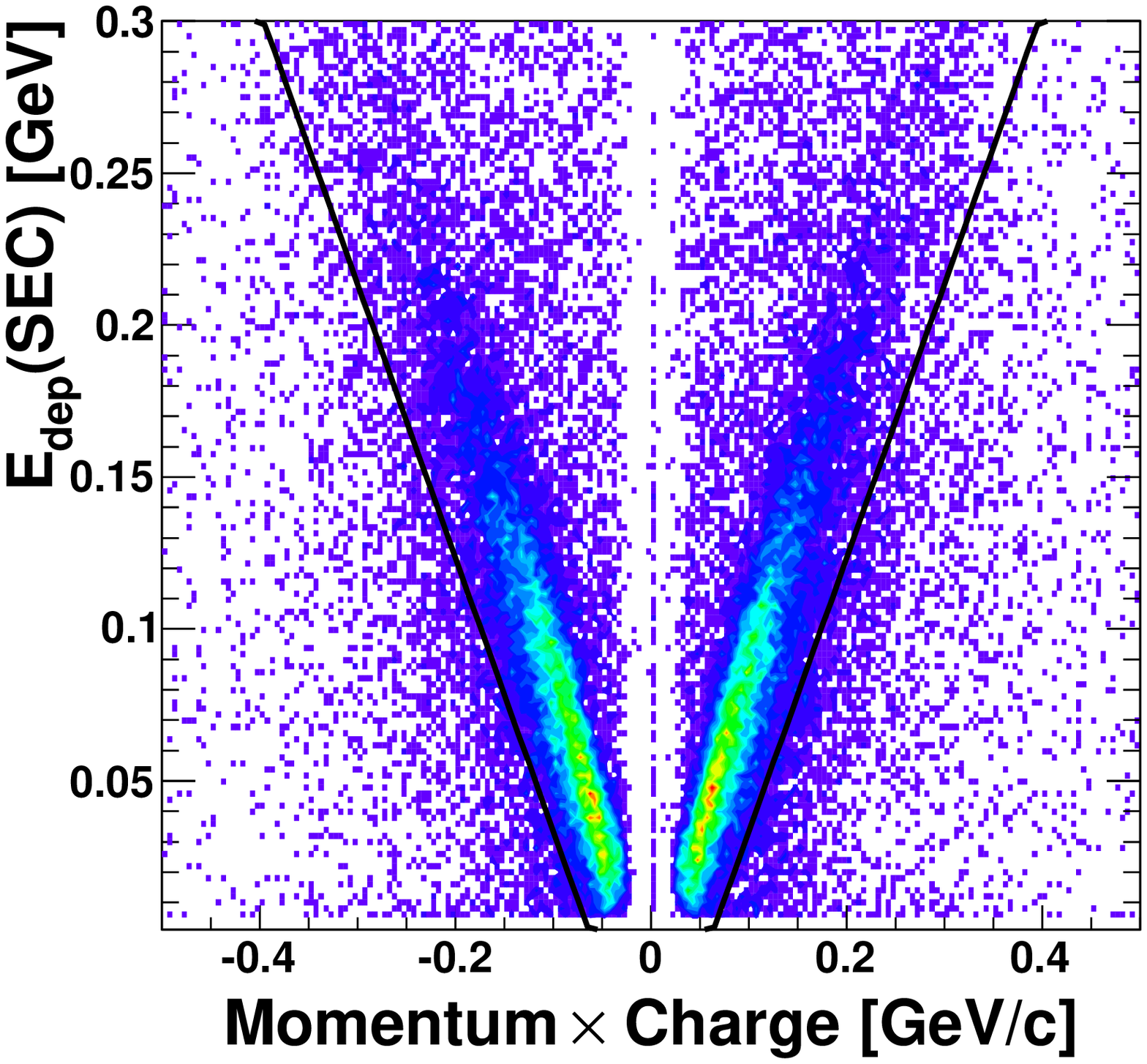}
  }
  \caption[Identification in SEC]{ Experimental spectra which are used to determine the area of leptons' occurrence.  Particles giving input to the region above solid lines shown in (\subref{fig:dIdentSEC_OA}), are considered as leptons. }
  \label{fig:dIdentSEC}
\vspace{-0.3cm}
 \end{figure}

However, this clear situation is observed in simulations only, where the number of simulated electrons and pions is similar (e.g. in $\eta \rightarrow e^{+}e^{-}\pi^{+}\pi^{-}$ or $\eta \rightarrow \pi^{+}\pi^{-}\pi^{0} \rightarrow \pi^{+}\pi^{-}(e^{+}e^{-}\gamma)$ reactions). In the analysis of experimental data, due to the higher number of pions with respect to electrons, pions' bands are shading electrons'. Such situation is seen even after \he{} and \g{} selection, when one demands exactly two, oppositely charged particles in the \cd{} correlated in time with the \he{} (see \fig{}\ref{fig:dIdentSEC_b}).
\begin{figure}[!hb]
	\includegraphics[width =0.5\textwidth,height=0.48\textwidth]{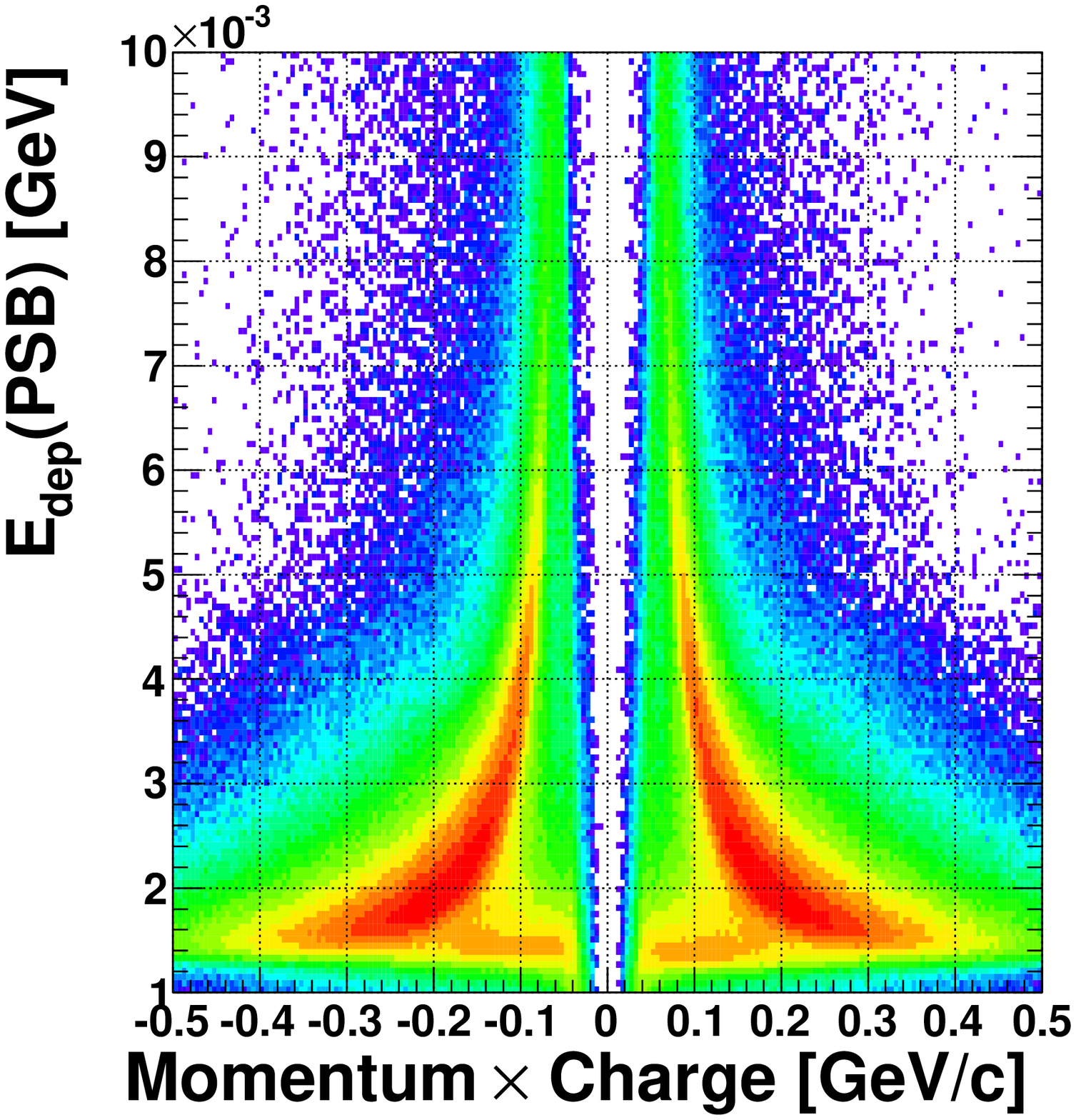}
	\includegraphics[width =0.5\textwidth,height=0.48\textwidth]{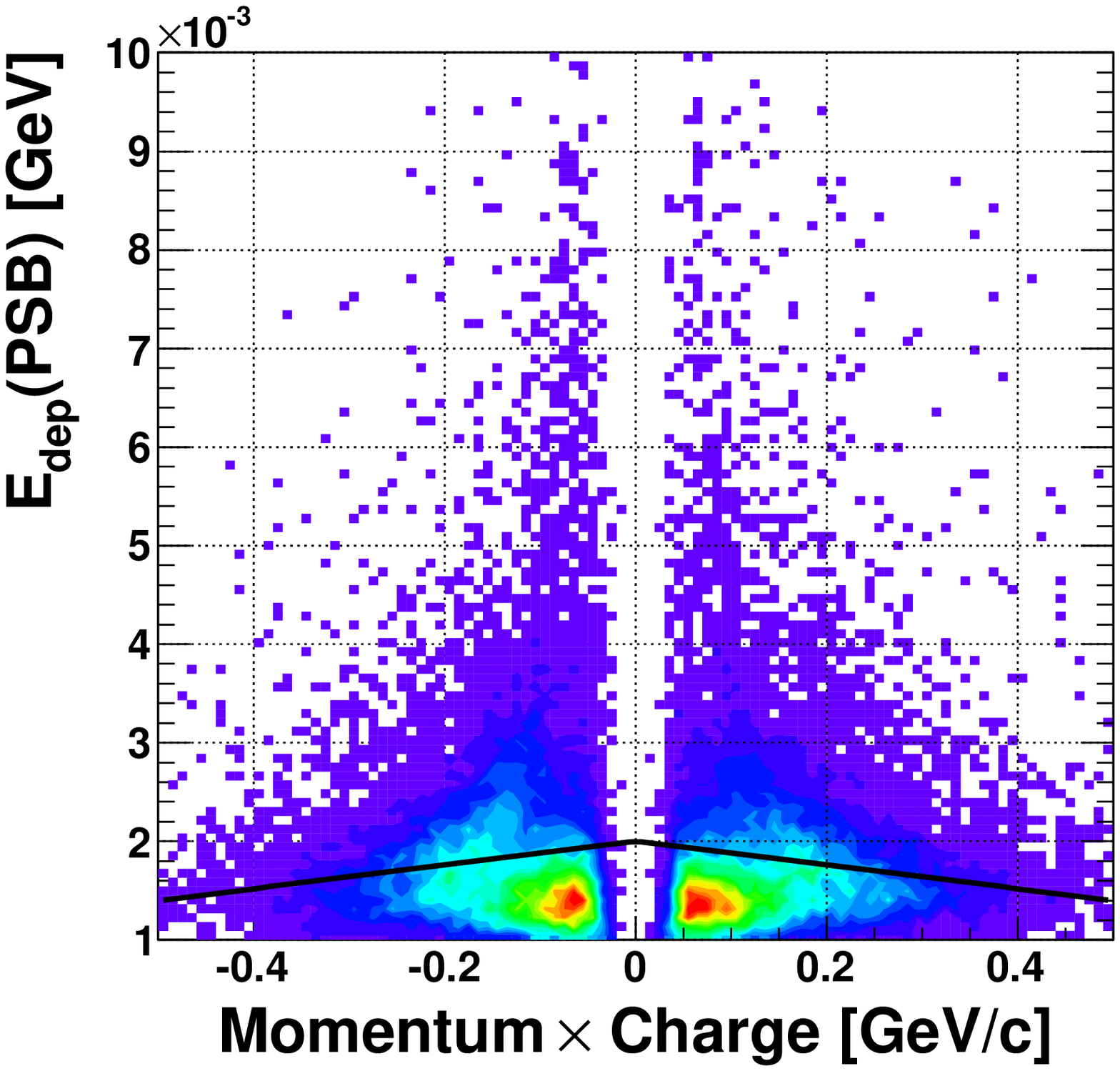}
  \caption[Identification in PSB]{ Energy deposited in the \psb{} as a function of the product of particles' momentum and charge from simulations (\underline {left}) and experiment (\underline {right}). The experimental spectrum was plotted after applying particle identification based on dependence of the energy deposition in the calorimeter on the particles momenta. Solid lines show the course of the cut.}
  \label{fig:dIdentPSB}
 \end{figure}

Therefore, to set up the region within which leptons are located, the identification plot is made with an additional restriction on the opening angle between positively and negatively charged particles $\Omega^{+,-}$. In the overwhelming majority, leptons from the ${\gamma^{*}\rightarrow e^{+}e^{-}}$ conversion create small opening angle \cite{Jarlskog:1967}.
Experimental spectrum in \fig{}\ref{fig:dIdentSEC_OA} is plotted under condition that $\Omega^{+,-}$ is less than $10^{\circ}$. This allows (on the basis of experimental data) to define the region where leptons are located. It is important to stress, that this restriction is not used in the further analysis.

Particles, identified in the calorimeter as $e^{+}$ and $e^{-}$, are further checked, if also in the \psb{} 
their energy losses are as expected for electrons. The simulated spectrum of the energy deposited in the \psb{} as a function of the product of particles' momentum and charge is shown in the left panel of  \fig{}\ref{fig:dIdentPSB}.
Below pion bands, electrons are forming stripes with almost a constant energy deposit in the whole range of the momentum.
The same distribution from experimental data, but containing only particles which fell into the lepton's region in the calorimeter, is presented in the right panel.
Misidentified particles in the \se{} are rejected in the \psb{} using the cut shown as a solid line in the right panel of \fig{}\ref{fig:dIdentPSB}.

\begin{figure}[!h]
\centering
	  \includegraphics[width =0.6\textwidth, height=0.45\textwidth]{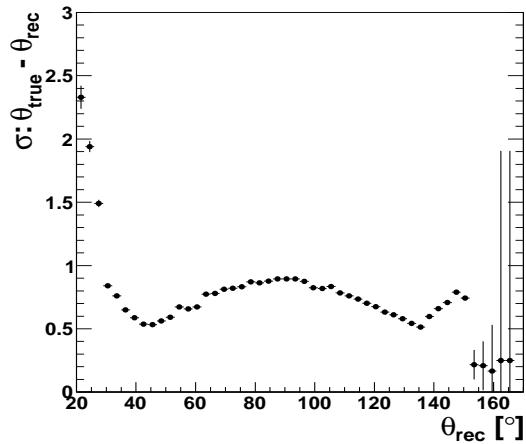}
	\caption{ The absolute error (standard deviation) of the leptons' polar angle reconstruction as a function of the reconstructed polar angle. At higher angles, due to the lower statistics, a good fit was not possible. For this plot $10^{6}$ of \eeeg{} events has been simulated.}
	\label{fig:pos_SigTh}
\end{figure}

The maximal geometrical acceptance of the \cd{} is $20-169$ degrees. Due to the lower granularity of the calorimeter crystals in the back part, and an exit cone in the front part of the \cd{}, the reconstruction of charged particles is worst in these regions. \fig{}\ref{fig:pos_SigTh} 
shows the standard deviation of uncertainty of the leptons' polar angle reconstruction as a function of reconstructed polar angle. A significant worsening of accuracy is observed for small and highest angles. Therefore, it is demanded in the further analysis that leptons were emitted in the range of the polar angle $\theta$ of $30-150$ degrees.

\clearpage

\outroformatting

 \section{Selection of the \eeeg}
\label{sec:IDsig}
\introformatting

\begin{figure}[!h]
\centering
	  \includegraphics[width =0.65\textwidth]{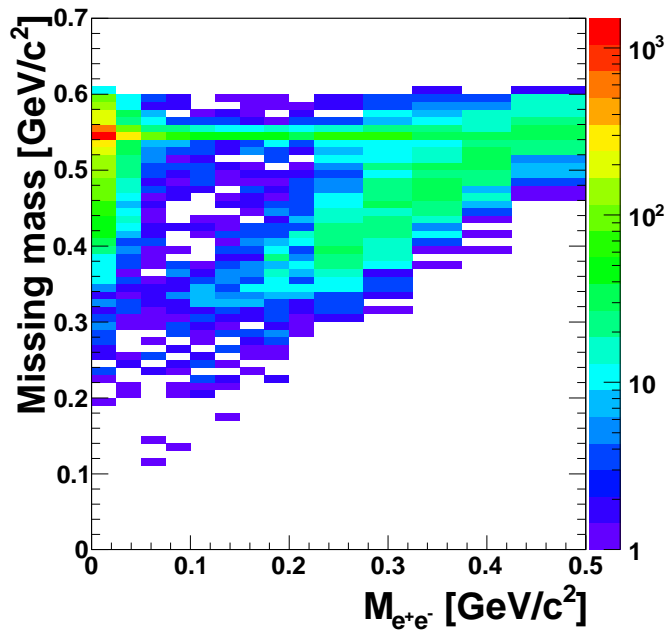}
	\caption[]{The experimental distribution of the missing mass for the ${pd \rightarrow {}^{3}HeX}$ reaction as a function of the invariant mass of the lepton pairs after particle selection described in the previous section.}
	\label{fig:d_b4_HeTMeBp}
\end{figure}

\begin{figure}[!h]
\centering
	  \includegraphics[width =0.65\textwidth]{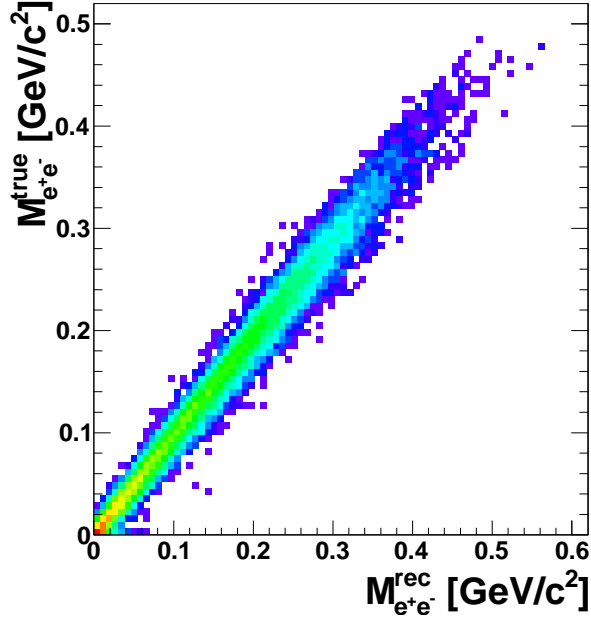}
	\caption[]{Distribution of generated $M_{e^+e^-}^{\text{true}}$ mass as a function of generated and then reconstructed $M_{e^+e^-}^{\text{rec}}$ mass of \ee{} pairs.}
	\label{fig:mc_lm_TvR}
\end{figure}

After the particle identification described above, one can plot the distribution of the missing mass for the $pd \rightarrow {}^{3}HeX$ reaction as a function of the invariant mass of the lepton pairs, \mee{}. This spectrum is shown in \fig{}\ref{fig:d_b4_HeTMeBp}. The width of intervals of \mee{} mass was chosen based on the \mee{} resolution and it is growing with the \mee{} mass. The corresponding spectrum of generated $M_{e^+e^-}^{\text{true}}$ mass as a function of generated and then reconstructed $M_{e^+e^-}^{\text{rec}}$ mass is shown in \fig{}\ref{fig:mc_lm_TvR}. One can see that the reconstruction resolution worsen for higher \mee{} masses. 

\begin{figure}[!h]
\centering
	  \includegraphics[width =0.65\textwidth]{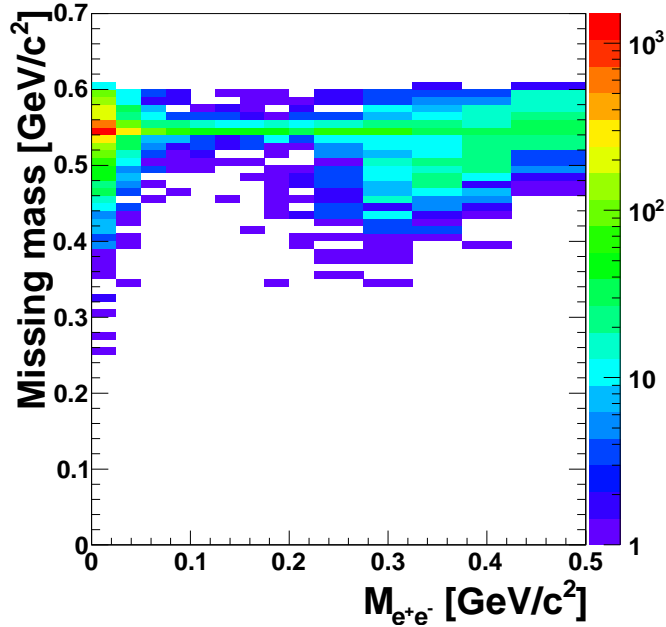}
	\caption[]{The experimental distribution of the missing mass for the ${pd \rightarrow {}^{3}HeX}$ reaction as a function of the invariant mass of the lepton pairs after particle selection described in the previous section and after cut on the \he{} scattering angle in laboratory frame ($\theta_{{}^{3}He}<11^{\circ}$).}
	\label{fig:d_b4_MeBp}
\end{figure}

In \fig{}\ref{fig:d_b4_HeTMeBp}, an enhancement in the number of entries is visible for the missing mass equal to the mass of the \e{} meson. However, this signal appears on a background caused by reaction in which only helium and pions were produced namely, the \Hpp{} reaction. In case of this events, the particle identification described in the previous section turned out to be not sufficient. Nevertheless, this background can be well suppressed in the region below the \e{} mass, by a cut on the scattering angle of helium ions, $\theta_{{}^{3}He}$. It is because in the case of the pion production, the maximal laboratory \he{} scattering angle increases as the missing mass decreases. \he{} particles produced in the \pdhe{} reaction at a beam momentum of $1.7~GeV/c$ are emitted up to the $\sim10^{\circ}$ of the \he{} scattering angle. Setting an upper limit on the $\theta_{{}^{3}He}$ makes the signal better visible. It is especially useful in the region of higher invariant masses of \ee{} pairs, where the 
statistics is decreasing.
The influence of this cut can be seen by comparing spectra in \fig{}\ref{fig:d_b4_HeTMeBp} (before applying this cut) and \fig{}\ref{fig:d_b4_MeBp} (after cut application).

\begin{figure}[!h]
\centering
	  \includegraphics[width =0.65\textwidth,height =0.6\textwidth]{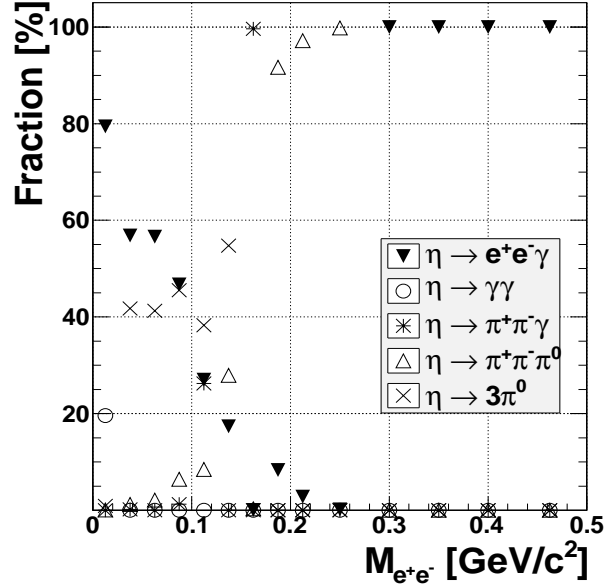}
	\caption[]{The percentage share of simulated \e{} decay channels as a function of the invariant mass of leptons pairs, plotted after selection described in the previous sections.}
	\label{fig:frac_c7}
\end{figure}

Direct production of pions results in a continuous missing mass distribution and can be subtracted from the signal by plotting a missing mass spectrum for each \ee{} invariant mass interval separately.
However, the signal on the missing mass spectrum corresponding to the mass of the \e{} meson may be not only due to the investigated \eeeg{} decay but also due to the  
\gaga{}, \eppg{}, \eppp{} and \etp{} decays.
The listed \e{} decays may still contribute to the signal, mainly because of the particle misidentification and the external conversion of photon in the detector material. As it will be shown in Sec.~\ref{s:back}, almost $3.5\%$ of the signal content at this stage of the analysis is due to the presence of \eppg{}, \eppp{} and, especially, \etp{} decays. The number of $3.5\%$ is very big since the background events contributing to it, are characterized by higher values of the invariant mass of misidentified leptons pairs and in this region the signal channel has a low cross section. The situation is shown in \fig{}\ref{fig:frac_c7}, where the percentage share\footnote{The {\it percentage share} is defined as the expected percentage contribution of considered \e{} decay channels to the 
total number of events reconstructed as \eeeg{}.}.
In the region of the invariant mass of leptons pairs from $\sim0.1$\um{} up to $\sim0.25$\um{}, the signal channel is drawn strongly from the events coming from the \e{} decay channels with pions. Below $0.1$\um{} the background comes mainly from the \etp{} and \gaga{} decays. 
In the very small invariant masses of leptons pairs, the \gaga{} decay channel is the dominating background contributor. It is due to the photons conversion at the beam pipe and will be described in Sec.~\ref{ssec:Bp}. 

\fig{}\ref{fig:frac_c7} shows, that there is essentially no background originating from the \e{} meson decays for \mee{} invariant masses higher than $0.3$\um{}. In this region, the main background is due to the direct pion production.

\subsection{Missing Mass for the $pd\rightarrow X\eta$ Reaction }
\label{ssec:MissEta}
\begin{figure}[!h]
	\centering
	\includegraphics[width =0.48\textwidth,height=0.48\textwidth]{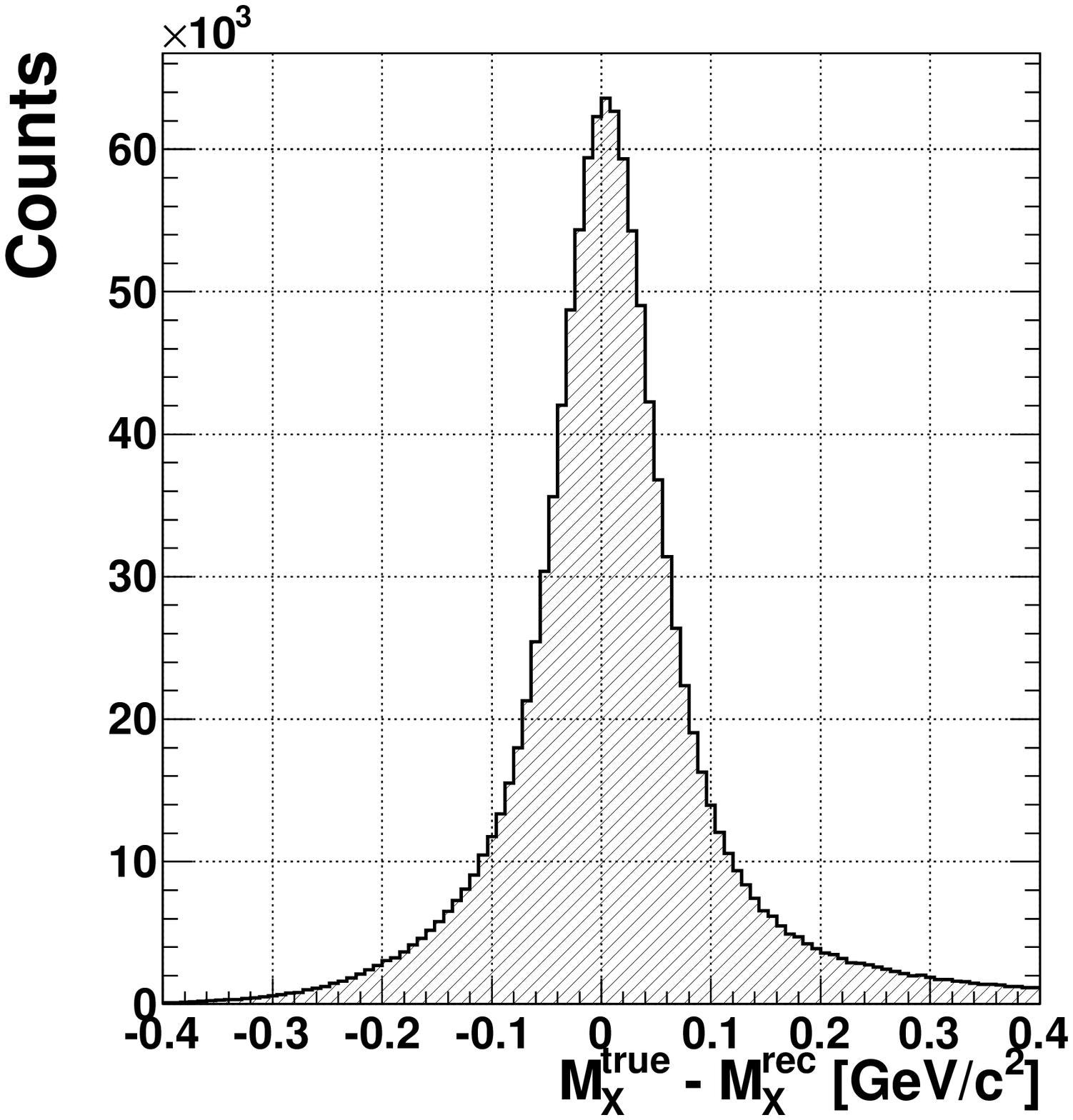}
	\includegraphics[width =0.48\textwidth,height=0.48\textwidth]{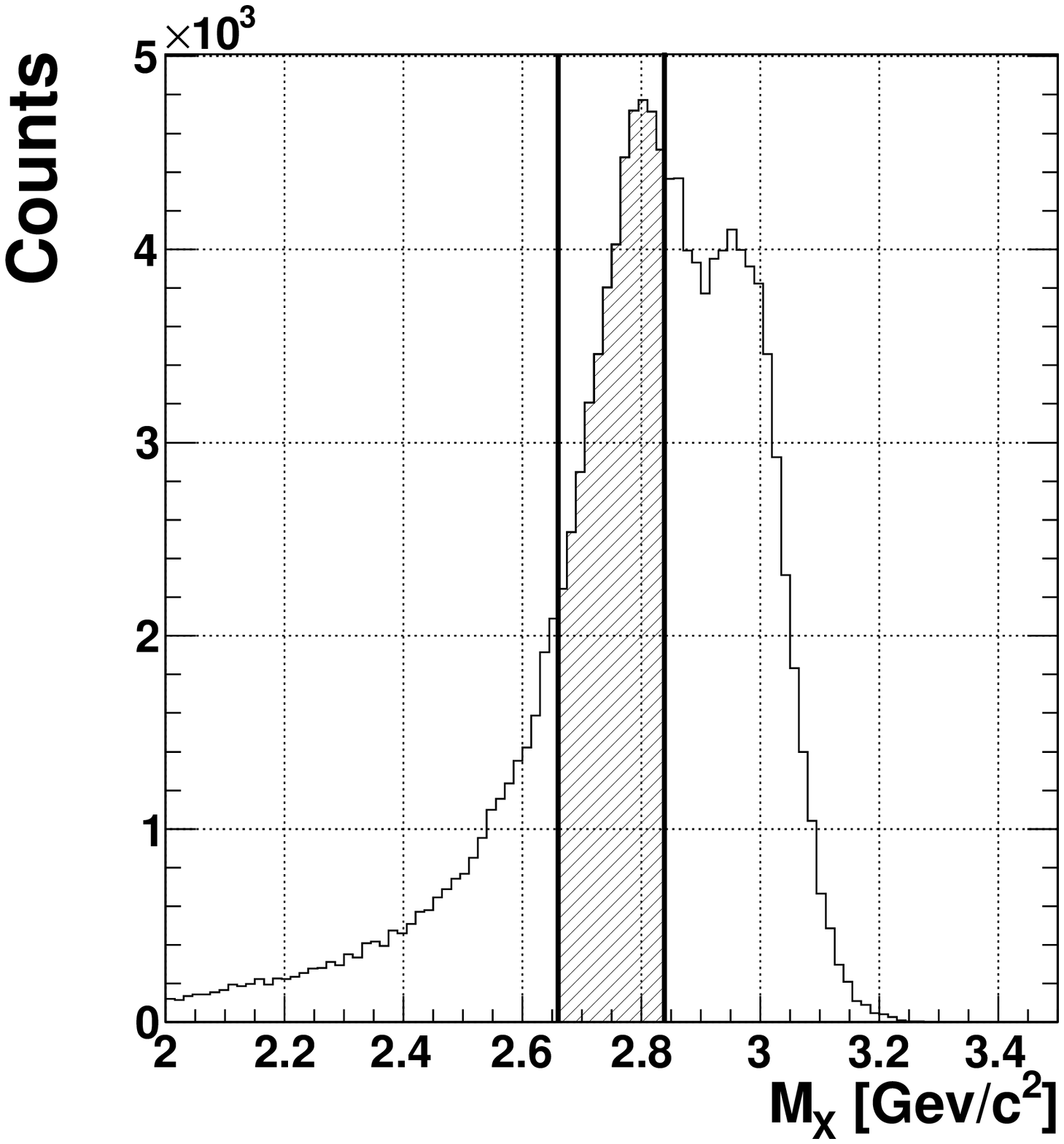}
	\caption[]{ \underline {Left}: The distribution of the difference between the true and the reconstructed value of the $M_{X}$. \underline {Right}: The experimental distribution of the $M_{X}$ mass. The cut was chosen in the range of $2.66-2.84$\um{} of $M_{X}$ as indicated by solid lines. }
  \label{fig:d_MissEtaMM}
\end{figure}

It is assumed in the analysis, that all charged particles reconstructed in the \cd{} have a mass of electron. Therefore, it is useful to look at the spectrum of the missing mass for the $pd \rightarrow Xe^{+}e^{-}\gamma$ reaction, calculated as:
\begin{equation}
\label{eq:MissEta}
(M_{X})^2c^4 = ( E_{beam} + M_{target}c^{2} - \sum E_{i} )^2 - c^2( \vec{p}_{beam} - \sum \vec{p}_{i} )^2,
\end{equation}
where mass of X should be equal to the mass of the \he{} particle ($M_{{}^{3}He}=2.809~\text{GeV/c}^2$), which has been already identified in the \fd{}, whereas index {\it i} runs over particles registered in the \cd{}.
\begin{figure}[!h]
 \subfigure[\mbox{$pd \rightarrow {}^{3}He\eta \to {}^{3}He \pi^+ \pi^- \gamma $} ]{
	\label{fig:ppgMissEta}
	\includegraphics[width =0.48\textwidth,height=0.48\textwidth]{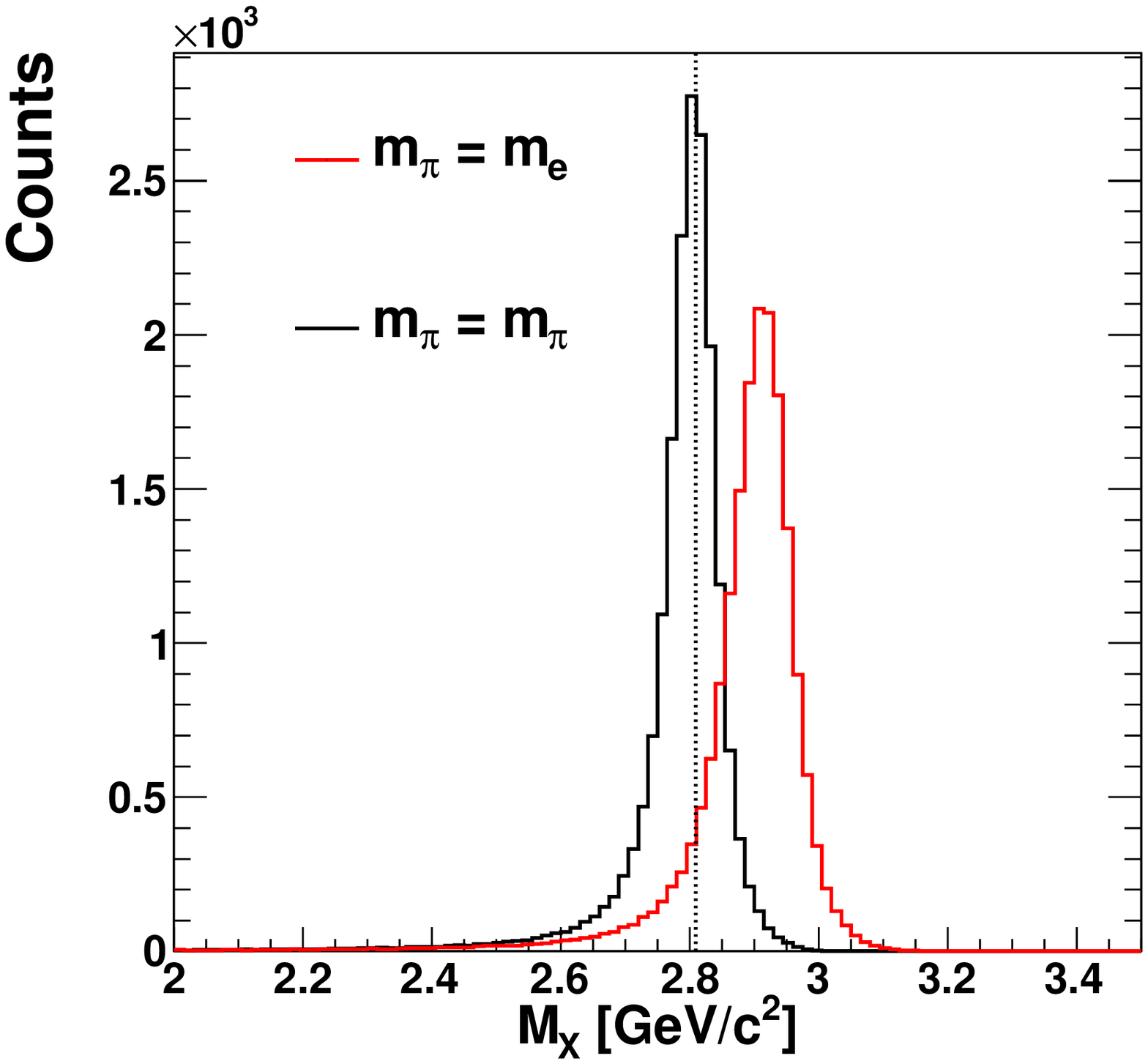}
  }
 \subfigure[\mbox{$pd \rightarrow {}^{3}He\eta \to {}^{3}He \pi^+ \pi^- \pi^0 $}]{
	\label{fig:pppMissEta}
	\includegraphics[width =0.48\textwidth,height=0.48\textwidth]{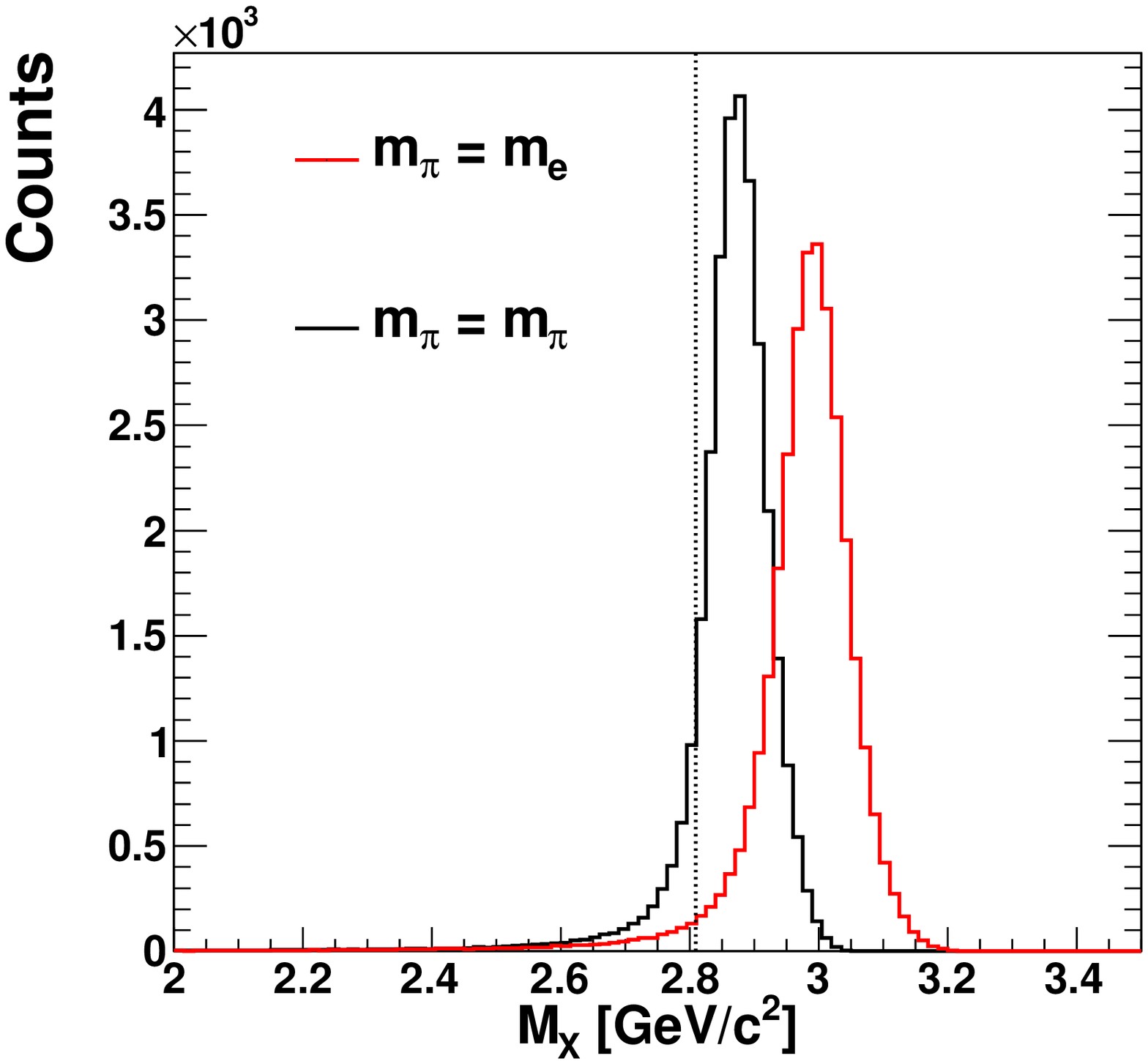}
  }
  \caption{ Simulated missing mass distribution, $M_{X}$, for two \e{} decay channels. The straight, dashed line shows the \he{} mass, $M_{{}^{3}He}=2.809~\text{GeV/c}^2$. The electron's mass assumption causes spectra shift towards higher masses. In case of the \eppp{} decay, the shift is even more pronounced since there is, additionally, one photon missing. }
  \label{fig:MissEta}
\end{figure}
If the mass assumption for particles identified as electrons is wrong and pions were misidentified as electrons, the missing mass $M_{X}$ will not be equal to the mass of the recoil particle. In case of \e{} decays into pions it will be shifted towards higher masses as can be seen in \fig{}\ref{fig:MissEta}. Experimental spectrum of the missing mass, $M_{X}$ is shown in the right panel of \fig{}\ref{fig:d_MissEtaMM}. 
Broadening of the peak on the right side is caused by the presence of events from such \e{} decay channels like \etp{}, \eppg{} and \eppp{}.
The range of accepted $M_{X}$ was chosen from $2.66$\um{} to $2.84$\um{}. The lower boundary condition was taken as $\sim(M_{{}^{3}He} - 3\sigma)$. In order to estimate the resolution of the $M_X$ determination, the distribution of the difference between true and reconstructed values of the $M_{X}$ was established. It is shown in the left panel of \fig{}\ref{fig:d_MissEtaMM}.
The upper limit was chosen more restrictively than lower one as $\sim(M_{{}^{3}He} + 1\sigma)$
in order to suppress the background from the \e{} decays into pions. 

\begin{figure}[!h]
  \centering
	\includegraphics[width =0.65\textwidth,height=0.55\textwidth]{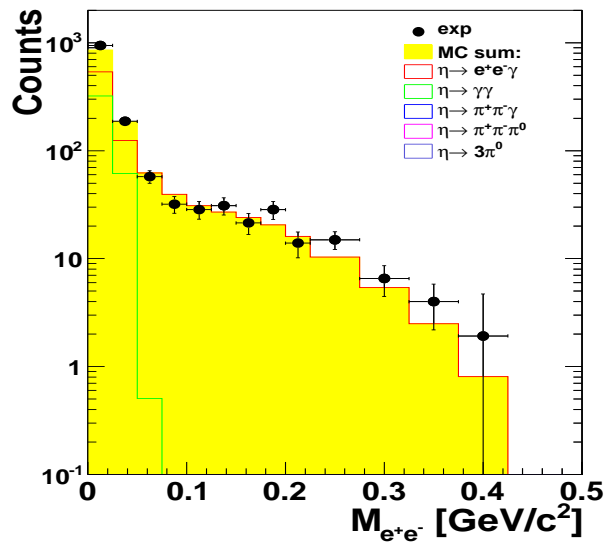}
    \caption{The spectrum of the \mee distribution from the experiment (black points) and the simulations (see the legend) after application of the cut on the missing mass for the $pd \rightarrow Xe^{+}e^{-}\gamma$ reaction.   }
  \label{fig:dNdM_me}
  \end{figure}

The influence of the cut on the missing mass for the $pd \rightarrow Xe^{+}e^{-}\gamma$ reaction is seen in \fig{}\ref{fig:dNdM_me}.
The previously very high contribution of \eppg{}, \eppp{} and \etp{} decay channels become insignificant.

\subsection{Photon's Conversion }
\label{ssec:Bp}
Despite of asking in the analysis for events with two oppositely charged particles, there is a large background coming from the \gaga{} decay channel, where only neutral particles are originally produced. It is due to the photons conversion at the beam pipe.
In this case, the \gaga{} decay channel has the same signature as the decay of interest.
Conversion events contribute strongly to the background in the low invariant mass of leptons pair as it is shown in \fig{}\ref{fig:frac_c7} and \ref{fig:dNdM_me}.

\begin{figure}[!h]
  \centering
  \includegraphics[width =0.65\textwidth]{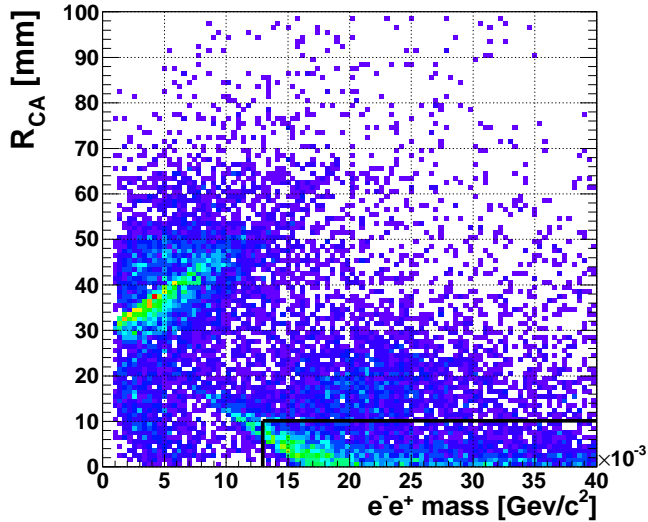}
  \caption[Experimental distribution of the $R_{CA}(M_{e^{+}e^{-}})$]{ The experimental distribution of the radius of the \ee{} closest approach as a function of the invariant mass of the \ee{} pair calculated assuming that electrons originate from the conversion at the beam pipe. }
  \label{fig:d_Bp}
\end{figure}

Lepton's pairs, having their origin in the material of the beam pipe, should create there small values of the $M_{e^+e^-}^{BP}$\footnote{The \mee{} is calculated as follows: i)particle azimuthal angle is evaluated at the beam pipe and used to calculate the momentum components, ii) assuming electron mass, the four-vectors are determined, iii) four-vectors are added and mass of created pair is calculated.}. Also, the distance, $R_{CA}$, between the points, where theirs paths are closest to each other, and the beam line, should be in order of the beam pipe radius\footnote{The beam pipe has a radius of $30$~mm and is made of $1.2$~mm thick beryllium \cite{Adam:2004ch}.}. The relevant distribution is shown in \fig{}\ref{fig:d_Bp}.
Conversion events populate the spectrum in the range of low mass, having $R_{CA}$ above $30$~mm. 
The condition used in the analysis was chosen so, that the $R_{CA}$ was lower than $10~$mm while the $M_{e^+e^-}^{BP}$, calculated assuming that electrons originate from the conversion at the beam pipe, is higher than $0.13\times10^{-3}~GeV/c^{2}$.

  \begin{figure}[!h]
  \centering
	\includegraphics[width =0.65\textwidth,height=0.55\textwidth]{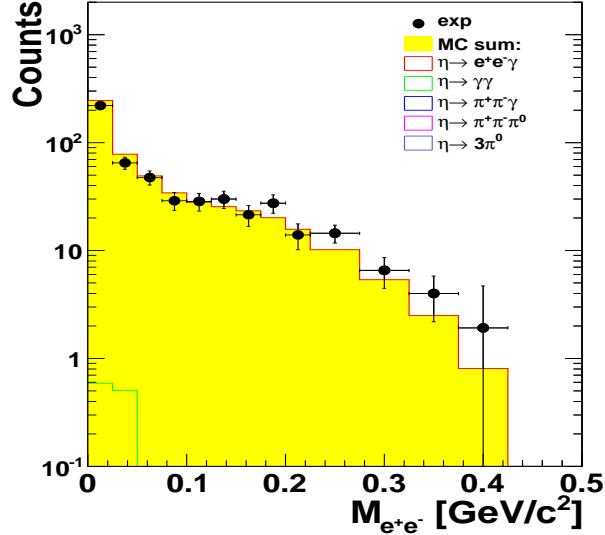}
  \caption{The spectrum of the \mee distribution from the experiment (black points) and the simulations (see the legend) after suppression of events originating from the $\gamma$ conversion at the beam pipe.}
  \label{fig:dNdM_bp}
  \end{figure}

In \fig{}\ref{fig:dNdM_me} one can see how the conversion influences the \mee{} distribution.
The spectrum is plotted after applying all the other cuts described previously. The green and red lines represent the simulated events of the \gaga{} and \eeeg{}, respectively. In yellow, the sum of the Monte Carlo generated decays (listed in the legend) is shown. The experimental points are plotted after the subtraction of the background from direct pions production and before efficiency correction.
The simulated events of the \gaga{} are entering in the \mee{} region up to $0.1$\um{}, contributing mostly up to the \ee{} mass of $0.06$\um{}. \fig{}\ref{fig:dNdM_bp} 
was plotted with the cut on the distribution of the radius of the closest approach as a function of the $M_{e^+e^-}^{BP}$ (see \fig{}\ref{fig:d_Bp}) added to the analysis.
The \gaga{} input to the signal has been reduced significantly.

\begin{figure}[!h]
  \centering
	\includegraphics[width =0.55\textwidth]{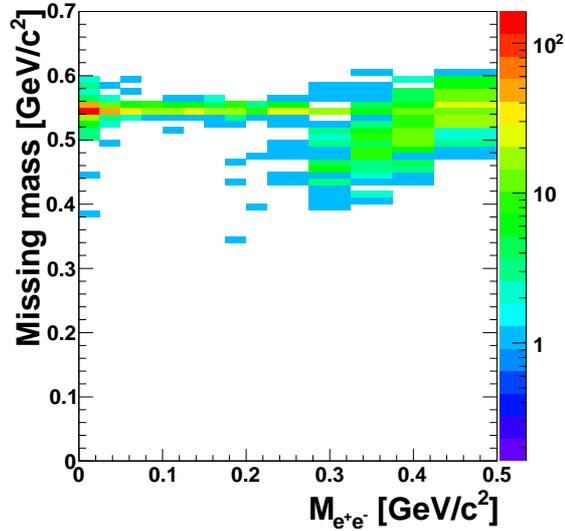}
	\caption{ Experimental distribution of the missing mass for the $pd \to {}^{3}He X$ reaction as a function of the invariant mass of the \ee{} pair after applying all cuts.}
	\label{fig:h_deeg0}
  \end{figure}

\begin{center}{\it Summary of Sec.~\ref{sec:IDsig} }\end{center}
After application of, above described, i) cut on the \he{} scattering angle, ii) the restriction on the missing mass for the $pd\rightarrow X\eta$ reaction and iii) the suppression of photons conversion at the beam pipe, one obtains the spectrum of the missing mass for the $pd\rightarrow {}^{3}He X$ as a function of the invariant mass of leptons pairs, \mee{}, as presented in \fig{}\ref{fig:h_deeg0}. 
\begin{figure}[!h]
 \subfigure[Projection of \fig{}\ref{fig:d_b4_HeTMeBp} onto the y-axis]{
	\label{fig:d_b4_MeBp_py}\centering
	\includegraphics[width =0.5\textwidth, height =0.45\textwidth]{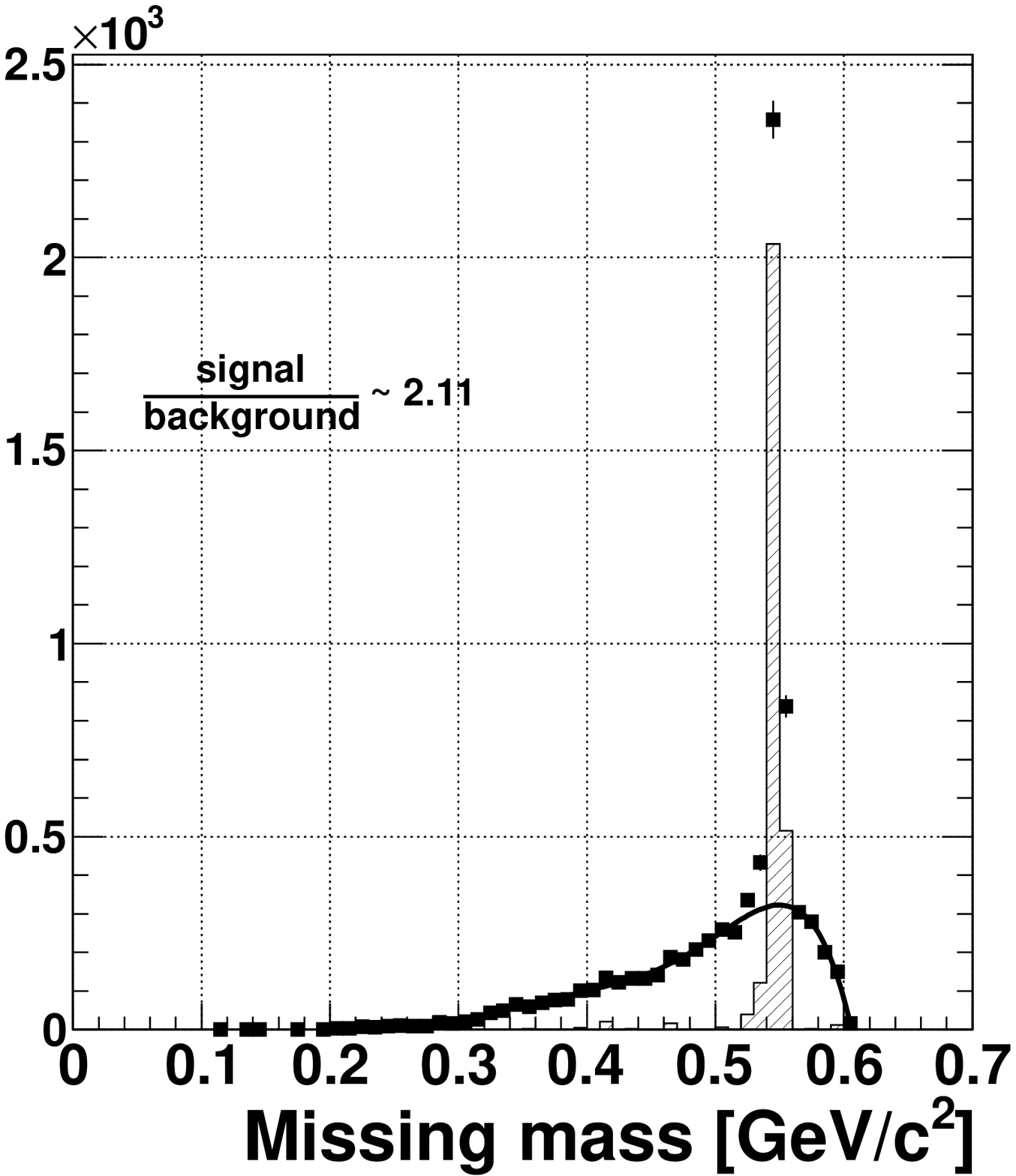}
  }
 \subfigure[Projection of \fig{}\ref{fig:h_deeg0} onto the y-axis]{
	\label{fig:h_deeg0_py}\centering
	\includegraphics[width =0.5\textwidth, height =0.45\textwidth]{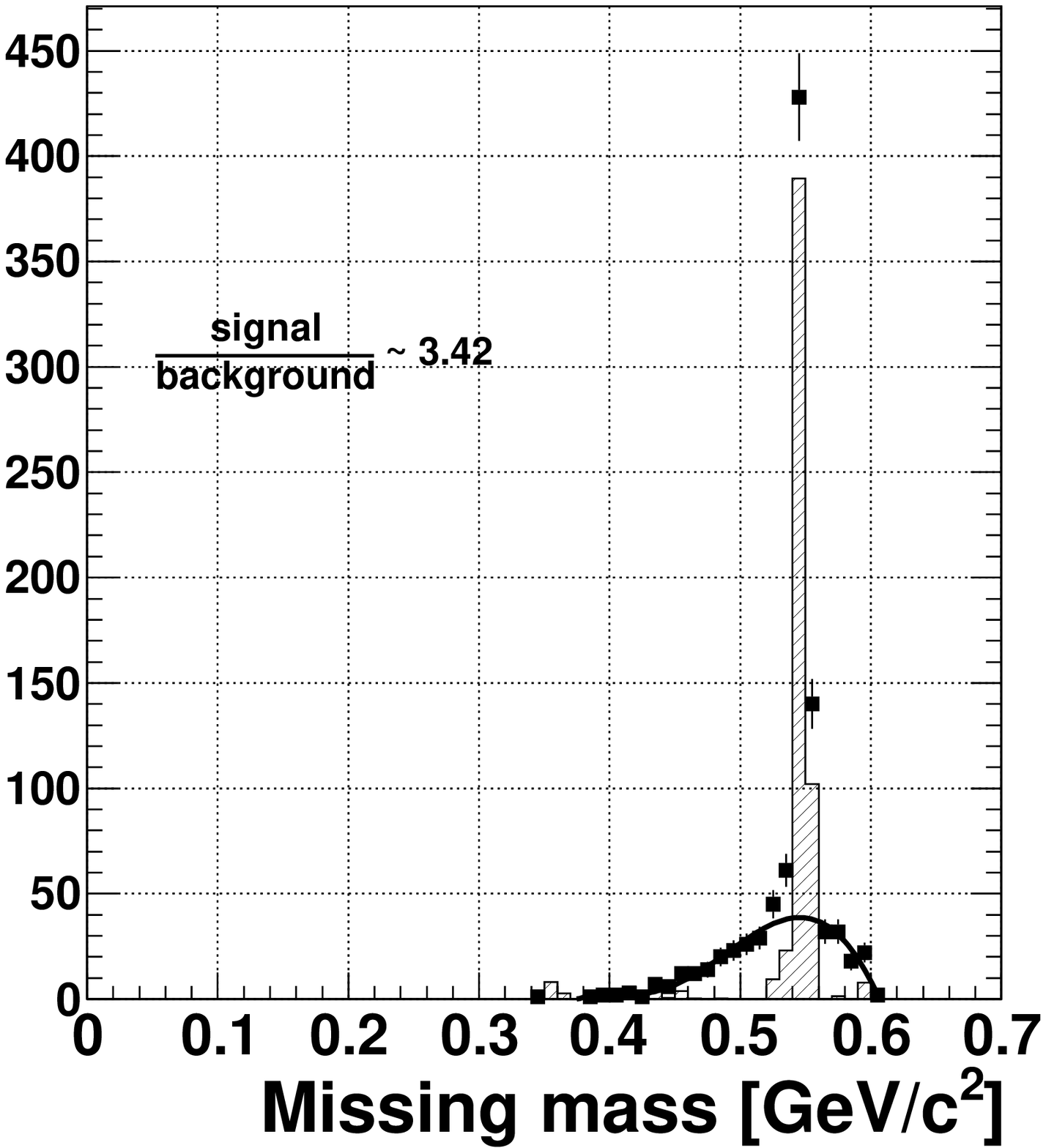}
  }
  \caption{Experimental distributions of the missing mass for the $pd \to {}^{3}He X$ reaction. Solid lines show the fit of the 5th order polynomial to the continuous pion background consisting of \Hpp{} and \Hppp{} reactions. Dashed histograms correspond to data after subtraction of this background.}
  \label{fig:MissMasses}
\end{figure}

It can be compared to \fig{}\ref{fig:d_b4_HeTMeBp} which shows the situation before applying above mentioned conditions. One can notice that the signal become more clear and that there is also less contribution from the \Hpp{} reaction. Corresponding missing mass spectra for the $pd \to {}^{3}He X$ reaction are shown in \fig{}\ref{fig:MissMasses}. The signal to background ratio has increased by $\sim60\%$.

\section{Consistency Check}
\label{sec:ccheck}

\begin{figure}[!b]
\centering
	  \includegraphics[width =0.49\textwidth,height=0.52\textwidth]{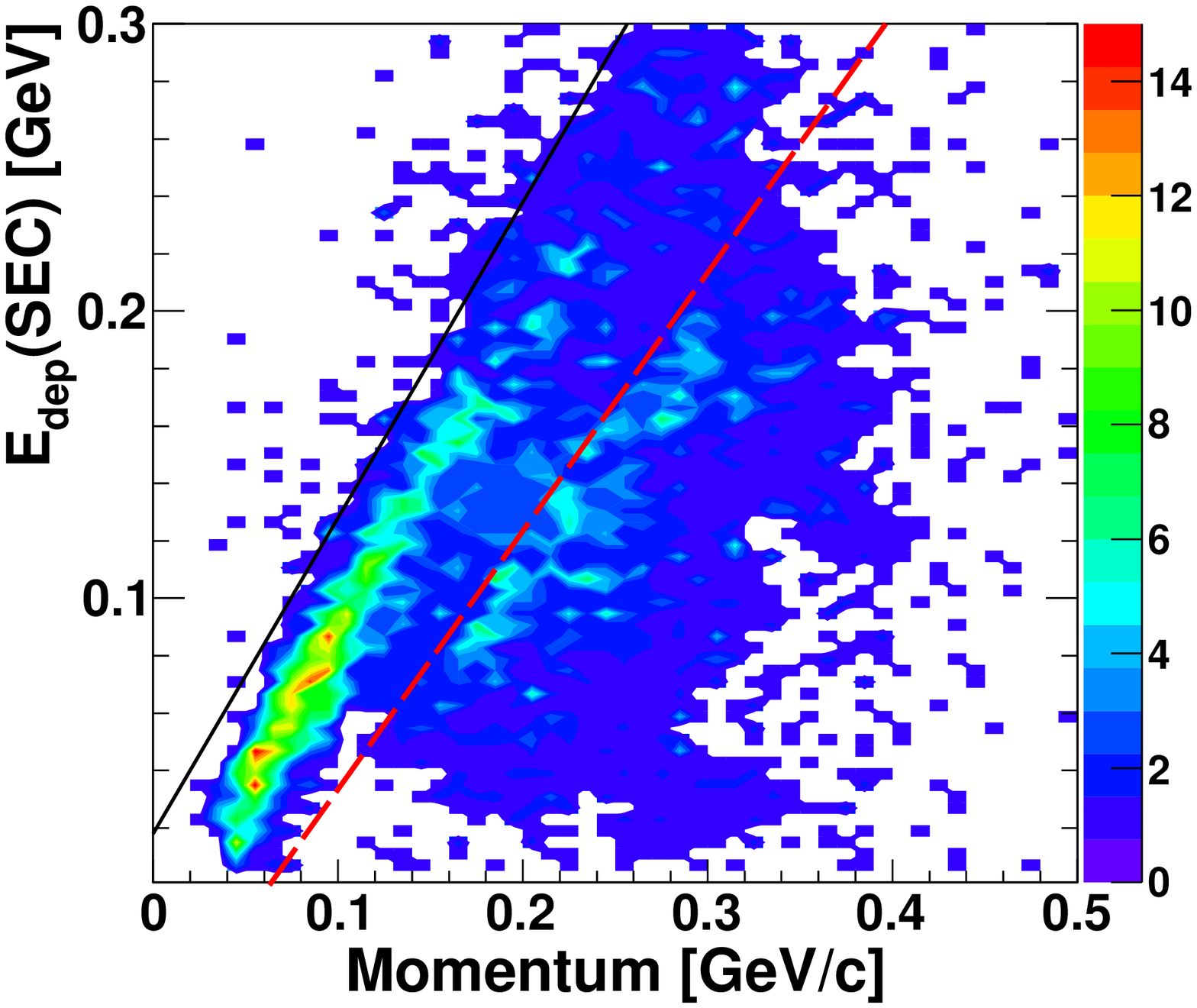}
	  \includegraphics[width =0.49\textwidth,height=0.52\textwidth]{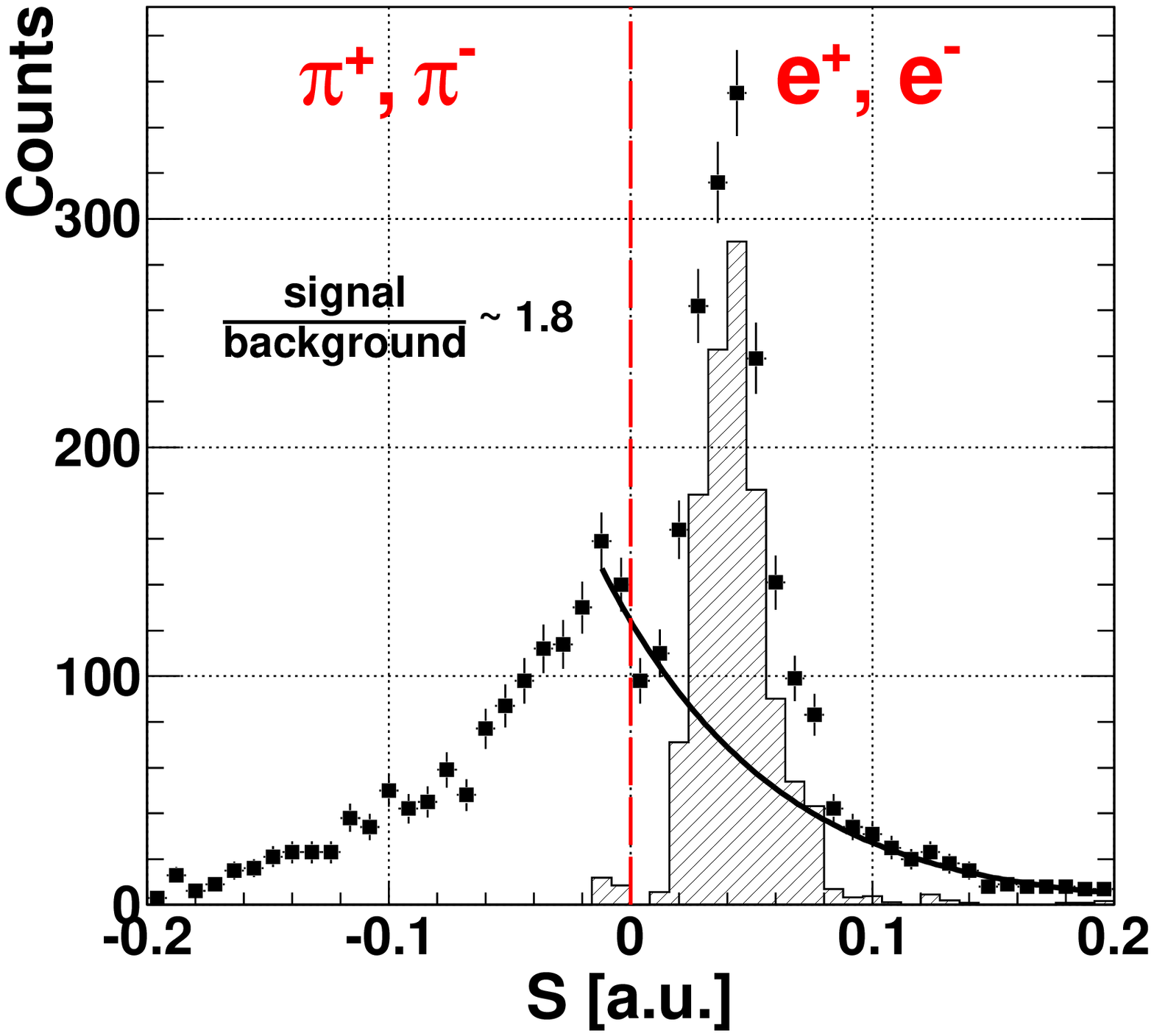}
	\caption[Consistency check in the SEC]{ \underline {Left} : experimental distribution of the energy deposited in the \se{} as a function of the momentum. \underline {Right}: experimental spectrum of the shortest distance, S, between points and the red line shown on the left plot. Points on the left side of the red line are assigned to the positive value of S and these on the right side to negative. The red line indicates the zero position. The black line indicates the fit of the 4th order polynomial function to the background and the dashed histogram is the signal after the background subtraction.}
	\label{fig:SEodl}
\end{figure}

It is necessary to check identification criteria after applying all the other cuts and to prove whether the areas of the occurrence of leptons on the identification plots were chosen properly. In the left panel of \fig{}\ref{fig:SEodl} the experimental distribution of the energy deposited in the \se{} as a function of the momentum is shown.

Plot was made after applying conditions described in previous sections (only identification in the \se{} was omitted). 
Additionally, only for the purpose of this check, in order to suppress contribution from direct pion production, cut on the missing mass was applied. 
The missing mass for the $pd\rightarrow{}^{3}HeX$ reaction was restricted to the range from $0.535$ to $0.56$\um{} (see \fig{}\ref{fig:MissMasses}). 

The shortest distance from each point on the spectrum to the solid line, is presented in the right panel of \fig{}\ref{fig:SEodl}. The peak centered at about $0.04$ corresponds to the leptons. In order to estimate the amount of misidentified pions, the signal to background ratio was calculated in the range of $S \in [0.008;0.08]$, with a resulting value of $\sim1.8$.

\outroformatting

 \section{Estimation of the Background Contribution}
\label{s:back}
\introformatting
To estimate the quality of the background suppression, a set of reactions have been simulated and studied. 
They are listed in \tab{}\ref{t:simul}, in the second column of which, the number of generated and analyzed events is given.
\begin{table}[!h]
  \centering
  \begin{tabular}{ l | c }
  \hline	\hline
  \multirow{2}{*}{Simulated background channels} & {Number of generated }      \\
                &   events $\, \times \, 10^6$	  \\ \hline	\hline \\
  \regg{}   	&	50  \\[0.4em]
  \retp{}   	&	43  \\[0.4em]
  \reppp{}	 	&	30  \\[0.4em]
  \reppg{} 		&  $~$7  \\ \\
  \hline	\hline
  \end{tabular}
  \caption[]{ A list of reactions which have been simulated and studied as a background for the \reac{} reaction.
 The relative number of generated events corresponds very 
roughly to the ratios of branching ratios of the studied channels (see \tab{}\ref{t:etaBR}).}
  \label{t:simul}
\end{table}
Each of \e{} decays may cause a signal contamination.

\begin{figure}[!hbt]
  \begin{tabular}{ p{1cm} p{8cm} p{0.1cm} p{2.5cm} }

  \multicolumn{4}{c}{{\centering \includegraphics[width =0.65\textwidth]{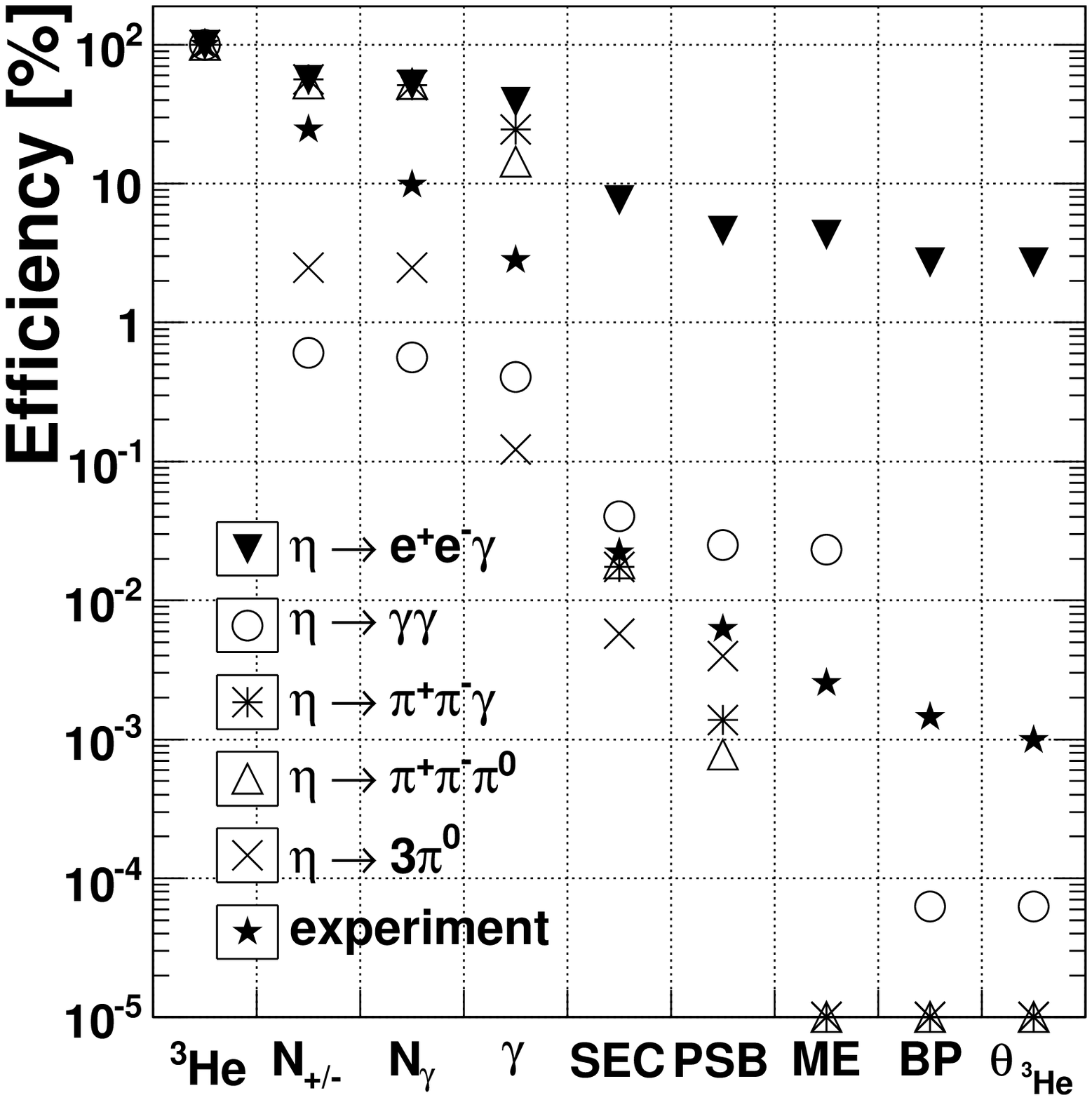} }}\\
  \multicolumn{4}{l}{Explanation of the x-axis abbreviations:}\\
  \multicolumn{4}{c}{}\\
\rowcolor{gray!40}
\multirow{2}{*}{${}^{3}He $} & selection of the \he{} in the \fd{}, &&  described in Sec.~\ref{sec:RecFD} \\
\rowcolor{gray!40}
   & time correlations between particles registered in the \cd{} and in the \fd{} && see \fig{}\ref{fig:LepTime} and \fig{}\ref{fig:TimeG} \\
$N_{+/-}$   & reconstruction of tracks corresponding to two particles with opposite charges &&   \\
\rowcolor{gray!40}
$N_{\gamma}$       &  requirement of at least one neutral particle with ${E>0.02}$~GeV &&   \\
$\gamma$           &  selection of the photon &&  described in Sec.~\ref{sec:IDgam}\\
\rowcolor{gray!40}
$SEC$              &  particle identification using the \se{} &&  see \fig{}\ref{fig:SEodl}\\
$PSB$              &  particle identification using the \psb{} &&  see \fig{}\ref{fig:dIdentPSB}\\
\rowcolor{gray!40}
$ME$               &  cut on the missing mass for the ${pd\rightarrow X\eta}$ reaction && [2.66,2.84]~GeV \\
$BP$               &  suppression of photons conversion && see \fig{}\ref{fig:d_Bp} \\
\rowcolor{gray!40}
$\theta_{{}^{3}He}$ & cut on the scattering angle of the \he{} && $\theta_{{}^{3}He}<11^{\circ}$  \\
  \end{tabular}
  \caption{Reconstruction efficiency for different \e{} decay channels after applying subsequent cuts.}
  \label{f:mcA}
\end{figure}

First, a study of the efficiency for the reconstruction of the \e{} decay channels under different 
conditions had been made as shown in \fig{}\ref{f:mcA}. 
The cuts were added subsequently to the analysis, one after the other, 
and the calculated percentages of events left, were plotted in corresponding bins named after 
\clearpage
the last added condition. The abbreviations used to describe the x-axis are explained below the plot. The last condition on the \he{} scattering angle is used for the reduction of the background from the direct pions production and here it is plotted to check if it doesn't influence the \e{} decay channels.
After applying all conditions, the final efficiency for the \eeeg{} decay is equal to $\sim 2.7\%$ while for the background channels the efficiency is negligible.
A non-zero efficiency is noted for the \gaga{} and \etp{} decay channels only.

\begin{figure}[!h]
 \begin{center}
\includegraphics[width =0.77\textwidth]{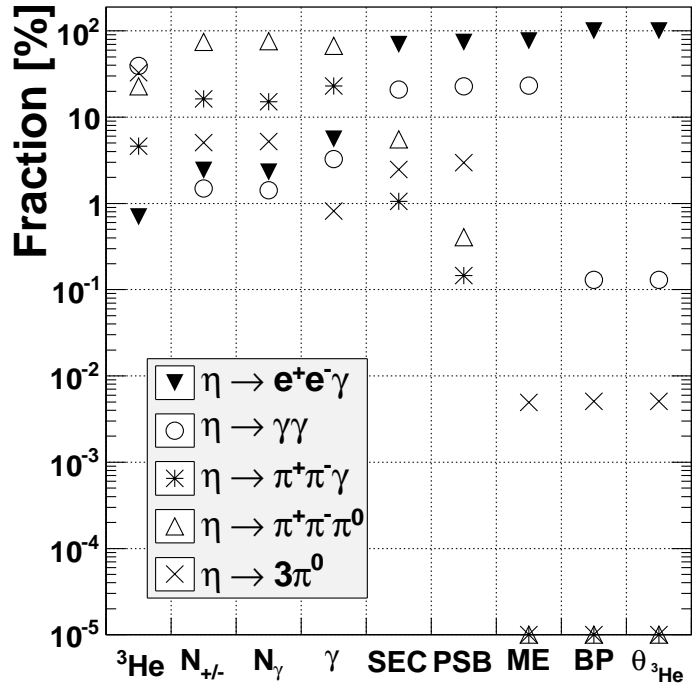} 
  \caption{Percentage share of different \e{} decay channels in the total number of reconstructed \eeeg{} events after applying subsequent cuts, obtained from the simulations. The cuts abbreviations are given in \fig{}\ref{f:mcA}. }
  \label{f:frac}
\end{center}
\end{figure}

One can notice that the efficiency for the signal channel decreases suddenly after application of the identification in the \se{}. There is however no damage in the percentage share of this channel, in the total number of \e{} mesons, shown in the \fig{}\ref{f:frac}. 
Moreover, a constant increase in the amount of the \eeeg{} events in the relation to other channels is observed. 

After applying in the analysis all conditions described above, almost $100\%$ of all events, reconstructed from studied \e{} decay channels, come from the \eeeg{} channel and the only significant background, comes from the direct $\pi$ meson production. This background creates a continuous shape over a wide range of the missing mass for the $pd \to {}^{3}He X$ reaction and can be easily subtracted using a polynomial function as was shown in \fig{}\ref{fig:MissMasses}.

\outroformatting

\outroformatting

\chaptertitle{Results}
\introformatting
\graphicspath{{Results/Rysunki/}}

\section{Calculation of the Transition Form Factor}
\label{s:ff}
\introformatting

\label{s:ff}

\begin{figure}[!h]
  \centering
	\includegraphics[width =0.55\textwidth]{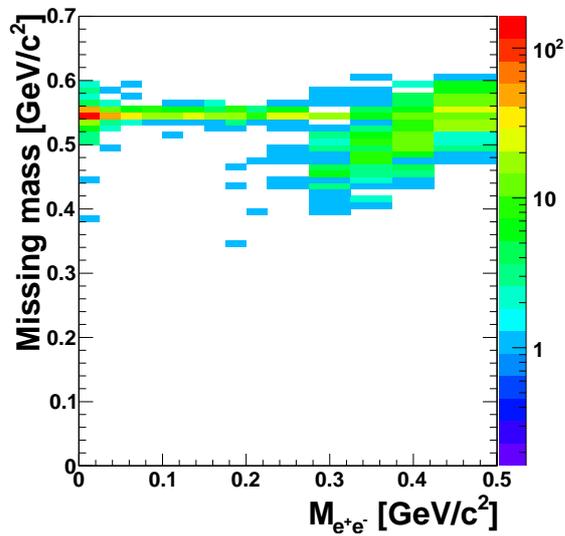}
	\caption{ Experimental distribution of the missing mass for the $pd \to {}^{3}He X$ reaction as a function of the invariant mass of the \ee{} pair after applying event selection described in previous chapters.}
	\label{fig:h_deeg}
  \end{figure}

The final spectrum of the missing mass for the $pd \to {}^{3}He X$ reaction as a function of the invariant mass of the \ee{} pair is shown in \fig{}\ref{fig:h_deeg}. 
\begin{figure}[!h]
  \begin{tabular}{r@{\begin{sideways}{$\xrightarrow{\hspace*{18cm}}$}\end{sideways}}l}

  \begin{sideways}\hspace{16.5cm} Counts\end{sideways}& \includegraphics[width =0.9\textwidth, height=18cm]{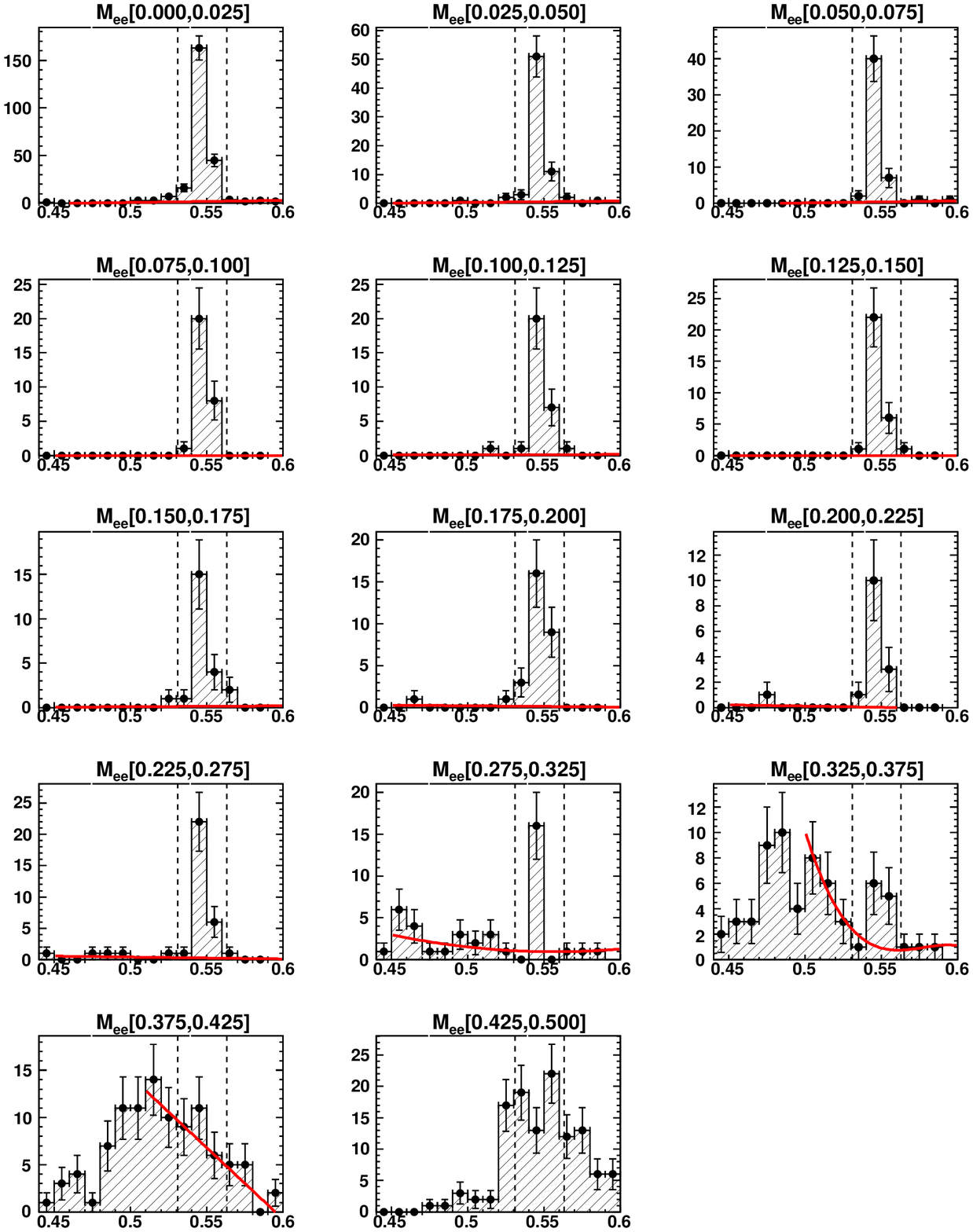}\\[-12pt]

  \multicolumn{2}{c}{$\xrightarrow{\hspace*{12.5cm}}$}\\[0pt]

  \multicolumn{2}{r}{Missing mass [$GeV/c^{2}$]}

  \end{tabular}
  \caption{Experimental distributions of the missing mass for the $pd \to {}^{3}He X$ reaction for fourteen \mee{} intervals ( x-bins of the distribution shown in \fig{}\ref{fig:h_deeg} ). With dashed lines, the signal region is marked. }
  \label{fig:FitBcg}
\end{figure}
\clearpage
In order to subtract the background coming from direct pion production, the missing mass for the $pd \to {}^{3}He X$ reaction was determined for each \mee{} interval separately.
The result is shown in \fig{}\ref{fig:FitBcg}.

The background was fitted with the first order polynomial, omitting the signal region of ${[0.531,0.563]}$\um{}. In the region of \mee{}$\in{[0.275,0.375]}$\um{} the background is not flat and was fitted with a higher order polynomial. 
In the range of \mee{} from $0.425$\um{} to $0.500$\um{} no signal is observed. Therefore, this interval of \mee{} has been excluded from further analysis.  

\begin{figure}[!h]
  \centering
	\includegraphics[width =0.65\textwidth]{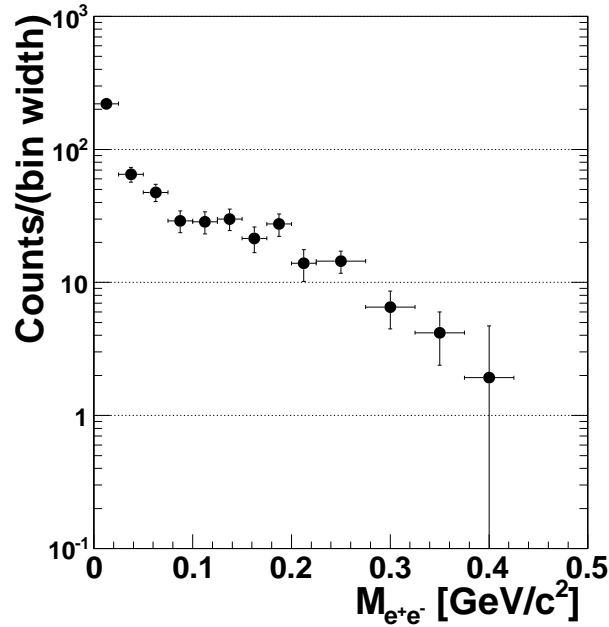}
	\caption{Experimental spectrum of the \mee{} distribution after the background subtraction.}
	\label{fig:dNdM}
\end{figure}

The \mee{} distribution, plotted after the background subtraction is shown in \fig{}\ref{fig:dNdM}. The obtained number of \eeeg{} events amounts to $525 \pm 26$.

\begin{figure}[!t]
    \centering
	\includegraphics[width =0.6\textwidth,height=0.55\textwidth]{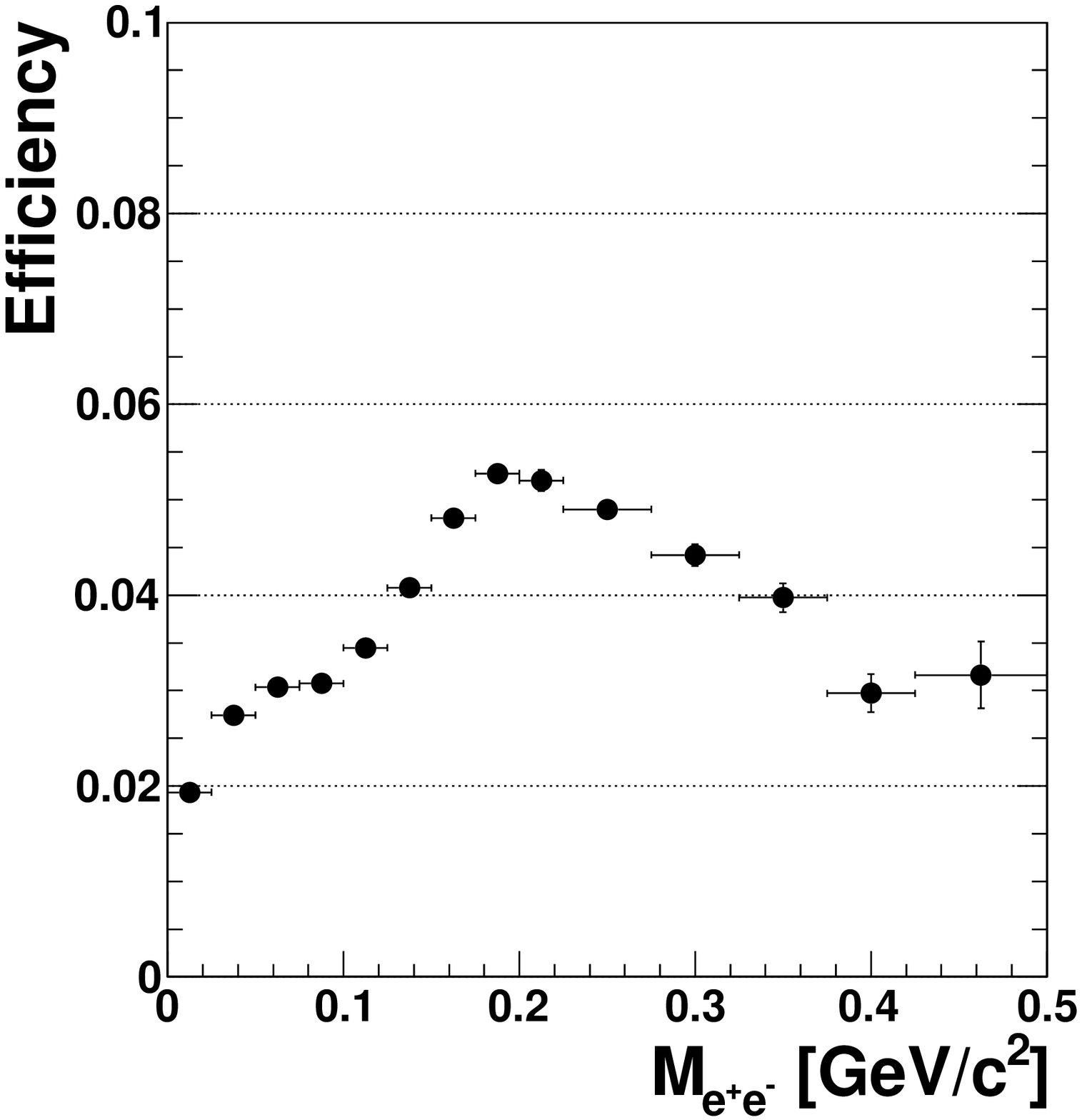}
	\caption{The efficiency for the reconstruction of the \reaction{} reaction as a function of the invariant mass of leptons pairs, \mee{}.}
	\label{fig:Acc}
\vspace*{-0.3cm}
\end{figure}

\begin{figure}[!h]
    \centering
	\includegraphics[width =0.6\textwidth,height=0.55\textwidth]{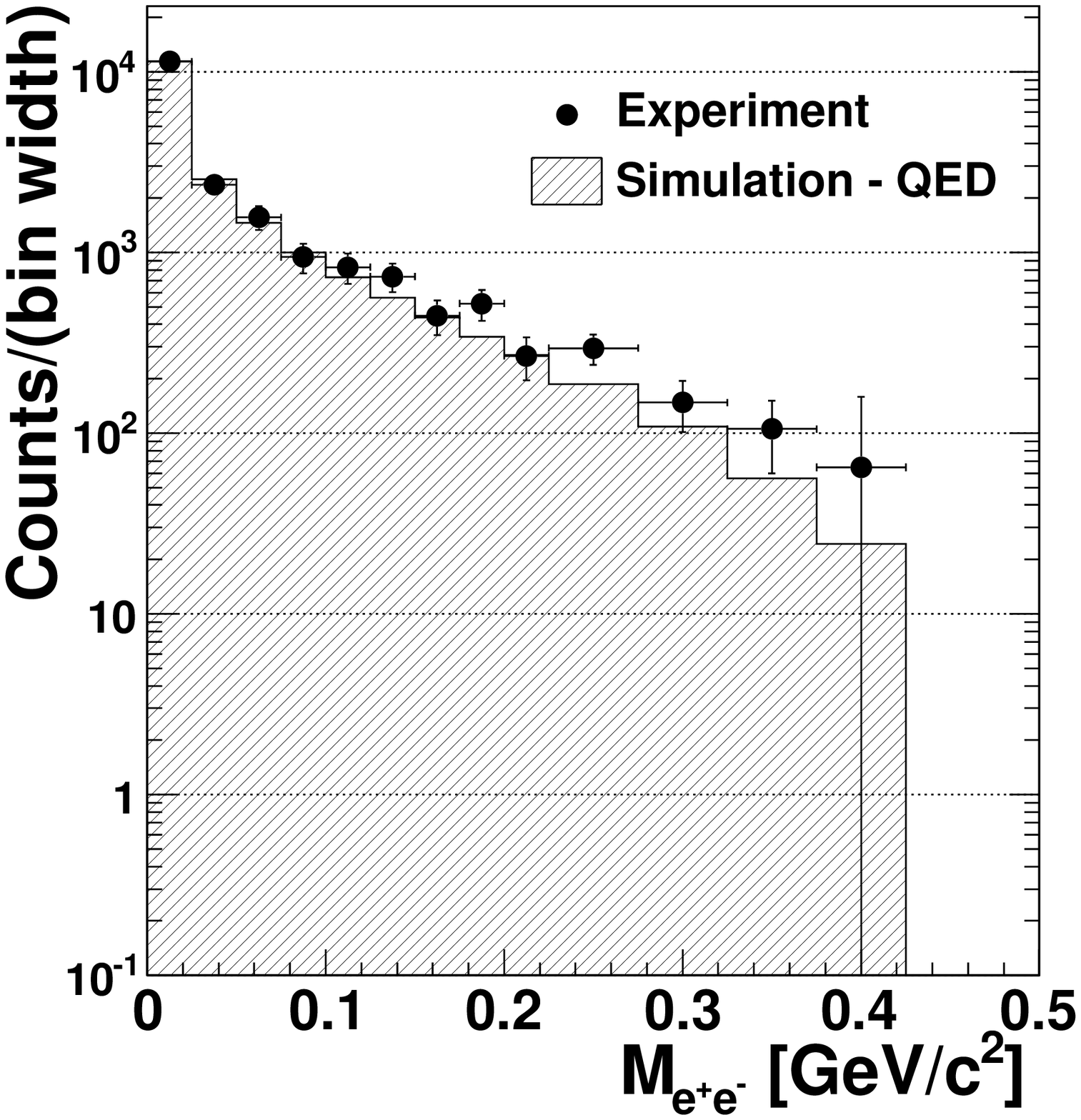}
  \caption{Experimental spectrum of the \mee{} distribution after background subtraction, efficiency corrected. Filled histogram corresponds to efficiency corrected data, simulated with the transition form factor equal to one.}
  \label{fig:dNdMa}
\vspace*{-0.3cm}
\end{figure}

In the next step, the \mee{} distribution was corrected for the reconstruction efficiency. 
Efficiences, calculated per bin of the \mee{}, consist of the trigger efficiency, the reconstruction and selection efficiencies and the geometrical acceptance. The efficiency as a function of the Mee is shown in \fig{}\ref{fig:Acc}.
Each of the \mee{} bin content was divided by the corresponding value of the efficiency. The outcome of this operation is presented in \fig{}\ref{fig:dNdMa}. 

\begin{figure}[!h]
  \centering
	\includegraphics[width =0.65\textwidth]{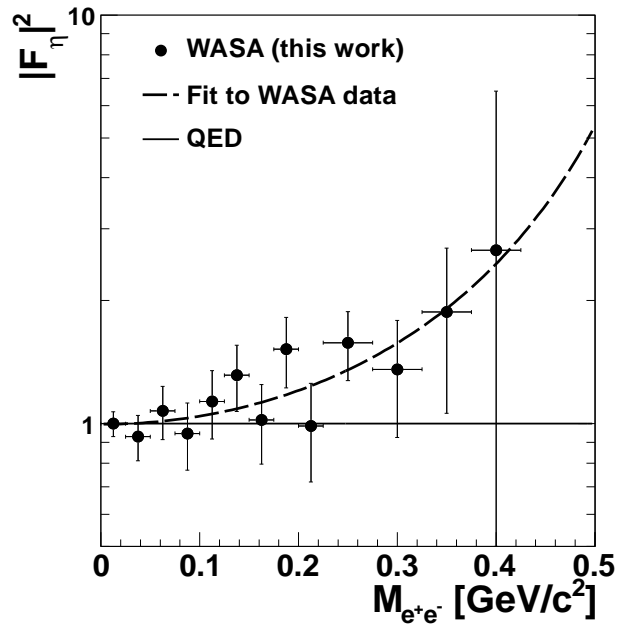}
	\caption{Experimental spectrum of the squared transition form factor, $|F_\eta|^{2}$, as a function of the \mee{}. The dashed line is the result of the fit of the single-pole formula to experimental points with $\chi^2/NDF = 0.43$. The solid line is the QED model assumption of a point-like meson. }
	\label{fig:FF}
\end{figure}

In order to obtain the transition form factor, experimental data points were divided by the Monte Carlo data, simulated with the transition form factor equal to one.
The QED model assumption of a point-like meson, superimposed over the experimental spectrum is shown in \fig{}\ref{fig:dNdMa}. It was normalized to the experimental data points using the $\chi^{2}$ method. The resulting distribution of the transition form factor squared, $|F_\eta|^{2}$, as a function of the invariant mass of the leptons pairs, \mee{} is shown in \fig{}\ref{fig:FF}.

The experimental data points were fitted with the single-pole formula:
	\begin{equation}
		\left[F_{P}(q^{2})\right]^{2} = \left[\alpha\left( 1 - \frac{q^{2}}{\Lambda_{P}^{2}} \right)^{-1}\right]^{2}.
        \label{eq:singlepole}
	\end{equation}
The extracted value of the normalization parameter $\alpha$ is $0.998 \pm 0.025$ whereas the fit parameter $\Lambda_{P}$ amounted to $(0.66 \pm 0.11)$~GeV.
Therefore the slope parameter $b_{P} \equiv 1/\Lambda_{P}^{2}$ is 
\large
	\begin{equation*}
	  {\Large \displaystyle 2.27 \pm 0.73}~\frac{\displaystyle 1 }{\displaystyle \text{GeV}^{2} }.
	\end{equation*}
\normalsize

\outroformatting

\section{Estimation of the Systematic Uncertainty}
\label{s:syst}
\introformatting

\newcommand{\sysSec}{$2.03$}
\newcommand{\sysPsb}{$2.31$}
\newcommand{\sysMx}{$2.02$}
\newcommand{\sysOm}{$2.37$}
\newcommand{\sysBp}{$2.27$}
\newcommand{\sysDp}{$2.41$}
\newcommand{\sysEOm}{$2.33$}
\newcommand{\sysBcg}{$2.33$}
\newcommand{\sysInt}{$2.04$}

To estimate the systematic uncertainty, the slope parameter, \bp{}, has been re-evaluated changing the initial condition of each selection criteria separately.
The systematic error evaluation was done in two ways. First, the systematic error, $\sigma_{syst}$, was calculated as the square root of the quadratic sum of all contributions:
\begin{equation}
\sigma_{syst} = \sqrt{\sum (x_{i} - x_{r})^{2}},
\label{e:ds_trad}
\end{equation}
where ${\it x_{i}}$ is the result obtained by analyzing data with change in the cut condition and ${\it x_{r} = \text{2.27}}$\ub{} 
is the result obtained in the previous section.
The second evaluation of the systematic uncertainty was done based on the method described in \cite{Barlow:2002yb}
and recommended by the WASA-at-COSY collaboration.

\begin{enumerate}
\item {\it Identification of charged particles (SEC)}

Charged particles are identified based on the energy deposited in the \se{} plotted as a function of the momentum. 
The relative distribution is shown in the left panel of \fig{}\ref{fig:SEodl} (page~\pageref{fig:SEodl}). A pair 
of charged particles is treated as leptons if both of them fall into the area limited by the cut lines. Lines can 
be parametrized by a function ${f(E_{dep})={\it A} \times Momentum + {\it B}}$. In order to estimate the influence 
of the identification of leptons, the red, dashed line visible in \fig{}\ref{fig:SEodl}, which separates pions from 
electrons, was moved closer to leptons by change of the {\it B} parameter. The allowed area was decreased by $5\%$.
The obtained value of the slope parameter \bp{} is~\sysSec{}\ub{}. The corresponding transition form factor 
distribution is shown in the left panel of \fig{}\ref{fig:sys_id}.
\begin{figure}[!h]
 \centering
	\includegraphics[width =0.49\textwidth, height =0.5\textwidth]{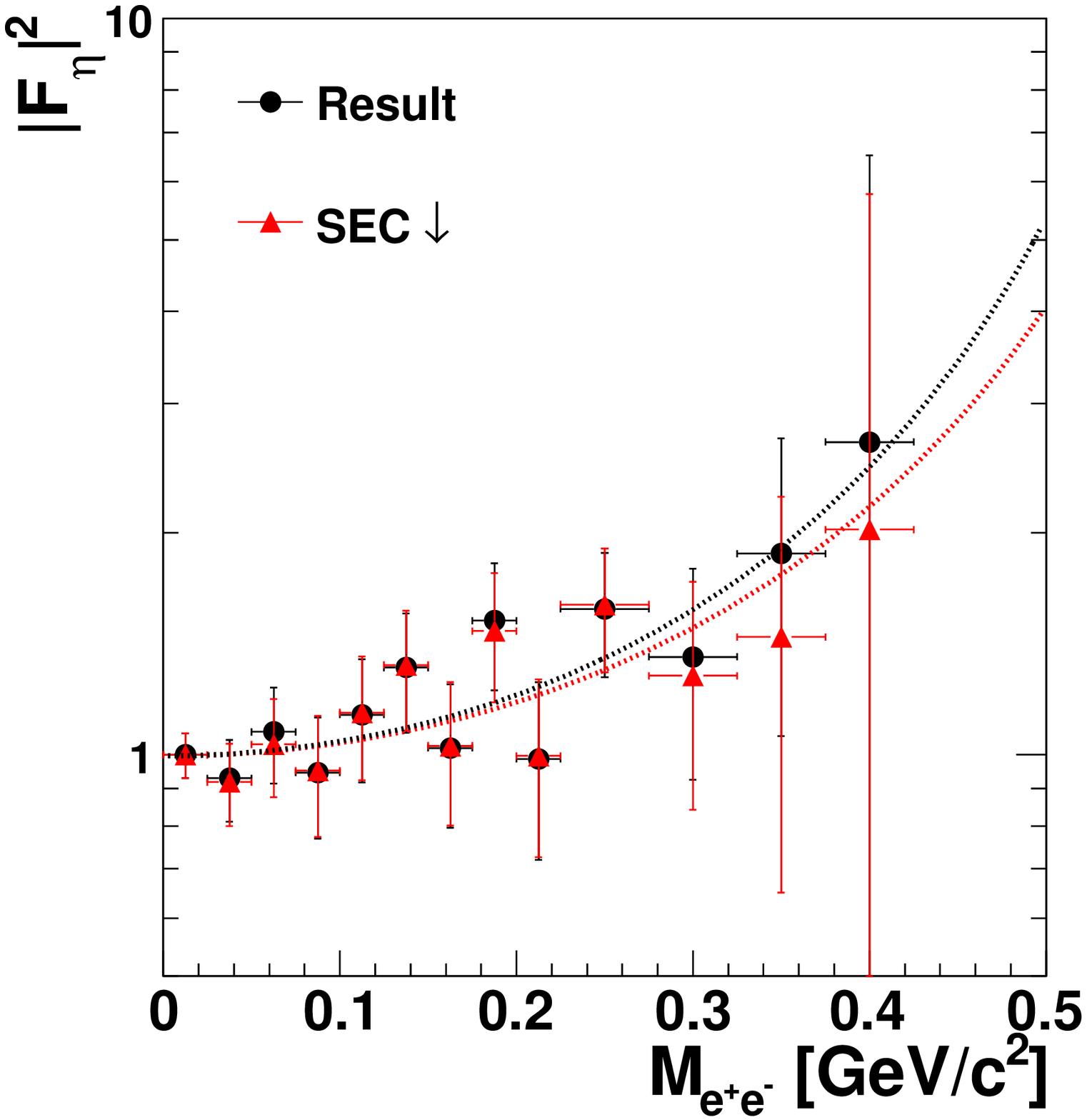}
	\includegraphics[width =0.49\textwidth, height =0.5\textwidth]{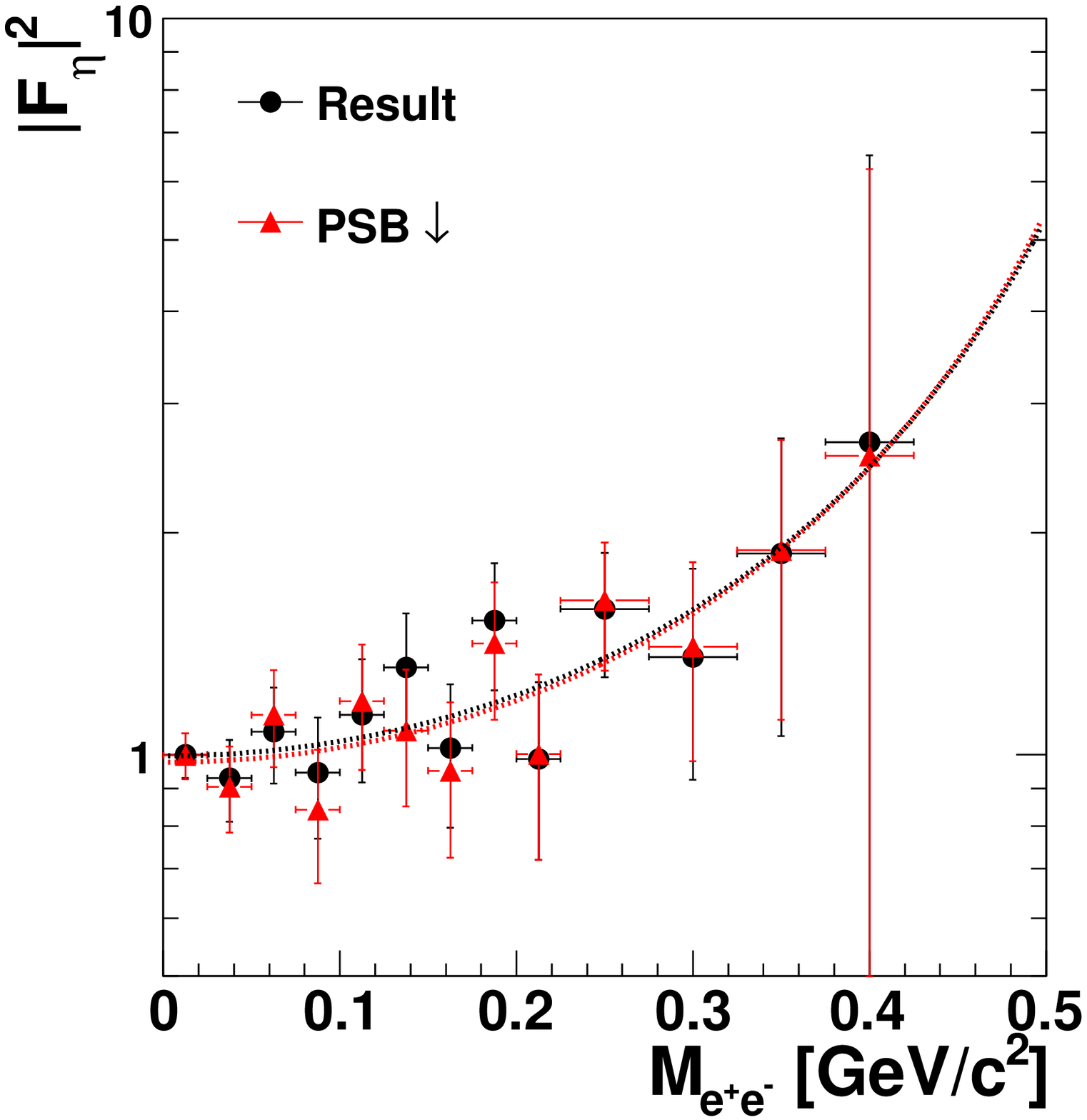}
  \caption{Experimental spectra of the transition form factor as a function of the invariant mass of \ee{} pairs. 
	The distribution obtained in Sec.~\ref{s:ff} (black points) is shown in comparison to distribution obtained 
	by changing the condition on the particle identification in the \se{} (\underline{left}) and in the \psb{} (\underline{right}) as it is described in the text. Additionally, results of two corresponding fits are plotted.}
  \label{fig:sys_id}
\end{figure}

\item {\it Identification of charged particles (PSB)}

The additional method to distinguish between pions and leptons in the \cd{} is to use the dependency of energy 
deposited in the \psb{} from particles momenta (see \fig{}\ref{fig:dIdentPSB}, page~\pageref{fig:dIdentPSB}). 
The originally accepted area shown in this figure was decreased by $5\%$
via the change of the angle of inclination of a cut line. The resulting value of the slope parameter is \bp{}~=~\sysPsb{}\ub{}. 
The corresponding form factor distribution is shown in the right panel of \fig{}\ref{fig:sys_id}.

\item {\it Cut on the missing mass for the $pd\rightarrow X\eta$ reaction, $M_{X}$}

The cut used in the analysis can be seen in \fig{}\ref{fig:d_MissEtaMM} (page~\pageref{fig:d_MissEtaMM}). 
In order to estimate the influence of this cut on the final result, the allowed window was increased by $5\%$ and the analysis was repeated. 
The obtained value of the slope parameter amounted to \bp{}~=~\sysMx{}\ub{}. The corresponding transition form factor distribution is shown in the left panel of \fig{}\ref{fig:sys_me_om}.
\begin{figure}[!h]
\centering
	\includegraphics[width=0.49\textwidth, height =0.5\textwidth]{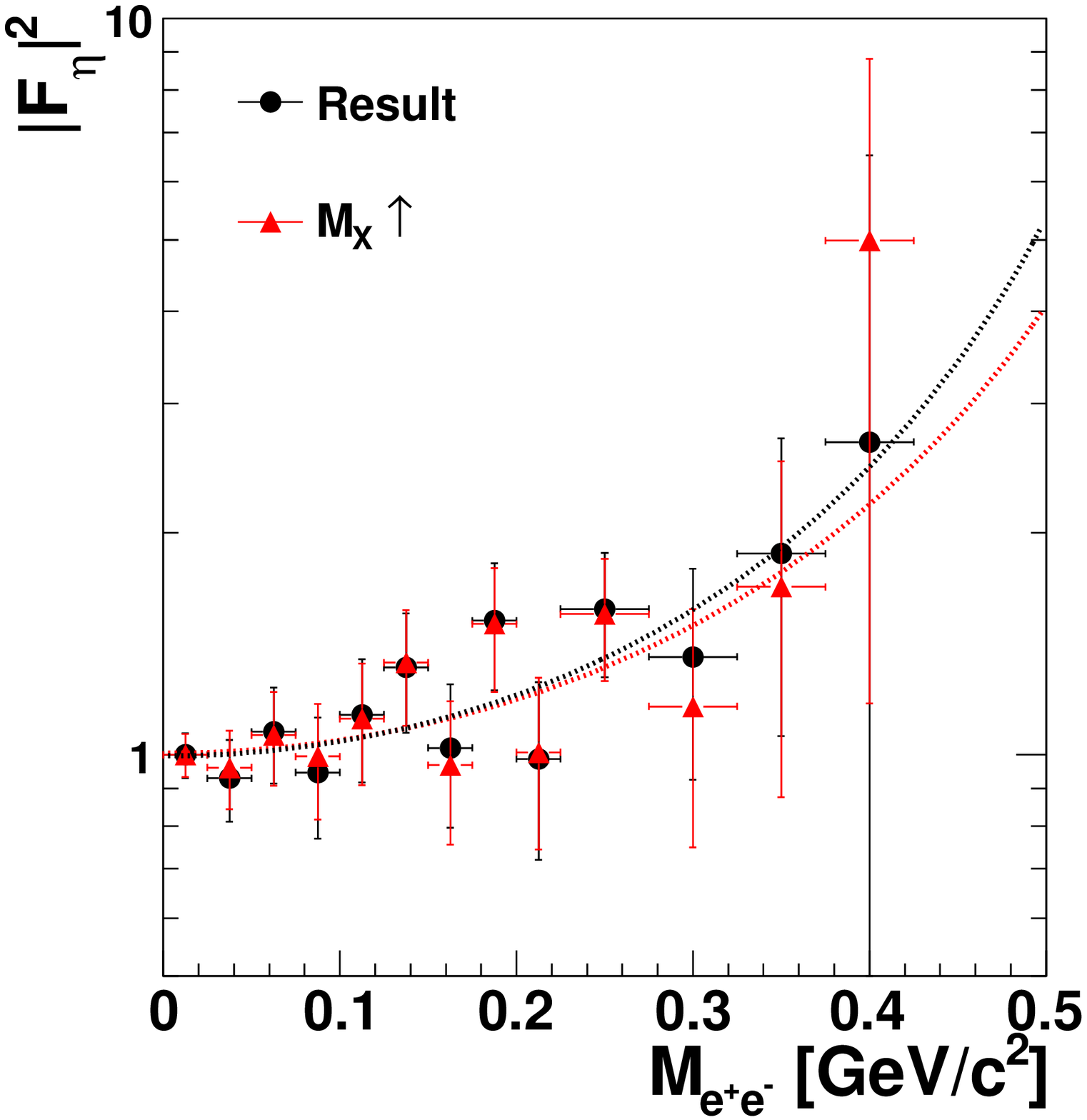}
	\includegraphics[width=0.49\textwidth, height =0.5\textwidth]{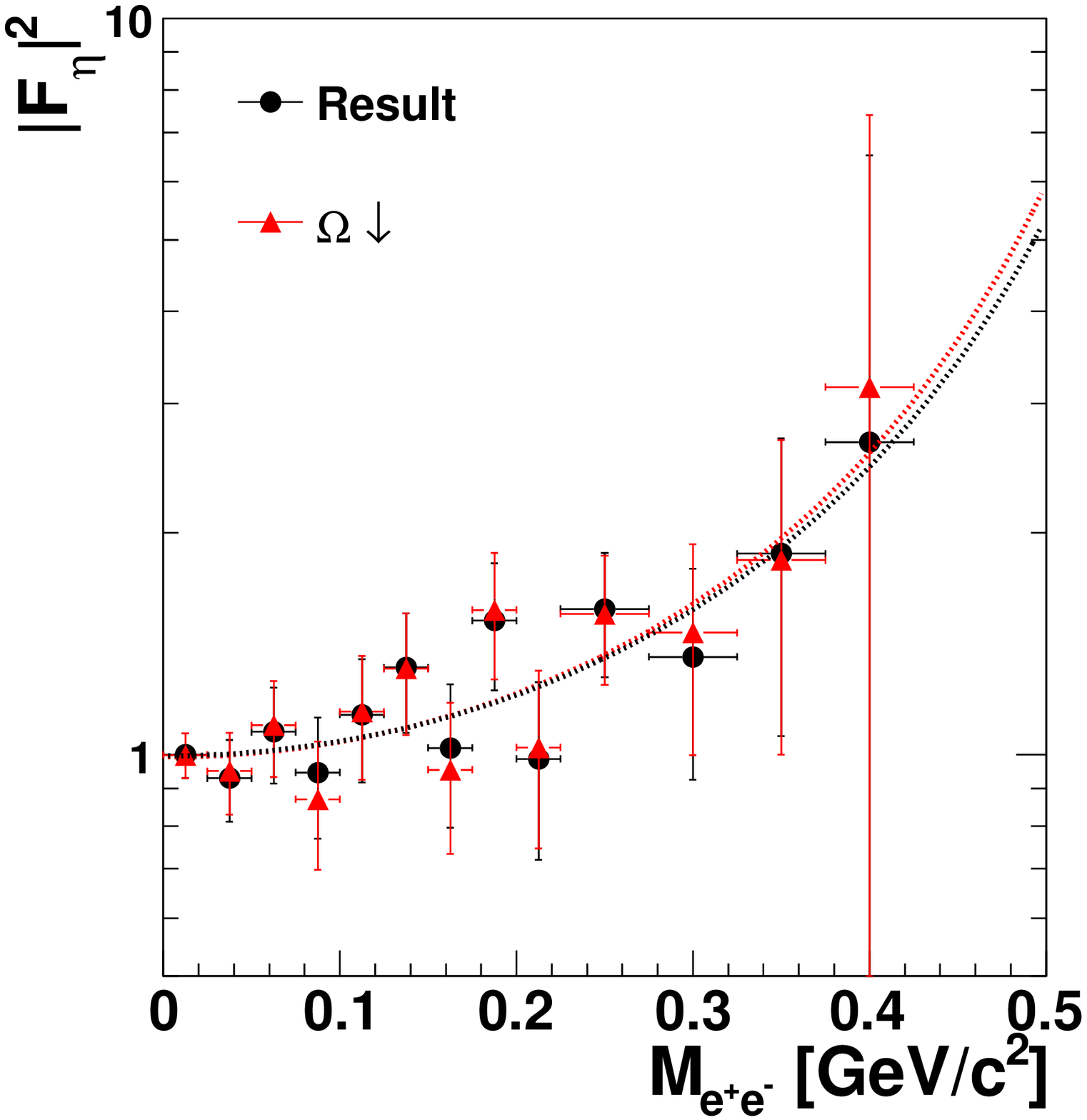}
\caption{Experimental spectra of the transition form factor as a function of the invariant mass of \ee{} pairs. 
	The distribution obtained in Sec.~\ref{s:ff} (black points) is shown in comparison to distribution obtained 
	by changing the condition on the missing mass for the $pd\rightarrow X\eta$ reaction (\underline{left}) and on the angle between neutral and closest charged particle (\underline{right}) as it is described in the text. Additionally, results of two corresponding fits are plotted.}
\label{fig:sys_me_om}
\end{figure}

\item {\it Cut on the angle between neutral and closest charged particle, $\Omega$}

In the analysis the minimal accepted angle $\Omega$ amounts to $60^{\circ}$ as depicted in \fig{}\ref{fig:SpOff} 
(page~\pageref{fig:SpOff}). The analysis was done also for the minimal $\Omega$ angle increased by $5\%$ resulting in the slope parameter 
\bp{}~=~\sysOm{}\ub{}. The corresponding transition form factor distribution is shown in the right panel of \fig{}\ref{fig:sys_me_om}.

\item {\it Cut suppressing photons conversion}

For the condition used in the analysis to minimize number of conversion events, the allowed area (see \fig{}\ref{fig:d_Bp}, 
page~\pageref{fig:d_Bp}) was increased by $5\%$. The resulting value of the slope parameter amounted to \bp{}~$= 2.24$\ub{}.

\item {\it Cut on the opening angle between virtual and real photon in the \e{} rest frame, $\Delta\phi_{\gamma\gamma*}$}

The $\Delta\phi_{\gamma\gamma*}$ distributions are shown in \fig{}\ref{fig:dP} (page~\pageref{fig:dP}). 
The original restriction of $\Delta\phi_{\gamma\gamma*}$ being in the range of $[70,290]$ degrees seems quite
loose and might give an impression that many right events are rejected due to the presence of fake signals caused 
by detector's noise and split-offs. To check that, the accepted cut window was decreased symmetrically by $20\%$ 
via change of the allowed range to $[92,268]$ degrees. The resulting slope parameter \bp{} amounts to  \sysDp{}\ub{}. 
Its deviation from the original result is small relative to the change of the cut made in this test.
It implicates that effects caused by fake signals have been already suppressed by former restrictions, especially by the cut 
on the photons energy as a function of the angle between the neutral and the closest charged particle, $E_{\gamma}(\Omega$). 

\item {\it Cut on photons energy, $E_{\gamma}(\Omega)$}

Cut on photons energy is shown in the same \fig{}\ref{fig:SpOff}. The minimal photons energy is chosen as a function 
of the angle between neutral and closest charged particle, $\Omega$. 
The influence of this condition on the final result was studied via decrease of accepted area as shown in the left panel of \fig{}\ref{fig:sys_Eo}.
The resulting value of the slope parameter amounted to \bp{}~=~\sysEOm{}\ub{}. The corresponding transition form 
factor distribution is shown in the right panel of \fig{}\ref{fig:sys_Eo}.
\begin{figure}[!t]
  \centering
	\includegraphics[width =0.45\textwidth,height=0.5\textwidth]{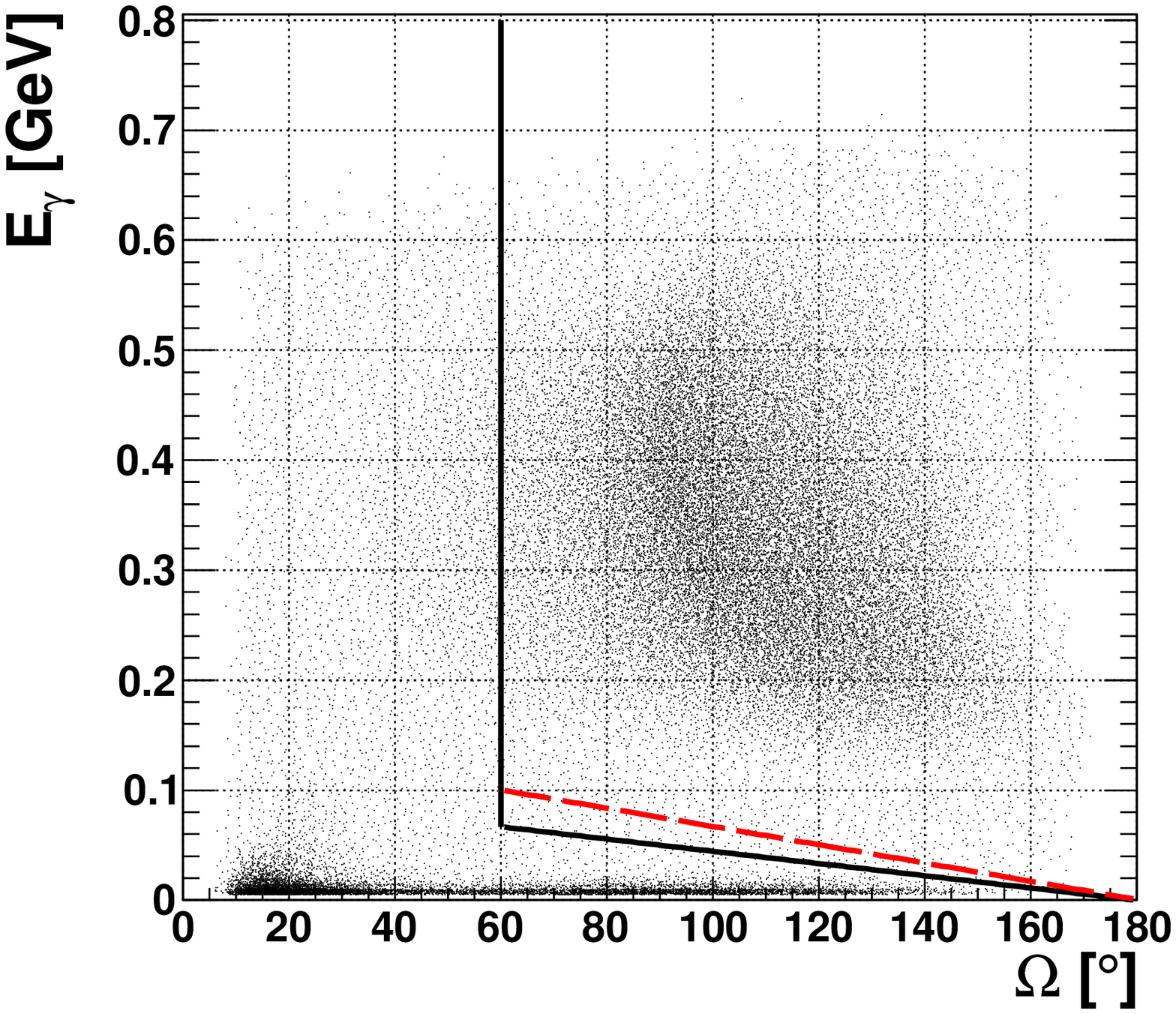}
	\includegraphics[width =0.45\textwidth,height=0.5\textwidth]{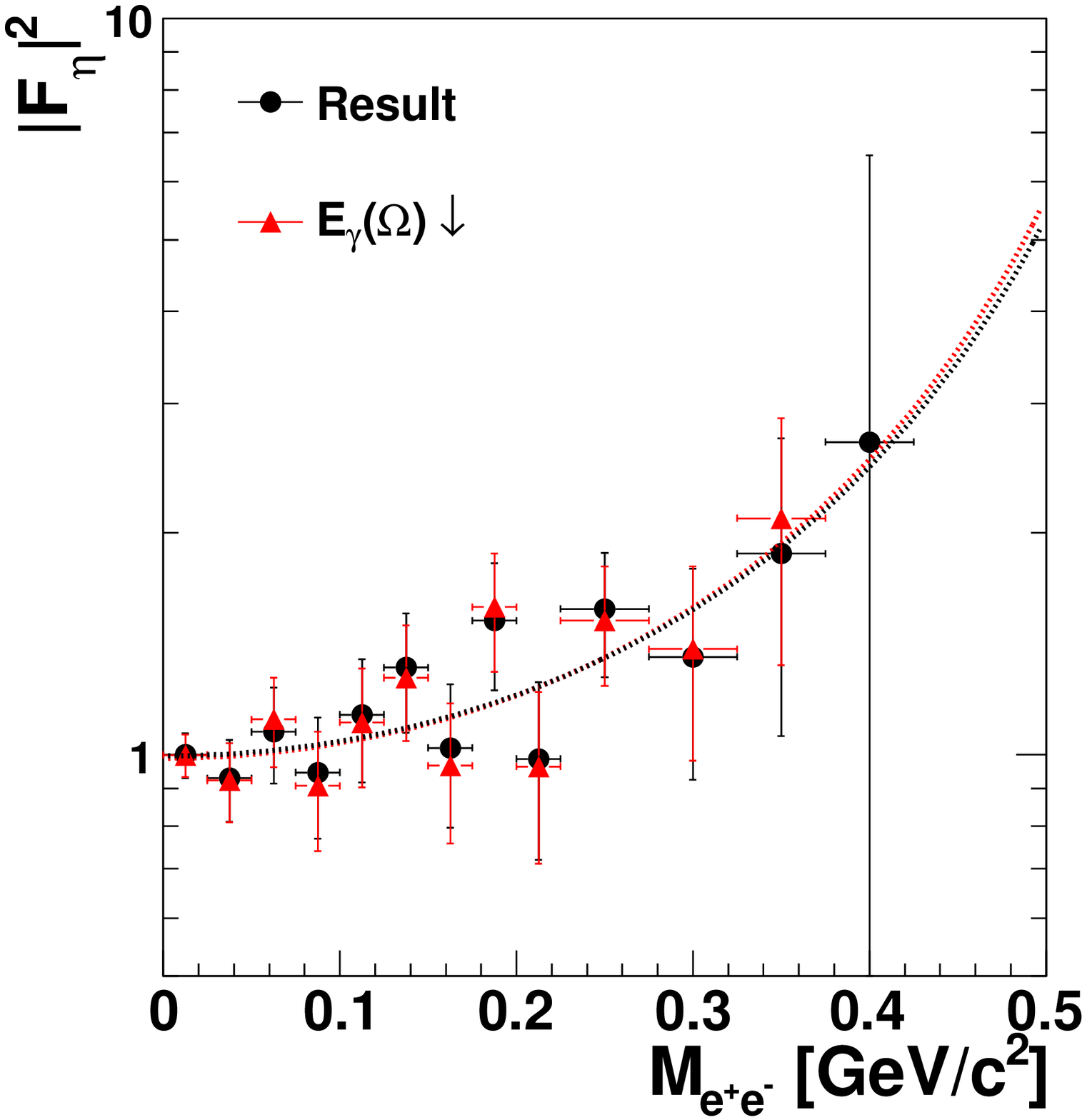}
	\caption{ \underline{Left:} Simulated spectrum of cluster's energy vs. the angle it creates with the nearest track for the \reac{} reaction. The red, dashed line shows the new position of the original cut marked by the solid, black line. \underline{Right:} Experimental spectrum of the transition form factor as a function of the invariant mass of \ee{} pairs. 
	The distribution obtained in Sec.~\ref{s:ff} (black points) is shown in comparison to distribution obtained by changing the condition on the $E_{\gamma}(\Omega)$ as  described in the text. Additionally, results of two corresponding fits are plotted.}
	\label{fig:sys_Eo}
  \end{figure}

\item {\it Shape of the multipion background}

The direct production of pions results in a continuous missing mass distribution. It can be described with a polynomial function and subtracted from the signal. 
In order to check how the choice of the fitting function influences the result, the order of the given polynomial was changed and the slope parameter was recalculated.
The resulting value is \bp{}~=~\sysBcg{}\ub{}.

\item {\it Size of the \mee{} bin intervals}

In order to check how the change of the binning of the \mee{} distribution changes the result, the size of the \mee{} bin intervals was set to $0.04$~MeV and the full analysis including multipion background subtraction was repeated.
With this new binning, the obtained value of the slope parameter amounted to \bp{}~=~\sysInt{}\ub{}.

\end{enumerate}

\begin{center}{\it Summary}\end{center}
\renewcommand{\arraystretch}{1.5}
The results of analyses are gathered in \tab{}\ref{t:syst}. In the last column, the difference between the result presented in the previous section 
and the value obtained with a different cut condition, is shown. 
\begin{table}[!h]
  \centering
  \begin{tabular}{| l |c | c|}
\hline	
\multicolumn{1}{|c|}{\multirow{2}{*}{Cut}}& \multirow{1}{*}{\bp{}[\ub{}]} & \multirow{1}{*}{$\Delta$\bp{}[\ub{}]}\\
&(${\it x_{i}}$)   & $|{\it x_{i} - x_{r}}|$\\	\hline		\hline

\multirow{1}{*}{Identification in the SEC} 	 &	\sysSec{} &       0.24   	\\ \hline
\multirow{1}{*}{Identification in the PSB} 	 &	\sysPsb{} &       0.04   	\\ \hline
\multirow{1}{*}{Missing mass for the
 $pd\rightarrow X\eta$ reaction} 		 &	\sysMx{}  &       0.25    \\ \hline
\multirow{1}{*}{$\Omega$} 			 &	\sysOm{}  &       0.10    \\ \hline
\multirow{1}{*}{Photons conversion} 		 &	\sysBp{}  &       0.00    \\ \hline
\multirow{1}{*}{$\Delta\phi_{\gamma\gamma*}$} 	 &	\sysDp{}  &       0.14    \\ \hline
\multirow{1}{*}{$E_{\gamma}(\Omega$)} 		 &	\sysEOm{} &       0.06    \\ \hline
\multirow{1}{*}{Shape of the multipion 
 background} 		 			 &	\sysBcg{} &       0.06    \\ \hline
\multirow{1}{*}{\mee{} intervals size} 		 &	\sysInt{} &       0.23    \\ \hline	
  \end{tabular}
  \caption{Values of the slope parameter, \bp{}, obtained as a result of analyses performed after change in a given cut parameter. 
	Details are given in the text.}
  \label{t:syst}
\end{table}

The value of the systematic uncertainty calculated using Eq.~\ref{e:ds_trad} amounts to:
\begin{equation*}
\sigma_{syst} = 0.46~ \text{GeV}^{-2}.
\end{equation*}

It is important to stress that all obtained values of systematic uncertainty
agree within one standard deviation of the statistical uncertainty. Therefore,
this preliminary estimation of the systematic error can be treated as conservative
taking into account performed tests.

In order to evaluate the systematic uncertainty as described in \cite{Barlow:2002yb},
the deviation of the original result from the one obtained after the change, denoted as $\Delta$\bp{},
is compared with $\Delta\sigma$, defined as:
\begin{equation}
\Delta\sigma = \sqrt{ \sigma_r^2 - \sigma_i^2 },
\end{equation}
where $\sigma_r$ and $\sigma_i$ denote statistical uncertainties of the $x_r$ and $x_i$ 
determinations respectively.\\

\begin{table}[!h]
  \centering
  \begin{tabular}{| l |c | c| c| c| c|}
\hline
 & \bp{} & $\sigma$ & $\Delta$\bp{} & $\Delta\sigma$ & $\frac{\Delta b_P}{\Delta\sigma}$ \\ \hline
\multirow{1}{*}{Result obtained in Sec.~\ref{s:ff}}   & 2.27  	& 0.73 & 	    &		&			\\ \hline
\multicolumn{1}{|c|}{\multirow{1}{*}{Cut}}	 &	\multicolumn{5}{c|}{}			\\ \hline
\multirow{1}{*}{Identification in the SEC} 	 &	\sysSec{}&       0.78    &       0.24    &       0.27    &       0.87 	\\ \hline
\multirow{1}{*}{Identification in the PSB} 	 &	\sysPsb{}&       0.72    &       0.04    &       0.12    &       0.33	\\ \hline
\multirow{1}{*}{Missing mass for the
 $pd\rightarrow X\eta$ reaction} 		 &	\sysMx{} &       0.78    &       0.25    &       0.27    &       0.91	\\ \hline
\multirow{1}{*}{$\Omega$} 			 &	\sysOm{} &       0.74    &       0.10    &       0.12    &       0.82 	\\ \hline
\multirow{1}{*}{Photons conversion} 		 &	\sysBp{} &       0.73    &       0.00    &       0.00    &       0.00	\\ \hline
\multirow{1}{*}{$\Delta\phi_{\gamma\gamma*}$} 	 &	\sysDp{} &       0.69    &       0.14    &       0.24    &       0.59	\\ \hline
\multirow{1}{*}{$E_{\gamma}(\Omega$)} 		 &	\sysEOm{}&       0.71    &       0.06    &       0.17    &       0.35	\\ \hline
\multirow{1}{*}{Shape of the multipion 
 background} 		 			 &	\sysBcg{}&       0.75    &       0.06    &       0.17    &       0.35 	\\ \hline
\multirow{1}{*}{\mee{} intervals size} 		 &	\sysInt{}&       0.81    &       0.23    &       0.35    &       0.66	\\ \hline	
  \end{tabular}
  \caption{Values of the slope parameter, \bp{}, obtained as a result of analyses performed after change in a given cut parameters. 
	Details are given in the text.}
  \label{t:systB}
\end{table}
\tab{}\ref{t:systB} summarizes the results of performed systematic checks for the slope parameter \bp{}, 
their deviations from the original result, $\Delta$\bp{}, and their correlations $\Delta$\bp{}$/\Delta\sigma$.
All performed checks give a non-significant deviation which manifest itself in $\Delta$\bp{}$/\Delta\sigma$ less than one. This indicates 
that the systematic error can be neglected.

To sum up foregoing tests,
the upper limit of the systematic uncertainty evaluation
can be accepted as $\sigma_{syst} = 0.46~ \text{GeV}^{-2}$ which is less
than the statistical error $\sigma_{stat} = 0.73~ \text{GeV}^{-2}$ obtained in Sec.~\ref{s:ff}.
It should be also noted, that although as many as nine possible
sources of systematic effects have been tested, there are still
checks which may be done in further studies. These are e.g. checks for the possible
systematic effects resulting from the accepted uncertainty of the \mee{} efficiency.

\outroformatting

\section{Charge Radius of the \e{} Meson}
\label{s:radius}
\begin{figure}[!h]
  \centering
	\includegraphics[width =0.7\textwidth]{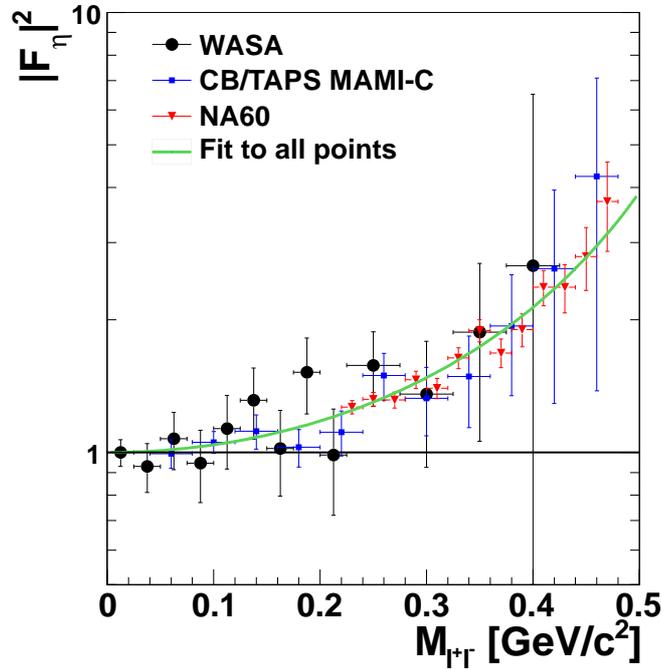}
	\caption{Experimental spectrum of the squared transition form factor, $|F_\eta|^{2}$, as a function of the $M_{l^{+}l^{-}}$. 
The green, solid line is the fit to all experimental points. The black, solid line is the QED model assumption of a point-like meson. }
	\label{fig:FFall}
\end{figure}

\introformatting
Having the slope parameter of the \e{} meson, one can attempt to evaluate its charge radius.
The charge distribution is related to the form factor by the Fourier transform 
\begin{equation}
 F(q^{2}) = \int d^{3}r \rho( {\bf r} )e^{-i {\bf q} \cdot {\bf r}} \simeq 1 - \frac{q^{2}}{6}<r^{2}> + ...,
\end{equation}
where $<r^{2}>/6 = b_{P}$. In order to obtain the charge radius of the \e{} meson, the most recent measurements of the \e{} 
transition form factor (this work, \cite{Berghauser:2011zz} and \cite{Arnaldi:2009wb}) have been gathered in \fig{}\ref{fig:FFall} 
and fitted with the single-pole formula given by Eq.~\ref{eq:singlepole} in the range of $M_{l^{+}l^{-}}$ from $0.$\um{} to $0.48$\um{}. 

The obtained value of the slope parameter amounts to
	\begin{equation*}
	  {\displaystyle 1.97 \pm 0.09}~\frac{\displaystyle 1 }{\displaystyle \text{GeV}^{2} }
	\end{equation*}
and therefore, the charge radius of the \e{} meson is equal to
	\begin{equation*}
	  {<r_{\eta}^{2}>^{1/2} = 0.68 \pm 0.02 ~{\text{fm}}}.
	\end{equation*}
Theoretical calculations within the framework of the VMD model (See \tab{}\ref{t:FF}) give the value of the charge radius 
$<r_{\eta}^{2}>^{1/2} = 0.64$~fm, which deviates from the experimental one, by more than two standard deviations.

Interestingly, similar situation has been already observed for pion. Precise calculations based on the chiral perturbation 
theory and data, give the radius of the pion meson equal to $<r_{\pi}^{2}>^{1/2} = (0.672 \pm 0.010)$~fm \cite{Bijnens:2002hp} 
while the measurement at MAMI-II led to the pion radius of $<r_{\pi}^{2}>^{1/2} = (0.74 \pm 0.03)$~fm \cite{Liesenfeld:1999mv}. 
The discrepancy is in the order of two standard deviation and to minimize it, it was proposed in \cite{Bernard:2000qz} to include 
corrections to the pion loops at the order where the radius appears.

Also in the case of a proton, the recently obtained result by Pohl et al. \cite{Pohl}, ${<r_{P}^{2}>^{1/2} = (0.84184 \pm 0.00067)}$~fm, 
is five standard deviations away from the one of the CODATA compilation of physical constants, $<r_{P}^{2}>^{1/2} = (0.8768 \pm 0.0069)$~fm 
\cite{Mohr:2008fa}. The CODATA values of the radius of the proton are determined via the Lamb shift in electronic \cite{Mohr:2008fa} 
hydrogen and via unpolarized \cite{Bernauer:2010wm} and polarized \cite{Zhan:2011ji} electron scattering. The discrepancy between the 
CODATA and the value extracted using the Lamb shift method in muonic \cite{Pohl} hydrogen is under a world-wide discussion with a tendency 
to see the cause of the problem in a not sufficiently exact QED calculations \cite{Distler:2010zq}.

\outroformatting

\outroformatting

\chaptertitle{Summary and Outlook}
\introformatting
\graphicspath{{SumAndOut/}}
At the turn of October and November 2008, the WASA-at-COSY collaboration performed an experiment
to collect data on the \e{} meson production and decays via the \pdhe{} reaction. The COSY facility provided a proton beam of momentum $1.7$~GeV/c which has been used to produce \e{} mesons by collisions with deuteron target. Decay products of short-lived meson
were registered in the central part of the WASA detector and \he{} ions in the forward part.
In this work, the conversion decay \eeeg{} has been investigated. It is a very interesting decay since the two electrons in the final state come from the conversion of a virtual $\gamma$ quantum and, therefore, they constitute a rich source of knowledge about the electromagnetic structure of decaying meson. 
It is a very important feature of this decay process since the \e{} meson is a short-lived neutral particle and it is not possible to investigate its structure via the classical method of particle scattering. 

The performed analysis allowed for the extraction of the \e{} transition form factor as a function of the \ee{} mass 
and, therefore, for the calculation of the slope parameter, related to the charge radius of the \e{} meson.

\begin{figure}[!t]
  \centering
	\includegraphics[width =0.48\textwidth,height=0.5\textwidth]{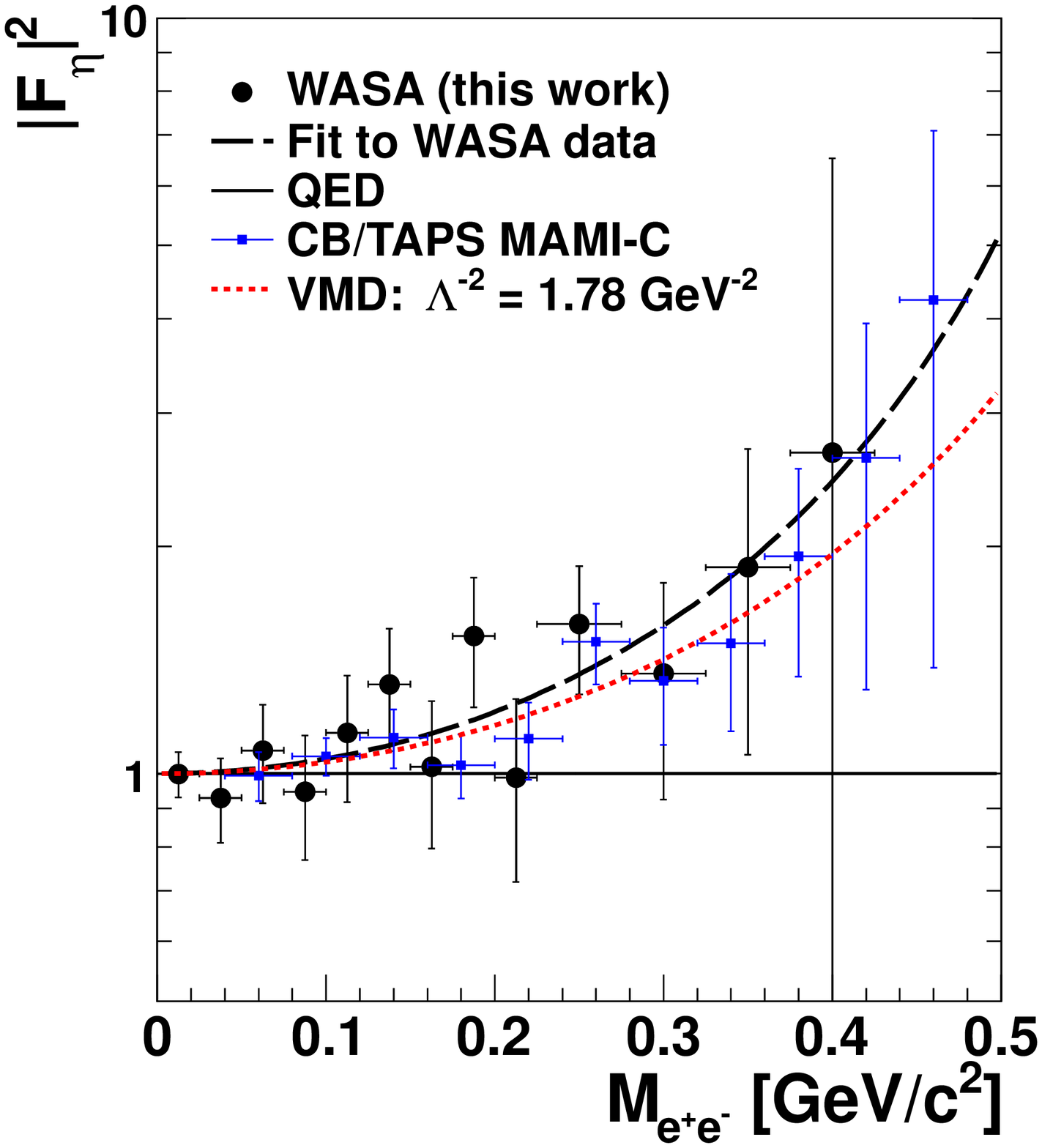}
	\includegraphics[width =0.48\textwidth,height=0.5\textwidth]{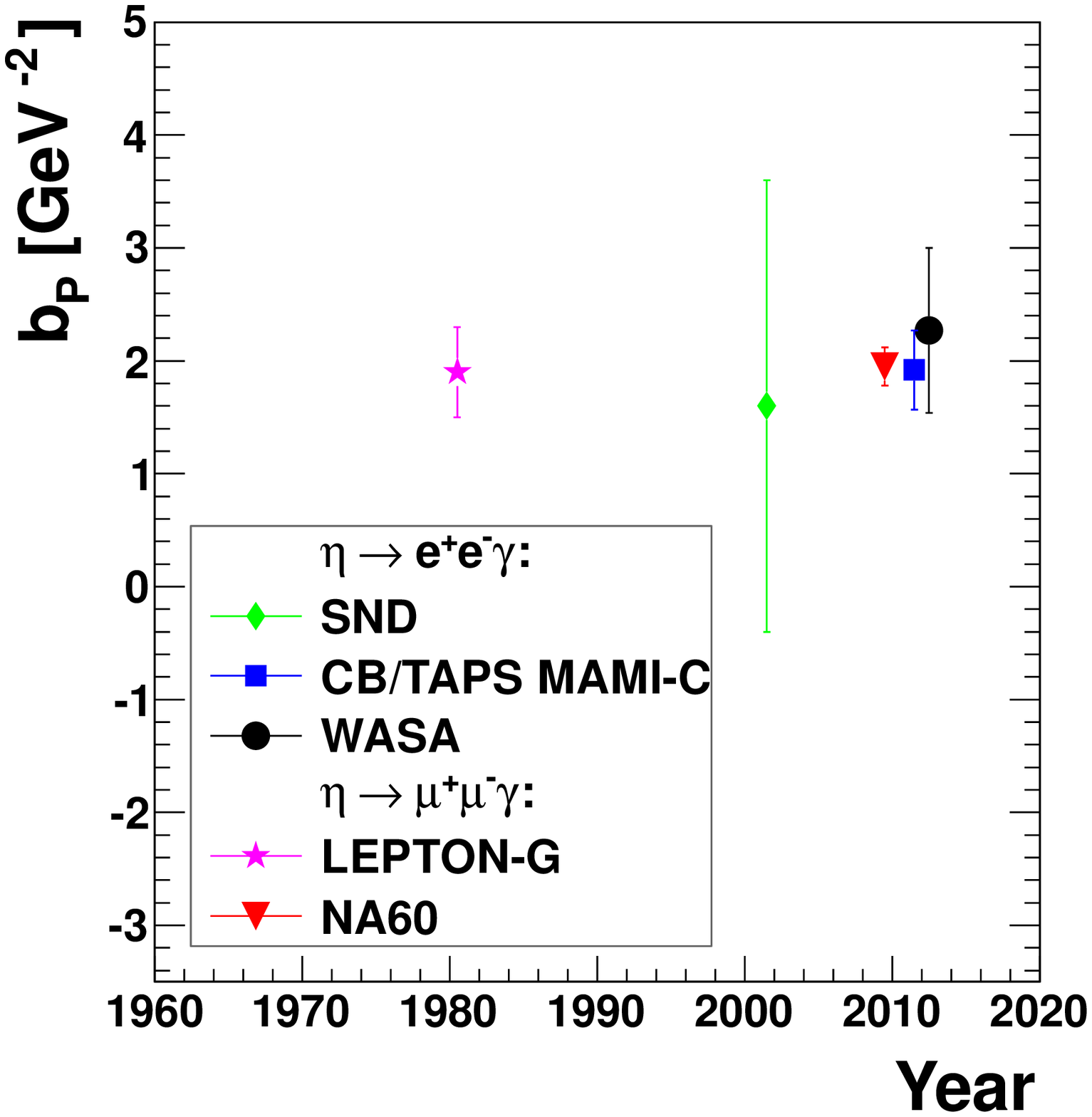}
	\caption{\underline {Left} : Experimental spectrum of the squared transition form factor, $|F_\eta|^{2}$, as a function of the $M_{e^{+}e^{-}}$ obtained in this work (black points) and by the CB/TAPS experiment (blue points). The black, solid line shows the QED calculations for a point-like meson while the red, dashed line is the prediction of the VMD model. \underline {Right} : The slope, \bp{}, of the \e{} transition form factor. The result of this work is shown in comparison with previous results. }
	\label{fig:FFwasa_taps}
\end{figure}

$525 \pm 26$ events of the \eeeg{} decay channel were reconstructed.
The applied restrictions allowed to suppress the background from the other \e{} decays to a negligible level and the multipion 
background was subtracted from the signal, based on missing mass distributions.
The analysis chain led to the determination of the value of the slope parameter, 
\bp, equal to (${2.27 \pm 0.73_\text{stat.} \pm 0.46_\text{sys.}}$)\ub{}, where the systematical uncertainty should be treated
as an upper limit only \cite{Barlow:2002yb}. This result is consistent with the one obtained by the CB/TAPS collaboration, ($1.92 \pm 0.35_\text{stat.} \pm 0.13_\text{sys.}$)\ub{} \cite{Berghauser:2011zz}. 
Distributions of the transition form factor  as a function of the  $M_{e^{+}e^{-}}$ mass extracted in both experiments are shown in the left panel of \fig{}\ref{fig:FFwasa_taps}.
Within the statistical uncertainty, the transition form factor distribution confirms the calculations of the Vector Meson Dominance (VMD) model \cite{Ametller:1991jv}. 
The result obtained in this work is also in agreement with the one obtained using the $\eta \to \mu^{+}\mu^{-}(\gamma)$ decays, studied
in the heavy ion experiment NA60, \bp=($1.95 \pm 0.17_\text{stat.} \pm 0.05_\text{sys.}$)\ub{}, in which photons were not registered \cite{Arnaldi:2009wb}. 

The three results mentioned above, enabled to estimate the charge radius of the \e{} meson: ${<r_{\eta}^{2}>^{1/2} = (0.68 \pm 0.02) ~{\text{fm}}}$. 
This value is in disagreement with the theoretical one ($<r_{\eta}^{2}>^{1/2} = 0.64$~fm), calculated using the slope parameter given in \cite{Ametller:1991jv},  
at the level of two standard deviations. It is also interesting to notice that the radius of the \e{} meson is smaller than the radius of the pion of
$<r_{\pi}^{2}>^{1/2} = (0.74 \pm 0.03)$~fm \cite{Liesenfeld:1999mv}.

It was shown that WASA-at-COSY is a suitable tool to study the \e{} transition form factor via the \eeeg{} decay with a negligible background from other \e{} decay channels. Therefore, it is tempting, in this context, to analyze the next data sample with much higher statistics. 
\outroformatting

\newpage
\begin{flushright}
\end{flushright}
\cleardoublepage
\newpage
\thispagestyle{empty}
\chapter*{Acknowledgments}
\introformatting
I would like to express my highest gratitude to my supervisor, Prof.~dr~hab. Pawel Moskal. 
Without his guidance, his knowledge and a wonderful attitude to science, this thesis
wouldn't come into existence. 	

I would like to thank Prof.~dr~hab.~Bogusław Kamys for the possibility to prepare this dissertation
in the Faculty of Physics, Astronomy and Applied Computer Science of the Jagiellonian
University and to Prof.~dr~James~Ritman for a great opportunity to work in the 
Institute f\"{u}r Kernphysik in Research Center J\"{u}lich.

I express my highest gratitude to WASA-at-COSY collaborators who made this work possible. 
I am especially grateful to Andrzej Kupsc, Susan Schadmand, Benedykt Ryszard Jany, Peter Vlasov, 
Christoph Florian Redmer, Leonid Yurev and Michal Janusz for their help provided 
at different stages of my work.

Deep gratitude to Eryk Czerwiński and Steven Bass for the IDEA :-)

I would particularly like to thank my dear c++ expert Pavel for dragging me away from work 
to let mind rest and to charge batteries :-)

Finally, I owe my Family a great debt of gratitude for all the support showed 
during work on this project.

\outroformatting

\newpage
\thispagestyle{empty}
\bibliographystyle{StylBib}
\bibliography{HodanasBiblias}{}

\cleardoublepage

\begin{LARGE}
\chapter*{List of Acronyms}
\label{c:loa}
\addcontentsline{toc}{chapter}{\protect\numberline{}List of Acronyms}
\chaptermark{}
\end{LARGE}
{\footnotesize
  \begin{tabular}{ l  @{ - }l }
	{\bf QED} & Quantum ElectroDynamics\\[0.5em]
	{\bf VMD} & Vector Meson Dominance\\[0.5em]
	{\bf COSY} & COoler SYnchrotron\\[0.5em]
	{\bf WASA} & Wide Angle Shower Apparatus\\[0.5em]
	{\bf CELSIUS} & \begin{tabular}[h]{@{}l@{}}Cooling with Electrons and Storing\\of
					Ions from Uppsala Synchrocyclotron\end{tabular}\\[0.5em]
	{\bf CD} & Central Detector  	\\[0.5em]
	{\bf FD} & Forward Detector  	 \\[0.5em]
	{\bf MDC}& Mini Drift Chamber  	\\[0.5em]
	{\bf PSB}& Plastic Scintillator Barrel  	\\[0.5em]
	{\bf SEC}& Scintillator Electromagnetic Calorimeter  	\\[0.5em]
	{\bf FWC}& Forward Window Counter  	\\[0.5em]
	{\bf FPC}& Forward Proportional Chamber  	\\[0.5em]
	{\bf FTH}& Forward Trigger Hodoscope  	\\[0.5em]
	{\bf FRH}& Forward Range Hodoscope  	\\[0.5em]
	{\bf FVH}& Forward Veto Hodoscope  \\[0.5em]
	{\bf FIFO}& First In First Out  \\[0.5em]
	{\bf CMS}& Center of Mass System\\[0.5em]

  \end{tabular}
}
\newpage
\thispagestyle{empty}
\begin{flushright}
\end{flushright}

\end{document}